\documentclass[a4paper,11pt]{article}
\usepackage[pdftex]{graphicx}
\usepackage[T1]{fontenc}
\usepackage[utf8]{inputenc}
\usepackage[english]{babel}
\usepackage{microtype}
\usepackage{cite}
\usepackage{amsmath}
\usepackage{amssymb}
\usepackage{amsfonts}
\usepackage[nottoc,notlot,notlof]{tocbibind}
\usepackage{upgreek}
\numberwithin{equation}{section}
\allowdisplaybreaks
\usepackage[all]{xy}
\usepackage{graphicx}
\graphicspath{{images/}}
\usepackage{geometry}
\usepackage[toc,page]{appendix}
\usepackage{hyperref}
\usepackage[normalem]{ulem}

\newcommand{\diff}{\mathrm{d}}

\newcommand{\ii}{\mathrm{i}}
\newcommand{\RR}{\bm R}
\newcommand{\rr}{\bm{\rho}}
\newcommand{\FF}{\bm F}
\usepackage{bm}
\usepackage{ragged2e}
\usepackage{appendix}
\usepackage{slashed}
\usepackage{bbold}
\usepackage{cancel}

\begin{document}
\begin{center}
{\bf\LARGE $\mathcal N=2$ AdS$_4$ supergravity, holography and Ward identities} \\
\vskip 8mm
{\bf \large L. Andrianopoli$^{1,2,3}$, B.L. Cerchiai$^{1,2,3,4}$, R. Matrecano$^{1,2}$, O. Miskovic$^5$, \\ R. Noris$^{1,2}$, R. Olea$^6$, L. Ravera$^{1,2}$, M. Trigiante$^{1,2,3}$}
\vskip 5mm
 \end{center}
    {\small $^1$ DISAT, Politecnico di Torino, Corso Duca
    degli Abruzzi 24, I-10129 Turin, Italy\\
    $^{2}$  Istituto Nazionale di
    Fisica Nucleare (INFN) Sezione di Torino, Italy
    \\
   $^{3}$ Arnold - Regge Center, Via P. Giuria 1, 10125 Torino, Italy \\
    $^{4}$ Centro Ricerche ``Enrico Fermi'', Piazza del Viminale, 1, I-00184 Rome, Italy\\
    $^{5}$ Instituto de F\'{\i}sica, Pontificia Universidad Cat\'{o}lica de Valpara\'{\i}so, %\\
    Casilla 4059, Valpara\'{\i}so, Chile\\
    $^{6}$ Departamento de Ciencias F\'{\i}sicas,
    Universidad Andres Bello, %\\
    Sazi\'{e} 2212, Piso 7, Santiago, Chile
    }

\vskip 6mm
\begin{center}
{\small {\bf Abstract}}
\end{center}

We develop in detail the holographic framework for an $\mathcal{N}=2$ pure AdS supergravity model in four dimensions, including all the contributions from the fermionic fields and adopting the Fefferman-Graham parametrization. We work in the first order formalism, where the full superconformal structure can be kept manifest in principle, even if only a part of it is realized as a symmetry on the boundary, while the remainder has a non-linear realization. Our study generalizes the results presented in antecedent literature and includes a general discussion of the gauge-fixing conditions on the bulk fields which yield the asymptotic symmetries at the boundary. We construct the corresponding superconformal currents and show that they satisfy the related Ward identities when the bulk equations of motion are imposed.
Consistency of the holographic setup requires the super-AdS curvatures to vanish at the boundary. This determines, in particular, the expression of the super-Schouten tensor of the boundary theory, which generalizes the purely bosonic Schouten tensor of standard gravity by including gravitini bilinears. The same applies to the superpartner of the super-Schouten tensor, the conformino. Furthermore, the vanishing of the supertorsion poses general constraints on the sources of the three-dimensional boundary conformal field theory  and requires that the super-Schouten tensor is endowed with an antisymmetric part proportional to a gravitino-squared term.

\vskip 6mm
\vfill
\noindent {\small{\it
    E-mail:  \\
{\tt laura.andrianopoli@polito.it}; \\
{\tt bianca.cerchiai@polito.it}; \\
{\tt riccardo.matrecano@polito.it}; \\
{\tt olivera.miskovic@pucv.cl}; \\
{\tt ruggero.noris@polito.it}; \\
{\tt rodrigo.olea@unab.cl}; \\
{\tt lucrezia.ravera@polito.it}; \\
{\tt mario.trigiante@polito.it}.}}
   \eject

\numberwithin{equation}{section}
\renewcommand\baselinestretch{1.2}

\tableofcontents

%%%%%%%%%%%%%%%%%%%%%%%%%%%%%%%%%%%%%%%%%%%

\section{Introduction}\label{AdS/CFT}
Since its inception, the anti-de Sitter (AdS) / Conformal Field Theory (CFT) holographic correspondence \cite{Maldacena:1997re,Gubser:1998bc,Witten:1998qj} has provided an important tool  to investigate the strong coupling regime of  field theories on a fixed background using classical supergravity on asymptotically anti-de Sitter (AAdS)  spacetimes in one dimension higher. This is a powerful framework since, being an intrinsically non-perturbative strong/weak coupling duality, it opens a window on aspects of the gauge theory which are otherwise not accessible.

In its original formulation, the duality was conjectured as a correspondence between the full type IIB superstring theory on its ${\rm AdS}_5\times {\rm S}^5$ solution and $\mathcal{N}=4$ four-dimensional Super Yang-Mills  theory on the boundary of the ${\rm AdS}_5$ spacetime. In the limit in which the classical effective low-energy description of the (super)gravity side can be trusted, the corresponding regime of the dual theory is strongly coupled.
The holographic correspondence has been extended to more general backgrounds of the form ${\rm AdS}_D\times \mathcal{M_{{\rm int}}}$, possibly with less supersymmetry, which can be embedded in other string theories or M-theory, such as  the maximally supersymmetric ${\rm AdS}_4\times {\rm S}^7$ and ${\rm AdS}_7\times {\rm S}^4$ solutions of $D=11$ supergravity and variants thereof. A valuable approach to the study of holography on a background of the form ${\rm AdS}_D\times\mathcal{M_{{\rm int}}}$   is to restrict to an effective $D$-dimensional low-energy supergravity originating from superstring/M-theory compactified on the internal manifold $\mathcal{M_{{\rm int}}}$. This supergravity admits the ${\rm AdS}_D$ part of the higher dimensional background as a vacuum and typically is of \emph{gauged} type. The geometry of  $ \mathcal{M_{{\rm int}}}$ determines the amount of supersymmetry preserved by this ${\rm AdS}_D$ vacuum and the general features of the effective theory. In this setting, the AdS/CFT conjecture can be restated as a holographic relation between the ${\rm AdS}_D$ supergravity and a $d=(D-1)$-dimensional superconformal field theory (SCFT) at the boundary of the AdS geometry\footnote{The spectrum of the fields in the $D$-dimensional supergravity theory corresponds only to a sector of the operators on the dual field theory side. }. Most interestingly, the duality has been extended, on the gravity side, from global AdS to backgrounds which have an AAdS geometry, reproducing the  renormalization group flow of the dual theory to an infrared (IR) conformal fixed point, the energy scale being fixed by the radial coordinate on the $D$-dimensional spacetime. Indeed, the essential ingredient for this correspondence is the conformal structure of the boundary of AAdS spaces. These are spacetimes with negative curvature and whose metric has a pole of order two in the asymptotic region  or, more precisely, conformally compact manifolds  \cite{Skenderis:2002wp,DeWolfe:2018dkl}. Supergravity solutions that are asymptotically (locally) AdS can be interpreted holographically generically either as explicit deformations of SCFTs or as models in which the superconformal symmetry is spontaneously broken.

Several important results have been obtained in the holographic study of strongly coupled quantum field theories, within the so-called {\it bottom-up} approach. This latter consists in crafting an appropriate $D$-dimensional AAdS gravity background of a suitably chosen gravity theory, which can reproduce interesting non-perturbative phenomena of a  boundary field theory, with some given general properties. In this approach emphasis is not given to the higher-dimensional ultraviolet (UV) completion of the (super)gravity theory, which typically has a minimal amount of supersymmetry, if any. Moreover, only certain features of the dual field theory are known, which are suitably fixed by the chosen background through the holographic correspondence.

As opposed to the bottom-up one, the so-called {\it top-down} approach is restricted to gravity theories whose higher-dimensional UV completions in superstring or M-theory are known. This has the advantage that the dual CFT is often known. In most cases supergravity models considered in this setting feature, in particular, an extended amount of supersymmetry (i.e. no less than eight supercharges), which makes them more constrained in field content and interactions and, therefore, more predictive.\footnote{For a discussion on bottom-up versus top-down approaches see, e.g., \cite{DeWolfe:2018dkl}.}
Generally inspired by the latter approach, the purpose of the present investigation is to generalize the holographic analysis of \cite{Amsel:2009rr}  to an extended supergravity, namely to a pure $\mathcal{N}=2$ model.  Some aspects of the minimal $\mathcal{N}=2$ gauged supergravity in the context of holography have been discussed in \cite{Papadimitriou:2017kzw}.

From a formal point of view, the AdS$_{d+1}$/CFT$_d$ correspondence states that the CFT$_d$ partition function is equal to the gravitational partition function in AAdS space in one dimension higher \cite{Gubser:1998bc,Witten:1998qj},
\begin{equation}
    Z_G[\hat{\Phi} \rightarrow\hat{\Phi}_{(0)}]=Z_{\mathrm{CFT}}[\mathcal{J}\equiv\hat{\Phi}_{(0)}]\,.\label{AdSCFTidentification}
\end{equation}
In the above formula,  $ Z_G[\hat{\Phi}_{(0)}]$ is the quantum partition function of the gravity theory in AAdS space, as a function of the boundary value $\hat{\Phi}_{(0)}$ of the bulk field $\hat{\Phi}$, while $Z_{\mathrm{CFT}}[\mathcal{J}]$ is the quantum partition function of the corresponding CFT, in which the source $\mathcal{J}$ of a local operator ${O}(x)$ dual to $\hat{\Phi}$ is identified with  $\hat{\Phi}_{(0)}$.

Let us recall the definition of  the quantum effective action $W[\mathcal{J}]$ for a $d$-dimensional CFT on $\partial \mathcal{M}$ in terms of the partition function $Z_{\mathrm{CFT}}[\mathcal{J}]$
\begin{equation}
Z_{\mathrm{CFT}}[\mathcal{J}]=\mathrm{e}^{\ii W[\mathcal{J}]}=\int \mathcal{D}\phi \,\mathrm{e}^{\ii I[\phi]+\ii \int_{\partial \mathcal{M}}\diff^{d} x\,O(\phi )\cdot \mathcal{J}}\,, \label{QEA}
\end{equation}
where the symbol $\phi(x)$ collectively denotes the fundamental fields of the CFT on which the functional integration is performed. The action $I[\phi ]$ should already be renormalized, that is, finite in the UV region. Even though $W$ is a (non-local) function of the external source $\mathcal{J}(x)$, the physical information of the theory is contained in the $n$-point functions of the operators $O(\phi (x))$,
\begin{equation}
\left\langle O(x_{1})\cdots O(x_{n})\right\rangle _{\mathrm{CFT}}=Z_{\mathrm{CFT}}^{-1}[0]\left. \frac{\delta ^{n}Z_{\mathrm{CFT}}[\mathcal{J}]}{\ii \delta \mathcal{J} (x_{1})\cdots \ii \delta \mathcal{J}(x_{n})}\right\vert _{\mathcal{J}=0}\,. \label{n-point}
\end{equation}
In particular, different correlators are related by Ward identities which express the symmetries in the CFT at the quantum level.

Let us expand on the identification (\ref{AdSCFTidentification}) in the special case of a pure  AdS gravity theory in which the only bulk field is the metric $\hat{g}_{\hat{\mu}\hat{\nu}}(x)$ defined on the AAdS spacetime, to be denoted by $\mathcal{M}^{d+1}$.
In this case the gravitational partition function has the form
\begin{equation}
Z_{\mathrm{G}}[g_{(0)}]=\int \mathcal{D}\hat{g}\,\mathrm{e}^{\ii I_{\mathrm{ren}}[\hat{g}]}\simeq \mathrm{e}^{\ii I_{\mathrm{on-shell}}[g_{(0)}]}\, \label{Zg0sim}.
\end{equation}
Up to a conformal transformation, $g_{(0)\mu \nu }$ is the value at the conformal boundary of the bulk field  $\hat{g}_{\hat{\mu}\hat{\nu}}(x)$, on which Dirichlet boundary conditions are imposed: $\left. \delta g_{(0)\mu \nu }\right\vert _{\partial \mathcal{M}}=0$.
The gravitational action $I_{\mathrm{ren}}[\hat{g}]$ has to be consistent with the boundary conditions and has to be finite in the asymptotic (IR) region. In equation (\ref{Zg0sim}) the classical approximation, for weak gravitational couplings, is performed, in which the partition function can be evaluated on the classical solution, by a saddle point approximation, giving rise to the on-shell action $I_{\mathrm{on-shell}}[g_{(0)}]$. The boundary metric $g_{(0)\mu \nu }$ becomes the source in the boundary CFT.

The AdS/CFT correspondence in the classical approximation of gravity identifies the quantum effective action $W$ as
\begin{equation}
W[g_{(0)}]\simeq I_{\mathrm{on-shell}}[g_{(0)}]\,,
\label{Iclass}
\end{equation}
where the boundary metric becomes a background field in the CFT, so that the energy-momentum tensor operator $T^{\mu\nu }(\phi, g_{(0)} )$ also depends on it. The expectation value of the latter can be calculated as the 1-point function from the effective theory,
\begin{equation}
\left\langle T^{\mu \nu }\right\rangle _{\mathrm{CFT}}=\frac{2}{\sqrt{|g_{(0)}|}}\frac{\delta I_{\mathrm{on-shell}}[g_{(0)}]}{\mathrm{i\,}\delta
g_{(0)\mu \nu }}=\tau ^{\mu \nu }\,,
\end{equation}
and $\tau ^{\mu \nu }$ is the holographic stress tensor in the gravity side.

The conformal Ward identities in the CFT have the form
\begin{equation}
\nabla _{(0)\mu }\tau ^{\mu \nu }=0\,,\qquad \tau _{\ \mu }^{\mu } =\mathcal{A}\,,
\end{equation}
where $\mathcal{A}$ is the Weyl anomaly \cite{Henningson:1998gx}. This quantum result can, therefore, be obtained in the classical regime of AdS gravity.

For the above formalism to be well-defined, a field theory has to be finite at short distances. However, a general feature of quantum field theory is that UV (and IR) divergences can appear at quantum level in the correlation functions. In order to guarantee the consistency of the theory, these unphysical effects are usually removed through the procedure of renormalization. In the framework of  the AdS/CFT correspondence, which is in fact a UV/IR duality, i.e. the ultraviolet regime of the field theory is related to the infrared one of the gravity side and vice versa, it is natural to think that the UV poles of CFT $n$-point functions \eqref{n-point} could be cancelled holographically, by adding appropriate boundary counterterms in the dual theory. Indeed, a first systematic method in this direction was implemented at the beginning of the century \cite{deHaro:2000vlm,Bianchi:2001kw} and was then  applied to various bosonic theories, in particular to gravitational actions coupled to bosonic  matter fields.\footnote{A very good review of the subject can be found in \cite{Skenderis:2002wp}.} Briefly, the procedure consists in regulating the bulk on-shell supergravity action by introducing a cut-off on the radial coordinate, adding appropriate boundary counterterms in order to eliminate the divergences,  and then removing the cut-off.\footnote{For completeness, let us mention that Gibbons and Hawking had already proposed to add a boundary contribution in 1977, namely the Gibbons-Hawking(-York) term, in order to have a well-defined variational principle for gravity theories \cite{Gibbons:1976ue,Brown:1992br}.}

In the subsequent years, the holographic renormalization scheme  was implemented also for   actions including fermionic fields. In \cite{Amsel:2009rr} the authors studied the case of $\mathcal{N}=1$ $D=4$ supergravity including contributions from the gravitini, while in \cite{Papadimitriou:2017kzw} the boundary counterterms for the minimal $\mathcal{N}=2$ gauged supergravities in $D=4$ and $D=5$ have been analysed, restricting to quadratic order in fermions in the action, by using a Hamiltonian approach.  Five-dimensional supersymmetric holographic renormalization has also been considered in \cite{An:2017ihs}.

A  different approach to the holographic renormalization was developed in \cite{Miskovic:2009bm}, where it was named topological regularization. It was proven to give the same results as the standard procedure in pure gravity, having  however the quality of giving a topological meaning to the resummation of the holographic counterterms series expansion. A detailed comparison of both counterterm series has been developed in pure AdS gravity in any dimension in \cite{Anastasiou:2020zwc}.
In particular, the topological counterterm needed to regularize four-dimensional gravity turns out to be the Gauss-Bonnet term and it is also able to restore the diffeomorphisms invariance, broken by the presence of the boundary \cite{Aros:1999id,Aros:1999kt,Olea:2005gb}. Moreover, the addition of this contribution allows to express the renormalized action in the MacDowell-Mansouri form \cite{MacDowell:1977jt}.

The above papers treat gravity in the second order formalism. However, an alternative formulation to the latter is the first order formalism, where the spin connection is considered as an independent field from the vielbein \cite{Banados:2006fe,Klemm:2007yu,Blagojevic:2013bu,Petkou:2010ve,Korovin:2016tsq,Korovin:2017xqu}. In this approach, the powerful tool of exterior calculus and the differential form language can be employed, yielding a geometrical description of gravity. The same approach was  used in \cite{Andrianopoli:2014aqa}  to extend the results of \cite{Aros:1999id, Aros:1999kt, Olea:2005gb} to supergravity and to find  the counterterms needed to restore the local supersymmetry, broken by the presence of a boundary, for the cases of    pure $\mathcal{N}=1$ and $\mathcal{N}=2$ AdS$_4$ supergravities. The boundary terms found in \cite{Andrianopoli:2014aqa} to restore supersymmetry (that is diffeomorphisms in the fermionic directions of superspace) are in fact the supersymmetric extension of the Gauss-Bonnet term, which was necessary to restore diffeomorphisms invariance in the case of gravity. Correspondingly, those boundary terms were precisely the ones needed to rewrite the total supergravity action in a supersymmetric MacDowell-Mansouri form.

However, while the topological regularization was shown to be able to renormalize the bulk action for the pure gravity case, the same has not been proven yet for its supersymmetric extension, in particular for the $\mathcal{N}=2$ AdS$_4$ supergravity. The present paper proceeds from the foregoing works to achieve this goal but, in contrast to \cite{Papadimitriou:2017kzw}, we consider the full contribution from the gravitini and start from a rather general setup in view of possible future developments. In order to do it, we show that the Ward identities of the dual field theory are satisfied, as expected for a SCFT in three dimensions.

It still remains as an open problem the question of rendering the AdS supergravity action finite in the presence of matter multiplets by adding topological bulk terms.

From a different, but complementary, point of view, we explore a relation between the classical local symmetries of an AdS gravity defined on the bulk manifold $\mathcal{M}^D$ and the quantum symmetries in a field theory defined on $\partial \mathcal{M}$. The latter match the \textit{asymptotic symmetries}, at radial infinity, of the gravitational background.
 In our approach, they appear as \textit{residual symmetries} left over after the gauge fixing of bulk local symmetries and whose parameters take value on $\partial \mathcal{M}$.
This matching of symmetries is justified from the group theoretical  point of view. Namely (we restrict here, for simplicity, the discussion to the bosonic sector, but its supersymmetric extension will be considered in the body of the paper) the isometries of the AdS vacuum in $D=(d+1)$-dimensional asymptotically AdS spaces are described by the $\mathrm{SO}(2,d)$ group whose generators are $\mathbf{J}_{ab}$, $\mathbf{J}_{a}$.
It is important to emphasize that gravity with negative cosmological constant is not invariant under local $\mathrm{SO}(2,d)$ transformations. Instead, general coordinate transformations, combined with a field-dependent local Lorentz transformations, acquire a locally gauge-covariant form.

The $d$-dimensional boundary breaks the bulk local symmetries in the $x^d$ (radial) direction that naturally leads to the $d+1$ decomposition of the Lorentz indices into $a=(i,d)$. In that way, the bulk isometry group
is isomorphic to the conformal group\footnote{For explicit construction of this conformal algebra in the context of AdS/CFT in pure AdS gravity, see \cite{Korovin:2017xqu}.}  with generators $\mathbf{J}_{ij}$, $\mathbf{P}_{i} =\mathbf{J}_{i}+\mathbf{J}_{id}$, $\mathbf{K}_{i} =\mathbf{J}_{i}-\mathbf{J}_{id}$, $\mathbf{D}=\mathbf{J}_{d}$.
Therefore, choosing  suitable boundary conditions for the AdS gravity fields in $D$-dimensional bulk, which are $\hat{\omega}^{ab}$ (along $\mathbf{J}_{ab}$) and $\frac{1}{\ell }\,V^{a}$ (along $\mathbf{J}_{a}$), we can identify its $d$-dimensional boundary field content that should be the one of the CFT. Using the isomorphism, the boundary background fields, i.e. sources $\mathcal{J}=\{ \omega^{ij},E^i,B,\mathcal{S}^i \}$ associated with the conformal generators, have the form
\begin{equation*}
\begin{array}{lll}
\mathbf{J}_{ij}:\medskip  & \omega ^{ij} & \sim \;\hat{\omega}^{ij}\,, \\
\mathbf{P}_{i}:\medskip   & E^{i} & \sim \;V_{+}^{i}=\dfrac{1}{2}\,\left(
\ell \hat{\omega}^{id}+V^{i}\right) \,, \\
\mathbf{D}:\medskip  & B & \sim \;V^{d}\,, \\
\mathbf{K}_{i}:\medskip \quad  & \mathcal{S}^{i} & \sim \;V_{-}^{i}=\dfrac{1}{2}\,\left( \ell \hat{\omega}^{id}-V^{i}\right) \,,
\end{array}
\end{equation*}
where `$\sim $' means that the identifications are valid only up to a global rescaling on the boundary, allowed in CFT. This near-boundary rescaling is the first step in removing the long-distance divergences present in (super)gravity theory in asymptotically AdS spaces, equivalent to renormalization of the holographic CFT. From this discussion, we draw the following conclusions.
First, a full linearly realized conformal group on the boundary can be made manifest only in first order formalism, where the spin connection is an independent field. Second, the conformal structure on the boundary naturally introduces two geometric quantities in $d$ dimensions, a dilatation gauge field $B$ and the Schouten tensor $\mathcal{S}^{i}$. They will play an important role in the analysis of symmetries of this holographic correspondence.

As far as the asymptotic symmetries and the gauge-fixing conditions defining them are concerned, we shall keep our analysis as general as possible. More precisely, we shall be taking a ``cautious approach'', only imposing gauge-fixing conditions which appear to be strictly necessary for the consistent definition of the asymptotic symmetries. The reason for this relies on one of the motivations which have inspired the present analysis, namely the application of the AdS$_4$/CFT$_3$ holographic approach to the study of the model, originally constructed in \cite{Alvarez:2011gd} (to be referred to as the AVZ model), which features \emph{unconventional supersymmetry}. The latter has been  eventually embedded, as a boundary theory, in pure $\mathcal{N}=2$ AdS$_4$ supergravity in \cite{Andrianopoli:2018ymh}, although a fully fledged holographic correspondence has not been developed yet.
The present work represents a preliminary investigation in this direction.
Having this in mind, we avoid imposing the constraint $\gamma^\mu\psi_\mu=0$ on the gravitino field at the boundary since, in the AVZ model, this condition has to be relaxed, as the dynamical fermion of the theory is identified with the contraction $\gamma^\mu\psi_\mu$ itself.
This fermion satisfies a Dirac equation and was shown to be well-suited for the description of the electronic properties of graphene-like materials \cite{Alvarez:2011gd,Andrianopoli:2019sip}. Holographically embedding the AVZ model in $\mathcal{N}=2$ AdS$_4$ supergravity and eventually in $\mathcal{N}>2$ theories paves the way for a top-down approach to the study of this condensed-matter system.
In this direction, in \cite{Andrianopoli:2019sqe}, a possible relationship between the construction in \cite{Andrianopoli:2019sip} and a generalization of the $d=3$ interface model of Gaiotto and Witten  \cite{Gaiotto:2008sd}
was presented. This hints towards the definition of the dual conformal class of theories, which will be the object of a future investigation.
\vskip 5mm

The rest of this work is organized as follows: In Section \ref{WithoutF}  we review the asymptotic symmetries in Einstein AdS$_4$ gravity for purpose of introducing the first order formalism and in Section  \ref{4d} we summarize  the geometric approach to pure $\mathcal{N}=2$ AdS$_4$ supergravity, in the presence of a boundary. Section \ref{nba} is devoted to the near-boundary analysis of supergravity fields and local parameters. Then, in Section \ref{SuperWard}, we write out the superconformal currents and Ward identities, proving that the latter are indeed satisfied off-shell on the curved background when the bulk equations of motion are imposed.  We conclude the paper with some final remarks. Useful formulas and conventions are gathered in Appendix \ref{conventions}, while details on calculations are collected in Appendix \ref{Appasexp} and Appendix \ref{appC}.

%%%%%%%%%%%%%%%%%%%%%%%%%%%%%%%%%%%%%%%%%%%%%%%%%%%%%%
\section{Asymptotic symmetries in Einstein AdS\texorpdfstring{$_4$}{4} gravity \label{WithoutF}}

We start our discussion with a review of the results in pure AdS gravity and then reformulating them in first order framework.

Asymptotically AdS spaces $\mathcal{M}^D$  in $D=d+1$ dimensions are conformally compact Einstein spaces that can be described by local   coordinates $x^{\hat{\mu}}=(x^{\mu },x^d)$, where $x^{\mu }$ ($\mu=0,\ldots d-1$) are local coordinates on the boundary $\partial \mathcal{M}$ and $z=x^d$ is the radial coordinate with the asymptotic AdS boundary located at $z=0$. In a neighborhood of $z=0$, they  admit a metric $\hat{g}_{\hat{\mu}\hat{\nu}}$ (with a mostly negative signature) in the Fefferman-Graham (FG) form,\footnote{
The most general asymptotically AdS metric contains also the subleading $\hat{g}_{z\mu }$ terms, in particular $\hat{g}_{z\mu }=\mathcal{O}(z)$ in three dimensions \cite{Brown:1986nw} and $\hat{g}_{z\mu}=\mathcal{O}(z^2)$ in four dimensions \cite{Henneaux:1985tv}.
They can always be set to zero by choosing FG coordinate frame on a patch near the boundary.
}
\begin{equation}
\diff s^{2}=\hat{g}_{\hat{\mu}\hat{\nu}}\,\diff x^{\hat{\mu}}\diff x^{\hat{\nu}}=\frac{\ell ^{2}}{z^{2}}\,\left( \rule{0pt}{13pt}-\diff  z^{2}+g_{\mu \nu }(x,z)\,\diff x^{\mu }\diff x^{\nu }\right) \,,
\label{FG}
\end{equation}
where $\ell $ is the AdS radius, $g_{\mu \nu }$ is regular on the boundary and it admits a power expansion in the radial coordinate $z$,
\begin{equation}
g_{\mu \nu }=g_{(0)\mu \nu }(x)+\frac{z^{2}}{\ell ^{2}}\,g_{(2)\mu \nu }(x)+\cdots \,.  \label{g expansion}
\end{equation}
Only even powers in $z$ appear in the series, until the order $z^{d-1}$.
This is particular to pure AdS gravity. In general, addition of matter fields, as is the case in supergravity, requires more general powers in the $z$-expansion of the metric, depending on the value of the AdS mass of the field. By solving order by order the Einstein equations, the corresponding coefficients in the expansion are determined as local functions of $g_{(0)\mu \nu }$. For example, $g_{(2)\mu \nu }$ depends linearly on the curvature in a combination that produces the boundary Schouten tensor $\mathcal{S}_{\mu \nu}(g_{(0)})$,
\begin{equation}
g_{(2)\mu \nu }=\ell ^{2}\mathcal{S}_{\mu \nu }=\ell ^{2}\left( \mathring{\mathcal{R}}_{\mu \nu } -\frac{1}{2(d-1)}\,g_{(0)\mu \nu }\,\mathring{\mathcal{R}} \right) \,,  \label{g(2)}
\end{equation}
where $\mathring{\mathcal{R}}^{\mu}_{\ \nu\lambda\sigma}(g_{(0)})$ is the boundary Riemann curvature  and $\mathring{\mathcal{R}}_{\mu \nu}$ and $\mathring{\mathcal{R}}$ are the corresponding Ricci tensor and Ricci scalar, respectively. The conventions we adopt on curvatures can be found in Appendix \ref{Rconventions}. On top of this, only in odd spacetime dimensions $D$, there is a term $z^{d}\log z$. In contrast, the mode $g_{(d)\mu \nu }$ cannot be fully resolved from the equations of motion (only its local part), as it is proportional to the holographic stress tensor of the theory \cite{deHaro:2000vlm,Skenderis:2002wp}.

The FG form of the metric (\ref{FG}) is obtained by gauge fixing of spacetime coordinate frame. The invariance of (\ref{FG}) under radial diffeomorphisms leads to the Penrose-Brown-Henneaux (PBH) transformations \cite{Imbimbo:1999bj}. The full set of residual symmetries includes, apart from the PBH transformations, also the boundary (transversal) diffeomorphisms.
They have the form of asymptotic
symmetries, that is, their parameters take value on $\partial \mathcal{M}$.
The AdS gravity is invariant under the action of these transformations at asymptotic infinity.

In an explicit form, using the Lie derivative $\delta \hat{g}_{\hat{\mu}\hat{\nu}}=\pounds_{\hat{\xi}}\hat{g}_{\hat{\mu}\hat{\nu}}$   for diffeomorphisms generated by parameters $\hat\xi^{\hat\mu}$,   the FG gauge fixing
implies
\begin{eqnarray}
\delta \hat{g}_{zz} &=&0\quad \Rightarrow \quad \hat{\xi}^{z}=z\sigma (x)\,,
\notag \\
\delta \hat{g}_{\mu z} &=&0\quad \Rightarrow \quad \hat{\xi}^{\mu }=\xi^{\mu }(x)+\frac{z^{2}}{2\ell }\,g_{(0)}^{\mu \nu} \mathcal{\partial }_{\nu}\sigma +\mathcal{O}(z^{4})\,,
\end{eqnarray}
where $\xi ^{\mu }(x)$ and $\sigma (x)$ are arbitrary local parameters on the boundary. From $\delta \hat{g}_{\mu \nu }=\frac{\ell ^{2}}{z^{2}}\,\delta g_{\mu \nu }$, we obtain the transformation law of first terms    in the asymptotic expansion of \eqref{g expansion}  as
\begin{eqnarray}
\delta g_{(0)\mu \nu } &=&\pounds _{\xi }g_{(0)\mu \nu }-2\sigma \,g_{(0)\mu
\nu }\,,  \notag \\
\delta g_{(2)\mu \nu } &=&\pounds _{\xi }g_{(2)\mu \nu }-\ell \nabla _{(\mu }^{(0)}\nabla _{\nu )}^{(0)}\sigma \,.  \label{PBH}
\end{eqnarray}
From the first equation it is clear that radial diffeomorphisms induce Weyl transformations on the boundary described by the parameter $\sigma(x)$. This purely kinematic treatment allows to determine the local part of the coefficients in the series (\ref{g expansion}) without resorting to the asymptotic resolution of the field
equations. This is done by integrating the Weyl parameter from the transformation law above.
This PBH approach, although a powerful tool to match symmetries in a holographic field theory, is not an alternative to holographic renormalization.

The asymptotic symmetries produce conservation laws which are mapped into holographic Ward identities for the boundary CFT.

%%%%%%%%%%%%%%%%%%%%%%%%%%%%%%%%%%%%%%%%%%%%%%%%%%%%%%%%%%%%%%%%%%
\subparagraph{Holographic gauge fixing in first order formalism.}

In what concerns us here, we work in first order formalism in $D=4$ where the independent fields are 1-forms on $\mathcal{M}^4$. Indeed, one has the vielbein
$V^{a}=V_{\ \hat{\mu}}^{a}(x)\,\diff x^{\hat{\mu}}$, stemmed from the metric $\hat{g}_{\hat{\mu}\hat{\nu}}=\kappa _{ab}\,V_{\ \hat{ \mu}}^{a}V_{\ \hat{\nu}}^{b}$ (with the Minkowski metric $\kappa_{ab}$), and the spin connection $\hat{\omega}^{ab}=\hat{\omega}_{\hat{\mu}}^{ab}(x)\, \diff x^{\hat{\mu}}$. World indices on four-dimensional spacetime are denoted by hatted Greek letters $\hat\mu,\hat\nu,\dotsc=0,1,2,3$ and the corresponding anholonomic tangent space indices are labeled by Latin letters  $a,b,\dotsc = 0,1,2,3$.

Apart from general coordinate transformations $\delta x^{\hat{\mu}}= - \hat\xi ^{\hat{\mu}}$, which define local translations with
parameters $p^{a}=\hat\xi^{\hat{\mu}}V_{\ \hat{\mu}}^{a}$, the theory is now endowed with local Lorentz invariance, whose parameters are $j^{ab}=-j^{ba}$.

The AdS gravity in first order formalism is invariant under the general transformations\footnote{\label{footnote1}
This transformation law is the local Lorentz transformation combined with the Lie derivative $\pounds _{p}A=D\left(i_{p}A\right) +i_{p}F\,$valid for any gauge field $A$ and its associated field strength $F$, where $p^{\hat{\mu}}=\hat{\xi} ^{\hat{\mu}}$.}
\begin{eqnarray}
\delta V^{a} &=&\mathcal{\hat{D}}p^{a}-j^{ab}V_{b}+i_{p}\hat{T}^{a}\,,
\notag \\
\delta \hat{\omega}^{ab} &=&\mathcal{\hat{D}}j^{ab}+\frac{2}{\ell ^{2}}\,p^{[a}V^{b]}+i_{p}\hat{R}^{ab},  \label{var}
\end{eqnarray}
where $\mathcal{\hat{D}(}\hat{\omega})$ is the Lorentz-covariant derivative, $\mathcal{\hat{R}}^{ab}(\hat{\omega})$ is the Lorentz curvature and $\hat{T}^{a}=\mathcal{\hat{D}}V^{a}$ is the torsion 2-form.  We have also introduced the AdS curvature $\hat{R}^{ab}=\mathcal{\hat{R}}^{ab}-\frac{1}{\ell ^{2}}\,V^{a}V^{b}=\frac{1}{2}\hat R^{ab}_{\hat\mu\hat\nu}\diff x^{\hat\mu}\diff x^{\hat\nu}$ and the contraction operator $i_{p}\hat{R}^{ab}=p^{c}V_{\  c}^{\hat{\nu}}\hat{R}_{\hat{\nu}\hat{\mu}}^{ab}$\textrm{d}$x^{\hat{\mu}}$ and similarly for $i_{p}\hat{T}^{a}$.
In the non-supersymmetric case we are discussing in this section,  we will assume that  the gravitational field is torsionless, thus $i_{p}\hat{T}^{a}=0$.

In order to extend the discussion for AAdS spacetimes from the metric formulation described above to first order formalism, we have to specify the form of $V^{a}$ and $\hat{\omega}^{ab}$. To this end, we have 10 local parameters $(p^{a},j^{ab})$ at our disposal to gauge fix.

This {\it holographic gauge fixing} has to provide the radial expansion of  gauge fields and parameters. In addition, the residual transformations (which leave invariant that gauge fixing) have to induce boundary Weyl dilatations. It also has to give rise to the transformation of the boundary fields, which lead to the conservation laws.

In this framework, the radial evolution of gravity considers the radial components of the gauge fields as Lagrange multipliers, similarly as the lapse and shift functions in the Arnowitt–Deser–Misner (ADM) formulation of gravity \cite{Arnowitt:1962hi}. The simplest choice $V_{\ z}^{a}=0$, $\hat{\omega}_{z}^{ab}=0$ leads to a trivial theory on the boundary. In particular, it does not have an invertible vielbein.

Radial expansion and holography in gravity in Riemann-Cartan space were developed in \cite{Banados:2006fe} and applied, for example, in three \cite{Banados:2006fe,Klemm:2007yu,Blagojevic:2013bu}, four \cite{Petkou:2010ve}, and five \cite{Banados:2006fe} bulk dimensions in different setups. In arbitrary dimension it was discussed in \cite{Korovin:2017xqu}.

A suitable gauge fixing for spacetime diffeomorphisms $p^{a}$ and Lorentz transformation $j^{ab}$ is
\begin{equation}
V_{\ z}^{a}=\dfrac{\ell }{z}\,\delta _{3}^{a},\qquad \hat{\omega}_{z}^{ab}=0\,.
\label{holo gauge Vw}
\end{equation}
These conditions, in principle, are sufficient to determine local symmetries. However, in AdS space, the vielbein should be chosen so that it reproduces the FG metric (\ref{FG}). For this reason, we assume an adapted frame where the boundary is orthogonal to the radial coordinate,
\begin{equation}
V_{\ \mu }^{3}=0\,.  \label{z-orthogonality}
\end{equation}

The last condition can be relaxed as long as the fall-off of the field $V_{\ \mu}^3(x)$ is consistent with the behaviour of AAdS spaces. As shown in \cite{Korovin:2017xqu,Ciambelli:2019bzz}, this field plays a role in the explicit construction of the gauged conformal algebra for a dual CFT. By setting $V_{\ \mu}^3$ to zero, the conformal symmetry of the boundary is still there, but its realization becomes non-linear, as the associated gauge field turns into a composite field.

As mentioned before, the choice (\ref{holo gauge Vw}) is holographic if it produces a radial expansion of the boundary fields.
Let us denote the $3+1$ decomposition of Lorentz indices as $a=(i,3)$ ($i=0,1,2$).
We use the following convention for the Levi-Civita tensor on $\mathcal{M}^{4}$ projected to the boundary $\partial \mathcal{M}$,
\begin{equation}
\epsilon ^{ijk 3}=-\epsilon ^{ijk} \,,\qquad
\epsilon ^{0123}=-\epsilon _{0123}=-1\,.  \label{3D epsilon}
\end{equation}
Then, in AAdS spaces, the vielbein behaves as
\begin{equation}
V_{\ \mu }^{i}=\frac{\ell }{z}\,\hat{E}_{\ \mu }^{i}(x,z)\,,  \label{V}
\end{equation}
where $\hat{E}_{\ \mu }^{i}$ is finite at the boundary $z=0$, so it can be expanded in a power series in its vicinity,
\begin{eqnarray}
\hat{E}_{\ \mu }^{i} &=&E_{(0)\ \mu }^{i}+\frac{z^{2}}{\ell ^{2}}\,E_{(2)\ \mu }^{i}+ \frac{z^{3}}{\ell ^{3}}\,E _{(3)\ \mu}^{i} +\mathcal{O}(z^{4})\,.
\end{eqnarray}
Because of physical implications it would have later, we rename the coefficients as $E_{(0)\ \mu }^i \equiv E_{\ \mu }^i$, $E_{(2)\ \mu }^i \equiv S_{\ \mu }^i$ and $E_{(3)\ \mu }^i \equiv \tau_{\ \mu }^i$. Then the expansion becomes
\begin{eqnarray}
\hat{E}_{\ \mu }^{i} &=&E_{\ \mu }^{i}+\frac{z^{2}}{\ell ^{2}}\,
S_{\ \mu }^{i}+ \frac{z^{3}}{\ell ^{3}}\,\tau _{\ \mu}^{i} +\mathcal{O}(z^{4})\,,  \notag \\
\hat{E}_{\ i}^{\mu } &=&E_{\ i}^{\mu }-\frac{z^{2}}{\ell ^{2}}\,S_{\ i}^{ \mu }- \frac{z^{3}}{\ell ^{3}}\,\tau _{i}^{\ \mu }+\mathcal{O}(z^{4})\,,  \label{E}
\end{eqnarray}
where $E_{\ i}^{\mu }$ is the inverse of the vielbein\footnote{Strictly speaking, the inverse vielbein $(E^{-1})_{\ i}^{\mu }\equiv E_{\ i}^{\mu }$ has the property $E^{\mu }{}_{i}=g_{(0)}^{\mu \nu }\eta_{ij}E^{j}{}_{\nu }=E_{i}{}^{\mu }$ following from the invertibility and symmetry of the metric. It implies that one can overlook the order of the indices in the vielbein and its inverse. The same argument holds for the bulk vielbein $V_{\ \mu }^{i}$ and its inverse $V_{\ i}^{\mu }$, but \textit{not} for the higher-order terms in the expansion that are not necessarily invertible.}  $E_{\ \mu }^{i}$. These tensors project the indices between the boundary spacetime  and its tangent space and we also have
\begin{equation}
e=\det [V_{\ \hat{\mu}}^{a}]=\frac{\ell ^{4}}{z^{4}}\,\hat{e}_{3}\,,
\quad\hat{e}_{3}=\det[\hat{E}_{\ \mu }^{i}]\,,\quad e_{3}\equiv \det [E_{\ \mu }^{i}]\,.  \label{e}
\end{equation}

Notice that we assumed that the linear terms in $z$ are absent in the induced vielbein $\hat{E}_{\ \mu }^{i}$, in order to reproduce the result $g_{(1)\mu \nu }=0$ in pure gravity. Furthermore, it is convenient to make use of the residual Lorentz transformations to get $S^{ij}=S_{\ \mu }^{i}E^{\mu j}$ and $\tau ^{ij}=\tau _{\ \mu }^{i}E^{\mu j}$ symmetric,
namely to set $S^{[ij]}=0$ and $\tau ^{[ij]}=0$ \cite{Amsel:2009rr}. If the Lorentz parameter at the boundary is expanded as
\begin{equation}
j^{ij}=\theta ^{ij}+\frac{z}{\ell }\,j_{(1)}^{ij}+\frac{z^{2}}{\ell ^{2}}\,j_{(2)}^{ij}+\frac{z^{3}}{\ell^{3}}\,j_{(3)}^{ij}+\mathcal{O}(z^{4})\,,
\end{equation}
from the Lorentz transformations (\ref{var}) we find $j_{(1)}^{ij}=0$ and
\begin{equation}
\begin{array}{llll}
\delta _{j}E_{\mu }^{i} & =-\theta ^{ij}E_{j\mu }\,,\qquad  & \delta_{j}S_{\ \mu }^{i} & =-\theta ^{ij}S_{j\mu} -j_{(2)}^{ij}E_{j\mu}\,,\medskip  \\
\delta _{j}E^{\mu i} & =-\theta ^{ij}E_{\ j}^{\mu }\,, & \delta _{j}\tau _{\ \mu}^{i} & =-\theta ^{ij}\tau _{j\mu }-j_{(3)}^{ij}E_{j\mu }\,.
\end{array}
\end{equation}
Here $\theta ^{ij}(x)$ is an asymptotic parameter which will become  a holographic symmetry. The antisymmetric parts are independent of $\theta ^{ij}$,
\begin{equation}
\delta _{j}S^{[ij]}=-j_{(2)}^{ij}\,,\qquad \delta _{j}\tau ^{[ij]}=-j_{(3)}^{ij}\,. \label{var_jS}
\end{equation}
Therefore, the antisymmetric parts of $S^{ij}$ and $\tau^{ij}$ are related only to the subleading Lorentz transformations and therefore they can always be set to zero,
\begin{equation}
S^{[ij]}=0\,,\qquad \tau ^{[ij]}=0\,. \label{symm_gauge}
\end{equation}
However, we will not assume yet that $j_{(2)}^{ij}$ and $j_{(3)}^{ij}$ vanish because they might not be independent parameters. We will come back to this issue later, after all independent asymptotic symmetries have been identified (see eq.~\eqref{deltaSantisym}).

In fact, the above procedure can be extended to make all coefficients  in  the expansion of $V_{\ \mu }^{i}$ symmetric. Without going into details, it can be shown that $\theta ^{ij}$ always decouples from the transformation of $E^{[ij]}_{(n)}\equiv E^{\mu [j}E_{(n)\mu }^{i]}$ and  we can always set $E_{(n)}^{[ij]}=0$ for $n\geq 0$. As a net result, all modes $E_{(n)\mu }^{i}$ are symmetric tensors,
\begin{equation}
 E_{(n)}^{[ij]}=0\,, \qquad n\geq 0\, . \label{gauge-fixing Lorentz}
\end{equation}

Thus, the expansion defined by the above considerations is consistent with the FG frame (\ref{FG}) and
\begin{eqnarray}
g_{(0)\mu \nu } &=&E_{i\nu }E_{\ \mu }^{i}\,,  \notag \\
g_{(2)\mu \nu } &=&2S_{\mu \nu }=\ell ^{2}\mathcal{S}_{\mu \nu }\,,  \notag \\
g_{(3)\mu \nu } &=&2\tau _{\mu \nu }\,.
\end{eqnarray}
Recalling the fact that in Einstein AdS gravity we know the solution of the coefficients $g_{(n)\mu \nu }$ ($n>0$) in terms of the source $g_{(0)\mu \nu }$ \cite{deHaro:2000vlm,Imbimbo:1999bj}, we identify $E_{\ \mu }^{i}$ as the vielbein at the conformal boundary, $S_{\ \mu }^{i}=\frac{\ell ^{2}}{2}\, \mathcal{S}_{\ \mu }^{i}$ as proportional to the Schouten tensor, and $\tau_{\ \mu }^{i}$ as the holographic stress tensor.

On the other hand, without supersymmetry, the torsion constraint  $\mathcal{\hat{D}}V^{a}=0$  determines the spin-connection to be (see \eqref{FVpostulate})
\begin{equation}
\hat{\omega}_{\hat{\mu}}^{ab}=V^{\hat{\nu} b}\left( -\partial _{\hat{\mu}}V_{\ \hat{\nu}}^{a}+\hat{\Gamma}_{\hat{\nu}\hat{\mu}}^{\hat{\lambda}}V_{\ \hat{\lambda}}^{a}\right) \,. \label{omega pure}
\end{equation}
In our notation,  $\hat{\Gamma}_{\hat{\nu}\hat{\mu}}^{\hat{\lambda}}$
is the affine connection in the bulk. In particular, the one appearing in \eqref{omega pure} is the Levi-Civita connection, that is, symmetric in $(\hat{\mu}\hat{\nu})$ and torsionless. The radial components of the spin-connection are consistent with the gauge fixing (\ref{holo gauge Vw}), assuming \eqref{gauge-fixing Lorentz} is satisfied. The boundary components of the spin connection become
\begin{eqnarray}
\hat{\omega}_{\mu }^{ij} &=&\hat{E}^{\nu j}\left( -\partial _{\mu }\hat{E}_{\ \nu }^{i}+\mathring{\Gamma}_{\nu \mu }^{\lambda }(g)\hat{E}_{\ \lambda }^{i}\right) = \mathring{\omega}_{\mu }^{ij}(x,z)\,,  \notag \\
\hat{\omega}_{\mu }^{i3} &=&\frac{1}{z}\,\hat{E}_{\ \mu }^{i}-\frac{1}{2}
\,k_{\mu \nu }\hat{E}^{\nu i}\,,  \label{tildeOmega}
\end{eqnarray}
where $\mathring{\omega}_{\mu }^{ij}(x,0)= \mathring{\omega}_{\mu }^{ij}(E)$ is the torsionless spin connection on the boundary, $\mathring{\Gamma} _{\nu \mu }^{\lambda }(g)$  is the affine Levi-Civita  connection at the boundary that depends on $z$ (in contrast to $\mathring{\Gamma} _{\nu \mu }^{\lambda }=\mathring{\Gamma} _{\nu \mu }^{\lambda }(g)|_{z=0}$) and we define the auxiliary tensor
\begin{equation}
k_{\mu \nu }\equiv \partial _{z} g_{\mu\nu} =\mathcal{O}(z)\,,\qquad
\partial _{z}g^{\mu \nu }=-k^{\mu \nu }\,.  \label{k}
\end{equation}
Both $\mathring{\Gamma} _{\nu \mu }^{\lambda }(g)$ and $k_{\mu \nu }$ are regular quantities at $z=0$. In a more explicit form,
\begin{eqnarray}
\hat{\omega}_{\mu }^{ij} &=&\mathring{\omega}_{\mu }^{ij}(x,z)=\mathring{\omega}_{\mu }^{ij}(x)+\frac{z^{2}}{\ell ^{2}}\,\omega _{(2)\mu }^{ij}(S,E)+%
\frac{z^{3}}{\ell ^{3}}\,\omega _{(3)\mu }^{ij}(\tau ,E)+\mathcal{O}(z^{4})\,,  \notag \\
\hat{\omega}_{\mu }^{i3} &=&\frac{1}{z}\,E_{\ \mu }^{i}-\frac{z}{\ell ^{2}}\,\tilde{S}_{\ \mu }^{i}-\frac{2z^{2}}{\ell^3}\,\tilde{\tau}_{\ \mu }^{i}+\mathcal{O}(z^{3})\,,
\label{spin conn_no fermions}
\end{eqnarray}
where
\begin{equation}
\tilde{S}_{\ \mu }^{i} \equiv S_{\mu }^{\ i} = S_{\ \mu }^{i}\,,\qquad \tilde{\tau}_{\ \mu }^{i}\equiv \frac{1}{4}\,\left( \tau _{\ \mu }^{i}+3\tau _{\mu }^{\ i}\right) =\tau _{\ \mu }^{i}\,, \label{pure equality}
\end{equation}
and the last step is only valid upon imposing the partial Lorentz gauge fixing \eqref{gauge-fixing Lorentz}. Thus, in pure AdS gravity, the tensors $\tilde{S}_{\ \mu }^{i}$ and $\tilde{\tau}_{\ \mu }^{i}$  can be chosen symmetric and equal to $S_{\ \mu }^{i}$ and $\tau _{\ \mu }^{i}$. We will see later (in eq.~(\ref{Shouten})) that the group theory definition of the boundary Schouten tensor is  $\mathcal{S}_{\ \mu }^{i}=\frac{1}{\ell ^{2}}\,(S_{\ \mu }^{i}+\tilde{S}_{\ \mu }^{i})$ and it reduces to $\frac{2}{\ell ^{2}}\,S_{\ \mu }^{i}$ only after using the above equality.

When the bulk torsion vanishes, we obtain at the first two orders ($z$ and $z^{2}$) near the boundary that the 1-forms $\omega
_{(2)}^{ij}=\omega _{(2)\mu }^{ij}\,\diff x^{\mu }$ and similarly for $\omega _{(3)}^{ij}$, are not arbitrary, but they can be expressed in terms of $S^{i}=S_{\ \mu }^{i}\,\diff x^{\mu }$ and $\tau ^{i}=\tau _{\ \mu }^{i}\,\diff x^{\mu }$ as
\begin{equation}
E_{j}\wedge \omega _{(2)}^{ij}=\mathcal{\mathring{D}}S^{i}\,,\qquad
E_{j}\wedge \omega _{(3)}^{ij}=\mathcal{\mathring{D}}\tau ^{i}\,,
\label{DS}
\end{equation}
where $\mathcal{\mathring{D}}$ denotes the covariant derivative with respect to the connection $\mathring{\omega}_{\mu }^{ij}(E)$. These equations can be explicitly solved in $\omega _{(2)}^{ij}$, $\omega
_{(3)}^{ij}$, as indicated by (\ref{spin conn_no fermions}).

Let us finally analyse the fall-off of the curvature. Asymptotically AdS spaces require the curvature to be asymptotically constant. Direct checkup confirms that the near-boundary form of the AdS curvature is
    \begin{equation}
    \begin{array}{llll}
    \hat{R}_{\mu \nu }^{i3} & = - z\,\mathcal{C}_{\ \mu \nu }^i
    +\mathcal{O}(z^{2})\,,
    & \hat{R}_{\mu z}^{i3} & =\dfrac{3z}{\ell ^{3}}\,\tau _{\ \mu }^{i}+\mathcal{O}(z^{2})\,, \medskip\\
    \hat{R}_{\mu \nu }^{ij} & =W_{\mu \nu }^{ij}-\dfrac{12z}{\ell ^{3}}\,
    E_{\ [\mu}^{[i}\tau _{\ \nu]}^{j]}+\mathcal{O}(z^{2})\,,\quad
    & \hat{R}_{\mu z}^{ij}
    & =-\dfrac{2z}{\ell^2}\, \omega_{(2)\mu}^{ij}-\dfrac{3z^2}{\ell^3}\, \omega_{(3)\mu}^{ij}+\mathcal{O}(z^3) \,,
    \end{array} \label{pure R}
    \end{equation}
where $\mathcal{C}^{i}=\frac{1}{2}\,\mathcal{C}_{\ \mu \nu }^{i}\,\diff  x^{\mu }\wedge \diff x^{\nu }=\mathcal{\mathring{D}S}^{i}$ is three-dimensional Cotton tensor.
In the above derivation of $\hat{R}^{i3}=-\frac{z}{\ell ^{2}}\left( \mathcal{\mathring{D}}\tilde{S}^{i}+E_{j}\wedge \omega_{(2)}^{ij}\right)  +\cdots $, the Cotton tensor appears after using \eqref{DS} to eliminate $\omega _{(2)}^{ij}$. This is because $\tilde{S}^{i}=S^{i}$ cannot be assumed directly under the derivative due to the relations (\ref{pure equality}) which involve the derivatives of the vielbein. Similarly, $\hat{R}^{ij}$ depends on the tensor $\tau ^{i}+2\tilde{\tau}^{i}$, but it reduces to the above result upon setting $\tau ^{i}=\tilde{\tau}^{i}$.
The Weyl tensor vanishes in three dimensions,
\begin{equation}
W^{ij}=\mathring{\mathcal{R}^{ij}}-2E^{[i}\wedge \mathcal{\mathring{S}}^{j]}=0\,,
\label{W=0}
\end{equation}
so that the
three-dimensional Bianchi identity can equivalently be written as
\begin{equation}
E^{[i}\wedge \mathcal{C}^{j]}=0\,,
\end{equation}
yielding that the Cotton tensor is traceless, $\mathcal{C}_{\ ij}^{i}=0$. It is also known that, in three dimensions, it is covariantly constant.\footnote{The dual of the Cotton tensor appears naturally --at the holographic order-- in the parity-odd sector of the the theory. This feature gives rise to a holographic stress tensor/Cotton tensor duality at the boundary \cite{deHaro:2008gp} which, in turn, is a consequence of an asymptotic  (anti-)self duality
condition for the Weyl tensor \cite{Miskovic:2009bm}.}
For more properties of the Cotton tensor in Riemannian geometries, see \cite{Garcia:2003bw}.

An important consequence of $W_{\mu \nu }^{ij}=0$ in three dimensions is that, from (\ref{pure R}), we get $\left.\hat{R}^{ab}\right\vert _{z=0}=0$.  Since the AdS boundary $\partial \mathcal{M}$ is located at constant radius $\diff z=0,z=0$, in supergravity the last condition can be relaxed to $\left. \hat{R}^{ab}\right\vert _{\diff z=0,z=0}=0$.

The fact that the curvature is constant at the conformal boundary does not guarantee --only by itself-- the regularity of the variation of the action and, therefore, a correct holographic description of the theory.

%%%%%%%%%%%%%%%%%%%%%%%%%%%%%%%%%%%%%%%%%%%%%%%%%%%%%%%%%%%%%%%%%%
\subparagraph{Residual symmetries.}

The gauge fixing adopted above leads to the asymptotic form of the boundary fields (\ref{V}), (\ref{E}) and (\ref{spin conn_no fermions}). We seek for transformations which do not change the frame choice (\ref{holo gauge Vw}). From (\ref{var}), it follows
\begin{eqnarray}
0 &=&\delta V_{\ z}^{3}=\partial _{z}p^{3}\,,  \label{3z} \\
0 &=&\delta V_{\ z}^{i}=\partial _{z}p^{i}+\frac{\ell }{z}\,j^{i3}\,,
\label{iz} \\
0 &=&\delta V_{\ \mu }^{3}=\partial _{\mu }p^{3}-\hat{\omega}_{\mu
}^{i3}p_{i}+j^{i3}V_{i\mu }\,,  \label{3m} \\
0 &=&\delta \hat{\omega}_{z}^{i3}=\frac{1}{\ell z}\,p^{i}+\partial
_{z}j^{i3}+i_{p}\hat{R}_{z}^{i3}\,,  \label{i3z} \\
0 &=&\delta \hat{\omega}_{z}^{ij}=\partial _{z}j^{ij}+i_{p}\hat{R}%
_{z}^{ij}\,.  \label{ijz}
\end{eqnarray}

In order to solve the above equations, we need the asymptotic expansion of the contraction of the AdS curvature (\ref{pure R})
\begin{eqnarray}
i_{p}\hat{R}_{z}^{i3} &=&p^{j}\,\left( \frac{3z^{2}}{\ell ^{4}}\,\tau_{\ j}^{i}+\mathcal{O}(z^{3})\right) \,,  \notag \\
i_{p}\hat{R}_{\mu }^{i3} &=&-p^{3}\left( \frac{3z^{2}}{\ell ^{4}}\,\tau_{\ \mu }^{i}+\mathcal{O}(z^{3})\right) + p^j
\left( \frac{z^{2}}{\ell }\,E_{\ j}^{\nu }\mathcal{C}_{\ \mu \nu }^{i}
+\mathcal{O}(z^3) \right)  \,,  \notag \\
i_{p}\hat{R}_{z}^{ij} &=&p^{k}\left( -\frac{2z^{2}}{\ell ^{3}}\,E_{\ k}^{\mu}\omega _{(2)\mu }^{ij}-\frac{3z^{3}}{\ell ^{4}}\,E_{\ k}^{\mu }\omega_{(3)\mu }^{ij}+\mathcal{O}(z^4)\right).
\end{eqnarray}
Then eqs.~(\ref{3z})--(\ref{ijz}) acquire the form
\begin{eqnarray}
0 &=&\partial _{z}p^{3}\,,  \label{3z'} \\
0 &=&\mathcal{\partial }_{z}j^{i3}+\frac{1}{\ell z}\,p^{i}+\frac{3z^{2}}{\ell ^{4}}\,p^{j}\left( \tau _{\ j}^{i}+\mathcal{O}(z)\right) \,,
\label{Vpm_b'} \\
0 &=&\mathcal{\partial }_{z}p^{i}+\frac{\ell }{z}\,j^{i3}\,,  \label{pi'} \\
0 &=&\partial _{\mu }p^{3}-\hat{\omega}_{\mu }^{i3}p_{i}+j^{i3}V_{i\mu }\,,
\label{3m'} \\
0 &=&\mathcal{\partial }_{z}j^{ij}+p^{k}\left( -\frac{2z^{2}}{\ell ^{3}}
\,E_{\ k}^{\mu }\omega _{(2)\mu }^{ij}-\frac{3z^{3}}{\ell ^{4}}\,E_{\ k}^{\mu}\omega _{(3)\mu }^{ij}+\mathcal{O}(z^4)\right) \,.  \label{ijz'}
\end{eqnarray}

The first equation (\ref{3z'}) can be readily solved as
\begin{equation}
p^{3}=-\ell \sigma (x)\,,  \label{p3}
\end{equation}%
with the boundary parameter $\sigma (x)$ introduced as an integration constant. The next two ones, (\ref{Vpm_b'}) and (\ref{pi'}), can be decoupled by eliminating $j^{i3}$ and finding the differential equation in $p^{i}$
\begin{equation}
0=\mathcal{\partial }_{z}^{2}p^{i}+\frac{1}{z}\,\mathcal{\partial }_{z}p^{i}-\frac{1}{z^{2}}\,p^{i}-\frac{3z}{\ell ^{3}}\,p^{j}\left( \tau _{\ j}^{i}+\mathcal{O}(z)\right) \,.
\end{equation}
The solution for both parameters reads
\begin{eqnarray}
p^{i} &=&\frac{\ell }{z}\,\xi ^{i}+\frac{z}{\ell }\,b^{i}+\frac{z^{2}}{\ell^{2}}\,\xi ^{j}\tau _{\ j}^{i}+\mathcal{O}(z^{3})\,,  \notag \\
j^{i3} &=&\frac{1}{z}\,\xi ^{i}-\frac{z}{\ell ^{2}}\,b^{i}-\frac{2z^{2}}{\ell ^{3}}\,\xi ^{j}\tau _{\ j}^{i}+\mathcal{O}(z^{3})\,,
\end{eqnarray}
where $\xi ^{i}(x)$ and $b^{i}(x)$ are new integration constants. Eq.~(\ref{ijz'}) then leads to the solution for the Lorentz parameter
\begin{equation}
j^{ij}=\theta ^{ij}+\frac{z^{2}}{\ell ^{2}}\,\xi ^{\mu }\omega _{(2)\mu}^{ij}+\frac{z^{3}}{\ell ^{3}}\,\xi ^{\mu }\omega _{(3)\mu }^{ij}+\mathcal{O}(z^4)\,, \label{j expansion}
\end{equation}
with $\theta ^{ij}(x)$ another arbitrary functions on the boundary, identified with the Lorentz parameter.

The last equation to be solved is the asymptotic condition (\ref{3m'}) which --with all solutions plugged in-- becomes
\begin{equation}
0=\delta V_{\ \mu }^{3}=-\ell \mathcal{\partial }_{\mu }\sigma +\frac{2}{\ell }
\,\xi _{i}S_{\ \mu }^{i}-\frac{2}{\ell }\,b^{i}E_{i\mu }+\mathcal{O}
(z^{2})\,,  \label{eq-b expanded}
\end{equation}
where the linear terms cancel out. At the leading order, \eqref{eq-b expanded} implies that the parameter $b^{i}$ is not independent, i.e.
\begin{equation}
b_{i}=-\frac{\ell ^{2}}{2}\,E_{\ i}^{\mu }\partial _{\mu }\sigma +S_{\ i }^{j}\xi _{j}\,.  \label{b(x)}
\end{equation}

Overall, the radial expansion of gauge parameters in absence of fermions has the form
\begin{eqnarray}
p^{3} &=&-\ell \sigma (x)\,,  \notag \\
p^{i} &=&\frac{\ell }{z}\,\xi ^{i}(x)+\frac{z}{\ell }\,b^{i}+\frac{z^{2}}{\ell ^{2}}\,\xi ^{j}\tau _{\ j}^{i}+\mathcal{O}(z^{3})\,,  \notag \\
j^{i3} &=&\frac{1}{z}\,\xi ^{i}(x)-\frac{z}{\ell ^{2}}\,b^{i}-\frac{2z^{2}}{%
\ell ^{3}}\,\xi ^{j}\tau _{\ j}^{i}+\mathcal{O}(z^{3})\,,  \notag \\
j^{ij} &=&\theta ^{ij}(x)+\frac{z^{2}}{\ell ^{2}}\,\xi ^{\mu }\omega _{(2)\mu}^{ij}+\frac{z^{3}}{\ell ^{3}}\,\xi ^{\mu }\omega _{(3)\mu }^{ij}+\mathcal{O}(z^4)\,,  \label{parameters_0}
\end{eqnarray}
where $b^{i}(\sigma ,\xi )$ is given by \eqref{b(x)}.
It is worth emphasizing that $\omega^{ij}_{(2)}$ and $\omega^{ij}_{(3)}$ satisfy eq.~\eqref{DS}. In particular, from $\mathcal{C}^i=\frac{2}{\ell^2}\, E_j\wedge\omega_{(2)}^{ij}$, we find that the component $\omega^{ij}_{(2)}$ is directly related to the Cotton tensor, and similarly for $\tau^i_{\ \mu}$, so that the higher-order spin connection components fulfill
\begin{equation}
\mathcal{C}^i_{\ \mu\nu}=\frac{4}{\ell^2}\,\omega^i_{(2)\,[\mu\nu]}\,,\qquad  \mathcal{\mathring{D}}_{[\mu}\tau^i_{\ \nu ]}=E_{j[\mu}\,\omega^{ij}_{(3)\,\nu]}\,. \label{C-omega}
\end{equation}

The independent boundary parameters
\begin{equation*}
\sigma (x),\;\xi ^{i}(x),\;\theta ^{ij}(x)\,
\end{equation*}
are associated with dilatations, diffeomorphisms, and Lorentz transformations, respectively. This can be seen from the change of the boundary fields found from the expansion of $\delta V_{\ \mu }^{i}$,
\begin{eqnarray}
\delta E_{\ \mu }^{i} &=&\mathcal{\mathring{D}}_{\mu }\xi ^{i}+\sigma E_{\ \mu }^{i}-\theta ^{ij}E_{j\mu }\,,  \notag \\
\delta S_{\ \mu }^{i} &=&\mathcal{\mathring{D}}_{\mu }b^{i}-\sigma S_{\ \mu }^{i}-\theta ^{ij}S_{j\mu }+\frac{\ell^2}{2}\,\xi^\nu \, \mathcal{C}^i_{\ \nu\mu}\,,  \notag \\
\delta \tau _{\ \mu }^{i} &=&\mathcal{\mathring{D}}_{\mu }\left( \xi ^{j}\tau _{\ j}^{i}\right) -2\sigma \tau _{\ \mu }^{i}-\theta ^{ij}\tau _{j\mu } +2\xi^\nu \mathring{\mathcal{D}}_{[\nu}\tau^i_{\ \mu ]} \,.  \label{delta}
\end{eqnarray}
In a similar fashion, the spin connection transforms as
\begin{equation}
\delta \mathring{\omega}_{\mu }^{ij}=\mathcal{\mathring{D}}_{\mu }\theta ^{ij} -2E^{\nu [ i}E_{\ \mu }^{j]}\partial _{\nu }\sigma +\frac{4}{\ell ^{2}} \,\left( -\xi _{k}E_{\ \mu }^{[i}S^{j]k}+\xi ^{[i}S_{\ \mu }^{j]}\right) \,.  \label{delta_w}
\end{equation}
This law is consistent with the torsionless boundary, which can be shown using (\ref{delta}) and the fact that the Weyl tensor (\ref{W=0}) vanishes identically in three dimensions.

In addition, it is straightforward to check that the obtained residual symmetries match the usual PBH transformations (\ref{PBH}) in the metric formalism, where the coefficient $g_{(d)\mu\nu} = g_{(3)\mu \nu }$ now transforms homogeneously,
\begin{equation}
\delta g_{(3)\mu \nu }=\pounds _{\xi }g_{(3)\mu \nu }-\sigma g_{(3)\mu \nu }\,,
\end{equation}
because it is proportional to the holographic stress tensor. In the proof, one has to use $\mathcal{\mathring{D}}_{[\mu }E_{\ \nu ]}^{i}=0$. \bigskip

Having the full set of asymptotic parameters \eqref{b(x)}--\eqref{parameters_0} and the transformation law of the boundary fields \eqref{delta}, we are ready  to return to the conditions \eqref{symm_gauge} and discuss their consistency with respect to the residual transformations.

As we can see from the expansion \eqref{j expansion}, if we restrict to Lorentz transformations only (namely we set $p^i=0$), then \eqref{symm_gauge} implies $j_{(2)}^{ij}=j_{(3)}^{ij}=0$ so that, according to eq.~\eqref{j expansion}, our choices $S^{[ij]}=0,\,\tau^{[ij]}=0$ are naturally preserved by the Lorentz part of the residual symmetry group. We need, however, to check the consistency of these conditions against a generic residual symmetry, including the diffeomorphisms on the boundary, parametrized by $\xi^i= E^i_{\ \mu}\,\xi^\mu$. For example, the condition $S^{[ij]}=0$ changes under these asymptotic symmetries as
\begin{eqnarray}
    \left.\delta S^{[ij]}\right\vert_{S^{[ij]}=0}&=&\left(-\theta^{i}{}_k\,S^{[kj]}+\theta^{j}{}_k\,S^{[ki]}- 2 \sigma S^{[ij]}+ E^{i\mu}\,S^{[jk]}\mathring{\mathcal{D}}_\mu\xi_k\right.\nonumber\\
    &&\left.\left.- E^{j\mu}\,S^{[ik]}\mathring{\mathcal{D}}_\mu\xi_k+\omega_{(2)}{}^{[i}{}_k{}^{j]}\,\xi^k-j_{(2)}^{ij}+\frac{\ell^2}{4}\,\mathcal{C}^{kji}\,\xi_k\right)\right\vert_{S^{[ij]}=0}\nonumber\\
    &=&-3\,\omega_{(2)}^{[ij|k]}\,\xi_k=-\frac{3\ell^2}{4}\,\mathcal{C}^{[i|jk]}\,\xi_k=0\,,\label{deltaSantisym}
\end{eqnarray}
where we have used the general transformation property of $S^{ij}$, see the second of eqs.~\eqref{delta}, and the identification $\mathcal{S}^{ij} = 2S^{ij}/\ell^2$ that  holds for $S^{[ij]}=0$, in light of which the components $\mathcal{C}^{i|jk}$ of the Cotton tensor are expressed in terms of $\omega_{(2)}$, as given by eq.~\eqref{C-omega}. In deriving \eqref{deltaSantisym}, we have also used the expression of $j_{(2)}^{ij}$ as $\omega_{(2) k}^{ij}\,\xi^k$, which follows from the expansion \eqref{j expansion}. Finally, the last equality in eq.~\eqref{deltaSantisym} follows from the property of the torsion-free Cotton tensor, $\mathcal{C}_{[i|jk]}=0$. As a result, we see that the condition $S^{[ij]}=0$ is consistent with having a generic Cotton tensor because its transformation law is proportional to $\mathcal{C}_{[i|jk]}$ which vanishes. A similar analysis can be made for the condition $\tau^{[ij]}=0$ and its consistency with the asymptotic symmetries.

In this way we have proven that having symmetric Schouten tensor and  identification $\tilde{S}_{ij}=S_{ij}=\frac{\ell^2}{2}\,{\cal S}_{ij}$ could be consistently imposed together with having a generic, non-vanishing Cotton tensor. This is not the case in $\mathcal{N}=1$ supergravity discussed in \cite{Amsel:2009rr} where, for the sake of simplicity, it was imposed  $j^{ij}_{(2)}=0$ to ensure symmetric Schouten tensor, which can be consistently implemented only in asymptotically conformally flat spaces.

%%%%%%%%%%%%%%%%%%%%%%%%%%%%%%%%%%%%%%%%%%%%%%%%%
\subparagraph{Conservation law for  conformal symmetry.}

In Riemann-Cartan AdS gravity, the leading orders of the bulk fields $E_{\ \mu }^{i}$, $\omega _{\mu }^{ij}$ remain arbitrary functions on the three-dimensional boundary: they act as sources in the dual field theory.
From (\ref{Iclass}), we can generalize the quantum effective action to first order formalism,
\begin{equation}
W[E,\omega ]=-\ii \ln Z[E,\omega ]\,,
\label{Z}
\end{equation}
in such a way that the (external) gravitational sources $E_{\ \mu }^{i}$ and $\omega_{\mu }^{ij}$\ are coupled to the external currents, namely the energy-momentum tensor $J_{\ i}^{\mu }$ and the spin current $J_{\ ij}^{\mu }$, written in differential form formalism on $\partial \mathcal{M}$ as
\begin{equation}
\delta W=\int \left( \delta E^{i}\wedge J_{i}+\frac{1}{2}\,\delta \omega
^{ij}\wedge J_{ij}\right) \,.
\label{varW}
\end{equation}
Here we have introduced the 2-form currents $J=\frac{1}{2}\,J_{\mu \nu }\, \diff x^{\mu }\wedge \diff x^{\nu }$ and the usual Noether currents
1-form $^{\ast }J=J_{\mu }\,\diff x^{\mu }$ are their Hodge star duals
\begin{equation}\label{Hodgestar}
J^{\mu }=\frac{1}{2e_{3}}\,\epsilon ^{\mu \nu \lambda }J_{\nu \lambda}\,.
\end{equation}
Both in the non-supersymmetric case discussed here and in the supersymmetric case discussed in the next sections, the spin connection is not an independent source and, therefore, $J_{ij}=0$. Here we assume that taking a variation commutes with setting $\mathcal{\mathring{D}}E^{i}=0$,
since in \cite{Korovin:2017xqu} it was proven that $\delta \mathring{\omega}^{ij}$ contributes to the stress tensor so that it becomes the symmetric Belinfante-Rosenfeld tensor. In our approach it will be a consequence of Lorentz symmetry.

Invariance of the action under the transformations (\ref{delta}), written in the differential form language on $\partial\mathcal{M}$, reads
\begin{equation}
\delta E^{i}=\mathcal{\mathring{D}}\xi ^{i}+\sigma E^{i}-\theta ^{ij}E_{j}\,.
\end{equation}
After partial integration where the boundary terms are neglected, we get
\begin{equation}
0=\delta W=\int \left[ -\xi ^{i}\mathcal{\mathring{D}}J_{i}+\left( \sigma
E^{i}-\theta ^{ij}E_{j}\right) \wedge J_{i}\right] \,.
\end{equation}
This implies the following \textit{classical} \textit{conservation laws} of conformal symmetry in $d=3$
\begin{equation}\label{Wardclassical}
\begin{array}{llll}
\xi ^{i}: & 0= & \mathcal{\mathring{D}}J_{i}\,,\medskip  & \text{(conserved } J_{\mu\nu}) \\
\sigma : & 0= & E^{i}\wedge J_{i}\,,\medskip  & \text{(traceless }J_{\mu \nu}) \\
\theta ^{ij}: & 0= & E_{i}\wedge J_{j}-E_{j}\wedge J_{i}\,. & \text{(symmetric }J_{\mu \nu })
\end{array}
\end{equation}
Note that we have the full Weyl symmetry on the boundary expressed in terms of the Belinfante-Rosenfeld tensor $J_{\ i}^{\mu }$, which is traceless. The field equations lead to $J_{\mu \nu }=-(3/\ell) \, \tau _{\mu \nu }$. The second relation is not modified at the quantum level because there is no conformal anomaly in three dimensions.

Finally, let us comment that the boundary 1-forms $E^i$ and $S^i$ transform under the $d=3$ diffeomorphisms as Lie derivatives $\pounds _{\xi }E^{i}$ and $\pounds _{\xi }S^{i}$, respectively. They are also Lorentz vectors. Using the   identity from footnote \ref{footnote1} and reabsorbing a part $i_{\xi }\mathring{\omega}^{ij}$ of the Lie derivative into the local Lorentz transformation $\theta ^{ij}$, the transversal diffeomorphisms in $\pounds _{\xi }E^i$
acquire the form of local AdS translations $\mathcal{\mathring{D}}\xi^i$, with $\xi^i=i_{\xi }E^i$, plus the term $i_{\xi }T^i=0$ that vanishes
on-shell. As a result, we obtain $\left( \pounds _{\xi }+\delta _{\theta }\right) E^{i} =\mathcal{\mathring{D}}\left( i_{\xi }E^{i}\right) -\theta ^{ij}E_{j}$, that is exactly the first line in (\ref{delta}) restricted to a subgroup with $\sigma =0$.  Similarly, in the second equation in (\ref{var}) we recognize the Lie derivative combined with a local Lorentz transformation, $\left(\pounds _{\xi }+\delta _{\theta }\right) S^{i} =\mathcal{\mathring{D}} b^{i} -\theta ^{ij}S_{j}+\frac{\ell ^{2}}{2}\,i_{\xi }\mathcal{C}^{i}$, where now the Cotton tensor $\mathcal{C}^{i}$ naturally appears as a term analogous to the contraction of the torsion for the bulk fields and $b^{i}=i_{\xi }S^{i}$. Thus, the transformation law of the Schouten tensor in (\ref{delta}) is expected to have the Cotton tensor contribution. Another way to see it is by using the group theory argument presented in the Introduction,
where the $d=3$ Schouten tensor as a gauge field comes from $V_{-\mu }^{i}=\frac{1}{2}(\ell \hat{\omega}_{\mu }^{i3}-V_{\ \mu }^{i})$ in the asymptotic sector, thus it involves both $\delta \hat{\omega}_{\mu }^{i3}$ and $\delta V_{\ \mu }^{i}$. Performing the expansion explicitly again gives rise to $i_{\xi }\mathcal{C}^{i}$.

It is also interesting to observe  that there is the full non-linearly realized conformal group on the boundary, where $\omega ^{ij}$ and $S^i=\frac{\ell^2}{2}\,{\cal S}^i$ are composite fields and the dilatation gauge field  $B =\frac{1}{\ell }\,V_{\ \ \mu }^3 \diff x^\mu$ is vanishing (and it transforms as (\ref{eq-b expanded})). We can go back to its linear realization by treating those three fields as independent. Then we have to add the special conformal current $J_{(K)i}$ and the dilatation current $J_{(D)}$ in the variation of the action (\ref{varW}) via the respective couplings $\delta {\cal S}^i \wedge J_{(K)i}$ and $\delta B\wedge J_{(D)}$ and also treat $b^i$ as an independent parameter.

As a result, we will obtain the generalized transformation laws (\ref{Wardclassical}) in the form
\begin{eqnarray}
\xi ^{i} &:&\quad \mathcal{D}J_{i}=B\wedge J_{i}+{\cal S}^{j}\wedge J_{ij}+\ell{\cal S}_i\wedge J_{(D)}\,,  \notag \\
\sigma  &:&\quad \ell \diff J_{(D)}=-E^i\wedge J_i+{\cal S}^i \wedge J_{(K)i}\,, \notag \\
\theta ^{ij} &:&\quad \mathcal{D}J_{ij}=2E_{[i}\wedge J_{j]}+2{\cal S}_{[i}\wedge J_{(K)j]}\,,  \label{full} \\
b^{i} &:&\quad \mathcal{D}J_{(K)i}= E^j\wedge J_{ij}-\ell \,E_i\wedge J_{(D)}-B\wedge J_{(K)i}\,,  \notag
\end{eqnarray}
where $\mathcal{D}$ is the covariant derivative with respect to the Lorentz connection $\left. \omega ^{ij}=\mathring{\omega}^{ij}-2B^{[i} \wedge E^{j]} \right. $. The torsion constraint in a local Weyl theory involves the dilatation field and it has the form $\mathcal{D}_{[\mu }E_{\ \nu ]}^i=E_{\ [\mu }^i B_{\nu ]}$.  The expressions \eqref{full} reduce to the previous conservation laws
\eqref{Wardclassical} after setting $J_{ij}=0$, $J_{(K)i}=0$ and $J_{(D)}=0$, because these currents correspond to $\omega ^{ij}$, ${\cal S}^i$ and $B$ which are not independent sources any longer, but composite fields. Thus, the full conformal symmetry is encoded in the previous conservation law (\ref{Wardclassical}), as also observed in \cite{Korovin:2017xqu}.
An extension of the FG formalism and enhancement of the boundary theory to include the Weyl current has been analysed in \cite{Ciambelli:2019bzz}.
The superconformal group approach to the holographic currents problem in $d=3$ is discussed in  Subsection \ref{SuperA}.

In the following sections, we will extend the above analysis to the supersymmetric case.

%%%%%%%%%%%%%%%%%%%%%%%%%%%%%%%%%%%%%%%%%%%%%%%%%%%%%%%%%%%%%%%%%%%%%%%%%%%%%

\section{Pure \texorpdfstring{$\mathcal{N}=2$ AdS$_4$}{N=2 AdS4} supergravity}
\label{4d}

Pure $\mathcal{N}=2$ supergravity in four-dimensional spacetime has a field content that, when expressed in terms of  spacetime quantities, is given by the vielbein $V^a_{\ \hat \mu}$, the gravitino $\Psi_{\hat \mu A}$ (we generally omit the spinor indices),  the $SO(1,3)$ spin connection $\hat\omega^{ab}_{\hat \mu}$ and the graviphoton $\hat A_{\hat\mu}$. The Latin ($a,b,\ldots$) and Greek ($\hat{\mu}, \hat{\nu}, \ldots$) indices are the same as before and $A,\ldots=1,2$ refers to indices in the fundamental representation of the R-symmetry group. Let us  recall that the R-symmetry group is $\rm{U}(2)$ for the ungauged theory, but the Fayet-Iliopoulos term, which depends on the AdS radius $\ell$ as $P\propto 1/\ell$ in the $\rm{SU}(2)$ sector, explicitly breaks the R-symmetry to SO(2) for AdS$_4$ supergravity.  The graviphoton is an Abelian gauge field and gravitini are Majorana spinors. The conventions on fermions can be found in Appendix \ref{gammaconventions}.

A geometric formulation of the theory in $\mathcal N =2$ {\em superspace}, in the presence of a negative cosmological constant and allowing for non-trivial boundary conditions, was given in \cite{Andrianopoli:2014aqa}.\footnote{We shall adopt the notation of \cite{Andrianopoli:2014aqa} where, in particular, the metric is mostly minus. With respect to that paper, however, here we made some changes which make the formulas more transparent and better adapted to match the notation in three dimensions. More precisely, the four-dimensional Lorentz spin connection and curvature are defined with different symbols and extra minus signs: $\omega^{ab}\rightarrow -\hat{\omega}^{ab},\,R^{ab}\rightarrow -\hat{\mathcal{R}}^{ab}$ and the graviphoton gauge connection with a prefactor, $A \rightarrow -\frac{1}{\sqrt{2}}\,\hat A$. We will use Majorana spinors both in four as well in three dimensions and redefine the constants appearing in the quoted paper as $L=\frac{1}{\sqrt 2}$ and
$\frac 1\ell=2e=\frac{P}{\sqrt 2}=\sqrt{ -\frac{\Lambda}{3}}$,
where $\Lambda$  is the cosmological constant and  $\ell$  is the AdS$_4$ radius.} In that setting, the field content is expressed in terms of 1-forms {\em{in superspace}}  $\mathcal{M}^{4|2}$, that is, by the supervielbein 1-form  $(V^a,\Psi_A)$, defining an orthonormal basis of $\mathcal{N}=2$ superspace, the Lorentz spin connection 1-form $\hat\omega^{ab}$ and the 1-form graviphoton gauge connection $\hat A$.

Let us remark that the whole analysis in the present paper is presented within a spacetime approach to supergravity and not in superspace. However, to make contact with the results of  \cite{Andrianopoli:2014aqa}, to which we generally refer for the description of the bulk setting, in this section we will first present the results in the geometric superspace approach and then translate them into the spacetime approach.

In the geometric approach \cite{Castellani:1991et}, the action is written as an integral of the Lagrangian 4-form over a bosonic subspace $\mathcal{M}^{4}\subset \mathcal{M}^{4|2}$, that is
\begin{equation}
I = \underset{\mathcal{M}^4\subset \mathcal{M}^{4|2}}{\int} \mathcal{L} \,.
\end{equation}
This is because, in the geometric framework  of \cite{Castellani:1991et}, the Lagrangian 4-form is invariant under general coordinate transformations {\em in superspace} and supersymmetry transformations on spacetime, which are associated, as we are going to discuss below, with {\em diffeomorphisms} in the fermionic directions of superspace;  one can thus exploit ``general super-coordinate transformations'' to freely choose, as the bosonic submanifold of integration  in superspace, any $\mathcal{M}^{4}\subset \mathcal{M}^{4|2}$  (see also \cite{D'AuriaGeomSugra} for details on this point and \cite{Castellani:2014goa,Castellani:2015paa} for a geometric formulation of supergravity based on integral forms which allows to write the superspace action as an integral on a supermanifold).

The bulk Lagrangian 4-form for the pure $\mathcal{N}=2$ theory is given by\footnote{The precise  definition of the hatted and tilded quantities in \eqref{Lbulk} can be found in equations  \eqref{curv} and \eqref{rhopara}.}  \cite{Andrianopoli:1996cm,Andrianopoli:2014aqa}
\begin{equation}\label{Lbulk}
\begin{split}
\mathcal{L}^{\text{bulk}}= & \frac{1}{4}\hat{ \mathcal R}^{ab}V^cV^d\epsilon_{abcd}+\overline\Psi ^A\Gamma_a\Gamma_5\hat \rho_{A}V^a+\frac{{\ii}}{2}\left(\hat F+\frac{1}{2} \,\overline\Psi^A\Psi^B\epsilon_{AB}\right)\overline\Psi^C\Gamma_5\Psi^D\epsilon_{CD} \\
&-\frac{{\ii}}{2\ell}\,\overline\Psi^A\Gamma_{ab}\Gamma_5\Psi_{A}V^aV^b-\frac{1}{8\ell^2}\,V^aV^bV^cV^d\epsilon_{abcd} \\
&+\frac{1}{4}\left(\tilde F^{cd}V^aV^b\hat F-\frac{1}{12}\,\tilde F_{lm} \tilde F^{lm}V^aV^bV^cV^d\right)\epsilon_{abcd} \,,
\end{split}
\end{equation}
where we will generally omit writing of the wedge product in long expressions to lighten the notation.
This Lagrangian is written in a first order approach for the gauge field $\hat A$.

A consistent definition of the action  in the presence of non-trivial boundary conditions requires the full Lagrangian to include a boundary contribution \cite{York:1972sj,Gibbons:1976ue},  that is
\begin{equation}
\mathcal{L}= \mathcal{L}^{\text{bulk}}+\mathcal{L}^{\text{boundary}} \,.
\end{equation}

The boundary term has to ensure both a well-defined action principle (for suitable AAdS boundary conditions) and the regularity of the full action in the asymptotic region. Holographic techniques renormalize a gravity theory in a covariant way by cutting of the spacetime at the finite radius $z$. The variation of the action is expressible in terms of the variation of the sources at the conformal boundary. Due to the asymptotic behaviour of the fields, the variational problem on the boundary sources induces infinities which have to be cancelled by the introduction of counterterms. Asymptotic regularity, then, is dictated by a well-posed variational principle \cite{Papadimitriou:2005ii}.  Holographic renormalization was first introduced in \cite{Henningson:1998gx} and further developed in \cite{Skenderis:2002wp,deHaro:2000vlm,Bianchi:2001kw}, while the counterterms for Einstein-Hilbert AdS gravity were obtained in \cite{Kraus:1999di,Myers:1987yn,Emparan:1999pm,Balasubramanian:1999re}\footnote{However, the counterterm prescription given in these references does not deal with the logarithmic divergence coming from the bulk action.}. The prescription has been applied to supergravity theories, as well, in particular for computation of the superconformal anomaly \cite{Papadimitriou:2017kzw}
(for computations in the field theory side see, e.g., \cite{Katsianis:2019hhg,Papadimitriou:2019gel}).

In our context, it is more convenient to adopt a geometric approach to the renormalization problem, originally formulated in \cite{Aros:1999id,Aros:1999kt,Olea:2005gb}, which considers the addition of the topological  Euler-Gauss-Bonnet term to the bulk gravity action.  The corresponding coupling is fixed by demanding the  vanishing of the AdS curvature on the boundary. In \cite{Miskovic:2009bm,Anastasiou:2020zwc} it was shown  that adding this topological term in four dimensions is equivalent to the holographic renormalization program.\footnote{This renormalization procedure also allows to make contact with the concept of Renormalized Volume
for asymptotically hyperbolic spaces in a more mathematical framework \cite{Anastasiou:2018mfk}.} Since the method is deeply rooted in first order formulation, clearly it is particularly suitable for  embedding holographic renormalization  in supergravity and specially within the geometrical approach in superspace.

A generalization of the approach to the supersymmetric case was given in \cite{Andrianopoli:2014aqa} and analogous results for the $\mathcal{N}=1$ case were previously obtained in \cite{Amsel:2009rr}.
The supersymmetric extension of the Euler-Gauss-Bonnet term is unique for a given theory with $\mathcal{N}$ supersymmetries, and it is a total derivative, corresponding to a boundary term  taking values in the fermionic directions of superspace. It is still an open question whether there is a topological index in the superspace associated to this invariant. A useful tool to face this problem could be the integral form approach in superspace developed in \cite{Castellani:2014goa,Castellani:2015paa}.

For the case at hand, the boundary Lagrangian is given by the supersymmetric generalization of the Euler-Gauss-Bonnet term,
\begin{equation}\label{Lbdy}
\mathcal{L}^{\text{boundary}}=-\frac{\ell^2}{8}\bigg(\hat{ \mathcal R}^{ab}\hat{ \mathcal R}^{cd}\epsilon_{abcd}+\frac{8{\ii}}{\ell}\,\hat{\overline\rho}^A\Gamma_5\hat\rho_{A}-\frac{2{\ii}}{\ell}\,\hat{ \mathcal R}^{ab}\overline\Psi^A\Gamma_{ab}\Gamma_5\Psi_{A}+\frac{4{\ii}}{\ell^2}\,\diff\hat A \, \overline\Psi^A\Gamma_5\Psi^B\epsilon_{AB}\bigg)\,.
\end{equation}
The supercurvatures appearing in \eqref{Lbulk} and \eqref{Lbdy} are defined by
\begin{eqnarray}
\hat{\mathcal{R}}^{ab}&=& \diff\hat\omega^{ab} +\hat\omega^{ac}\, \wedge \, \hat{ \omega}_c{}^b \,, \\
\hat\rho_A&=& \hat{\mathcal{D}}\Psi_A - \frac 1{ 2\ell} \, \hat A \epsilon_{AB} \wedge \Psi^B=\diff\Psi_A +\frac 14 \, \Gamma_{ab} \, \hat\omega^{ab} \wedge \Psi_A- \frac 1{ 2\ell} \,\hat A \epsilon_{AB} \wedge \Psi^B \,, \\
F&=&  \diff \hat A - \,\overline\Psi^A \wedge \Psi^B\epsilon_{AB} \,. \label{curv}
\end{eqnarray}

Most notably, the same full Lagrangian can be equivalently rewritten in terms of the $\rm{OSp}(2|4)$ curvatures, which are defined as
\begin{eqnarray}
{\bf \hat{R}}^{ab}&=&\hat{\mathcal{R}}^{ab} - \frac 1{\ell^2}\, V^a V^b - \frac 1{2\ell}\,\delta^{AB} \overline\Psi_A \Gamma^{ab}\Psi_B \,,\nonumber\\
{\bf \hat{{R}}}^{a}&=&\hat{\mathcal{D}} V^a -\frac{\ii}{2}\, \overline\Psi^A \Gamma^a \Psi_A \,,\label{lagsuper}\\
\boldsymbol {\hat{\rho}}_A&=&\hat \rho_A  -    \frac{\ii}{2\ell}\,\delta_{AB}\Gamma_a \Psi^B V^a \,,\nonumber\\
\bf{\hat F}&=&  F \,\nonumber.
\end{eqnarray}
\par
When expressed in terms of the supercurvatures \eqref{lagsuper}, apart from subtleties related to the extension of the action integral to superspace (see \cite{Castellani:2014goa,Castellani:2015paa}), the full Lagrangian acquires the following form \`a la MacDowell-Mansouri \cite{MacDowell:1977jt}, that is quadratic in the super AdS curvatures $F^\Lambda=\left({\bf\hat{R}}^a,{\bf\hat \RR}^{ab},\boldsymbol{\hat\rho}_A,\hat \FF\right)$, \begin{align}\label{Lfull}
\mathcal{L}&=-\frac{\ell^2}{8}\hat\RR^{ab}\wedge \hat\RR^{cd}\epsilon_{abcd}-\ii\ell\hat{\overline\rr}^A\Gamma_5\wedge \hat\rr_A+\frac{1}{4}\hat\FF\,\wedge{}^*\hat\FF \nonumber\\
&= \frac 12 F^\Lambda \wedge\eta_{\Lambda\Sigma} F^\Sigma \,.
\end{align}
The quantity ${}^*\hat\FF$ denotes the Hodge-dual on spacetime of the field strength $\hat\FF$, namely
\begin{equation}
^{\ast }\mathbf{\hat{F}}=\frac{1}{2}\,^{\ast }\mathbf{\hat{F}}_{\hat{\mu}\hat{\nu}}\,\diff x^{\hat{\mu}}\wedge \diff x^{\hat{\nu}}=\frac{e}{4}\,\epsilon _{\hat{\mu}\hat{\nu}\hat{\rho}\hat{\sigma}}\,\mathbf{\hat{F}}^{%
\hat{\rho}\hat{\sigma}}\,\diff x^{\hat{\mu}}\wedge \diff x^{\hat{\nu}}\,,  \label{Hodge}
\end{equation}
and the 4-form Lagrangian \eqref{Lfull}  depends on the fields $\hat\Phi^\Lambda=(V^a,\hat \omega^{ab}, \Psi_A, \hat A)$ only through their field strengths $F^\Lambda$. The matrix $\eta_{\Lambda\Sigma}$, in the last line of \eqref{Lfull}, can be schematically written as $\eta_{\Lambda\Sigma}=\mathrm{diag} ( 0,-\frac {\ell^2}4\,\epsilon_{abcd}-2 \ii \ell C\Gamma_5,^*)$ and it is a Lorentz invariant (but not $\rm{OSp}(2|4)$ invariant) tensor.

It is worthwhile emphasizing that, because of this (as observed in \cite{MacDowell:1977jt} for the case of AdS$_4$ gravity), the action \eqref{Lfull} is not invariant under local $\rm{OSp}(2|4)$ transformations, even though the super AdS curvatures \eqref{lagsuper} are covariant with respect to that supergroup.
This is in fact the supersymmetric extension of what was found for AdS$_4$ gravity in \cite{Miskovic:2009bm}, where the  topologically renormalized action including the Euler-Gauss-Bonnet term was cast in the MacDowell-Mansouri form \cite{MacDowell:1977jt}.

The super AdS curvatures \eqref{lagsuper} satisfy on-shell the Bianchi ``identities''
\begin{align}\label{bianchi4d}
\begin{split}
\mathcal{\hat D}\hat\RR^{ab}&=\frac{2}{\ell^2}V^{[a}\hat\RR^{b]}+\frac{1}{\ell}\overline\Psi^A\Gamma^{ab}\hat\rr_A \,,\\
\mathcal{\hat D} \hat\RR^a&=\hat\RR^a{}_bV^b+\ii\overline\Psi^A\Gamma^a\hat\rr_A \,,\\
\mathcal{\hat D} \hat\rr^A&=\frac{1}{2\ell}\hat A\epsilon^{AB}\hat\rr_B-\frac{\ii}{2\ell}\Gamma_aV^a\hat\rr^A+\frac{1}{4}\hat\RR_{ab}\Gamma^{ab}\Psi_A-\frac{1}{2\ell}\hat\FF\epsilon^{AB}\Psi_B+\frac{\ii}{2\ell}\Gamma_a\Psi^A\hat	\RR^a \,, \\
\diff F&=2\epsilon^{AB}\overline\Psi_A\hat\rr_B \,.
\end{split}
\end{align}
Let us recall, here, some basic facts about the geometric approach to supergravity introduced in \cite{Castellani:1991et}, also known as ``rheonomic approach'' to supergravity.
First of all, it is a superspace approach, which means that the fundamental forms are given in terms of superfields that are functions of all the coordinates  of superspace $\mathcal{M}^{4|2}(x^{\hat\mu},\theta^{ \alpha A})$, where  $x^{\hat\mu}$ are commuting bosonic coordinates while  $\theta^{ \alpha A}$ are fermionic Grassmann coordinates ($\alpha=1,\ldots ,4$ denoting spinor indices), namely
\begin{align}
\begin{split}
V^a(x,\theta) & = V^a_{\ \hat{\mu}} (x,\theta)\diff x^{\hat{\mu}} + V^a _{\ {\alpha}A} (x,\theta)\diff\theta^{{\alpha}A} \,, \\
\hat{\omega}^{ab}(x,\theta) & = \hat{\omega}^{ab}_{\hat{\mu}} (x, \theta)\diff x^{\hat{\mu}} + \hat{\omega}^{ab}_{{\alpha}A} (x,\theta)\diff \theta^{{\alpha}A} \,, \\
\Psi^A_\alpha(x,\theta) & = \Psi^A_{\alpha\hat{\mu}} (x,\theta)\diff x^{\hat{\mu}} + \Psi^A_{{\alpha}|\beta B} (x,\theta)\diff\theta^{{\beta}B} \,, \\
\hat{A} (x,\theta) & = \hat{A}_{\hat{\mu}} (x, \theta)\diff x^{\hat{\mu}} + \hat{A}_{{\alpha}A} (x,\theta)\diff \theta^{{\alpha}A} \,.
\end{split}
\end{align}
They are related to the corresponding spacetime quantities
\begin{equation*}
V^a(x) = V^a_{\ \hat{\mu}} (x)\diff x^{\hat{\mu}} \,, \quad
\hat{\omega}^{ab}(x) = \hat{\omega}^{ab}_{\hat{\mu}} (x)\diff x^{\hat{\mu}} \,, \quad
\Psi^A (x) = \Psi^A_{\hat{\mu}} (x)\diff x^{\hat{\mu}} \,, \quad
\hat{A} (x) = \hat{A}_{\hat{\mu}} (x)\diff x^{\hat{\mu}} \,,
\end{equation*}
by the restrictions
\begin{align}
\begin{split}
V^a(x) & =V^a(x,\theta)|_{\theta=\diff\theta=0}= V^a_{\ \hat{\mu}} (x,0)\diff x^{\hat{\mu}} \,, \\
\hat{\omega}^{ab}(x) & = \hat{\omega}^{ab}(x,\theta)|_{\theta=\diff\theta=0}= \hat{\omega}^{ab}_{\hat{\mu}} (x,0)\diff x^{\hat{\mu}} \,, \\
\Psi^A (x) & = \Psi^A (x,\theta)|_{\theta=\diff\theta=0}=\Psi^A_{\hat{\mu}} (x,0)\diff x^{\hat{\mu}} \,, \\
\hat{A} (x) & = \hat{A} (x,\theta)|_{\theta=\diff\theta=0}=\hat{A}_{\hat{\mu}} (x,0)\diff x^{\hat{\mu}} \,.
\end{split}
\end{align}
Given the above setting, the theory on superspace can in principle contain extra dynamic information  with respect to its projection on spacetime.

For  the theory extended to superspace to have the same physical content as   the theory on spacetime, some constraints have to be imposed on the superspace supercurvatures. This  is what in \cite{Castellani:1991et} was named a set of {\em{rheonomic constraints}} to be imposed on the parametrization of the supercurvatures.

To clarify this point, let us first emphasize that, since the supersymmetry algebra closes only on-shell on the supergravity multiplet (we are not including auxiliary fields in the supermultiplet), then the Bianchi identities  (\ref{bianchi4d})  are not, in fact, identities, but have instead to be understood as relations among the superfields and their curvatures, which are satisfied on-shell. This is realized by requiring that the supercurvatures, which are defined off-shell by (\ref{lagsuper}), have to be identified on-shell as particular 2-forms on superspace, i.e. they get a parametrization on a basis of 2-forms in superspace, whose expression is uniquely determined by requiring that the relations (\ref{bianchi4d}) are satisfied.
In the expansion of the curvature 2-forms in superspace along the supervielbein basis, the  rheonomic prescription requires that the outer components of the supercurvatures must be expressed, on-shell, as linear tensor combinations of the inner components (the ``outer'' components are defined as those having at least one index along the $\Psi^A$ direction of superspace while, when the only non-vanishing components are along the bosonic vielbein, they are called ``inner''). From the physical point of view, this means that the outer components do not contain extra degrees of freedom besides those already present in the  spacetime description.
The constraints discussed above turn out to be physically equivalent to the on-shell constraints, that is to say, to the equations of motion.
This is the way in which the on-shell closure of the supersymmetry algebra is implemented within this approach through the Bianchi identities.

Let us emphasize that in this approach, which is the one adopted in \cite{Andrianopoli:2014aqa}, it turns out, as shown in \cite{Castellani:1991et}, that the supersymmetry transformations on spacetime of the fields correspond to diffeomorphisms in the fermionic directions of superspace, which can be expressed through Lie derivatives in those directions (a very nice recent review of the geometric approach to supergravity can be found in \cite{D'AuriaGeomSugra}).
In the explicit evaluation, one should keep in mind that the expressions for the curvatures which hold on-shell, where supersymmetry is realized as a symmetry of the theory, are the (rheonomic) parametrizations. A short account of the prescriptions on the supercurvatures in the geometric approach can also be found in Appendix A of \cite{Andrianopoli:2014aqa}.

In the case at hand, the on-shell (rheonomic) parametrization of the supercurvatures (\ref{lagsuper})  results to be given by the following expressions,
\begin{eqnarray}
\hat{\mathbf{R}}^{a} & = & 0 \,, \notag \\
\hat{\mathbf{F}} & = & \tilde F_{ab} V^a V^b \,, \notag \\
\hat\rr^A & = & \tilde \rho^A_{ab} V^a V^b - {\frac{\ii}{2}} \, \Gamma^a \Psi^B V^b \tilde F_{ab} \epsilon^{AB} - {\frac{1}{2}}  \Gamma_5 \Gamma^a \Psi^B V^b \, {}^* \tilde F_{ab} \epsilon^{AB}  \,, \label{rhopara} \\
\hat{\mathbf{R}}^{ab} & = & {\tilde R^{ab}}_{cd} V^c V^d - \overline{\Theta}^{ab}_{A \vert c} \Psi_A V^c - {\frac{1}{2}} \, \overline{\Psi}_A \Psi_B \epsilon_{AB} \tilde F^{ab} -  {\frac{\ii}{2}}  \overline{\Psi}_A \Gamma_5 \Psi_B \epsilon_{AB}\, {}^*\tilde F^{ab} \, , \notag
\end{eqnarray}
where the spinor-tensor $\Theta^{ab|c}_A$ is given by eq.~\eqref{thetadef}.

Note that the quantities ${\tilde R^{ab}}_{cd}$, $\tilde \rho^A_{ab} $ and $\tilde F_{ab}$, appearing in the parametrizations (\ref{rhopara}), are the so-called {\em supercovariant field strengths} and they differ in general from the spacetime projections of the supercurvatures, that is ${\hat {\mathbf{R}}}^{ab}_{\hat{\mu}\hat{\nu}}\neq 2{\tilde R^{ab}}_{cd}\, V^c_{\ \hat\mu} V^c_{\ \hat\nu}$, $\hat{\rr}^A_{\hat{\mu}\hat{\nu}} \neq 2\tilde \rho^A_{ab}\, V^a_{\ \hat\mu} V^b_{\ \hat\nu} $. However, since in the present case the parametrization of $\hat {\mathbf{F}}$ takes contribution only from the 2-bosonic vielbein sector, we have $\hat {\mathbf{F}}_{\hat{\mu}\hat{\nu}}=2\tilde F_{ab}\, V^a_{\ \hat\mu} V^b_{\ \hat\nu}$.

Taking the above discussion into account, the transformation laws of the bulk fields with respect to the symmetries of the action, which are diffeomorphisms, local Lorentz transformations, supersymmetry and $\rm{U}(1)$ gauge transformations, whose corresponding parameters are $p^a$, $j^{ab}$, $\epsilon^A$ and $\lambda$ respectively, read
\begin{eqnarray}
\delta V^a &=& \hat{\mathcal{D}} p^a- j^{ab}V_b+ {\ii} \, \overline\epsilon_A\Gamma^a\Psi^A \,,\notag\\
\delta\hat\omega^{ab} &=& \hat{\mathcal D}j^{ab} +\frac{2}{\ell^2}\,p^{[a}V^{b]}+2 \, \tilde R^{ab}{}_{cd}p^cV^d +\overline\Theta^{ab}_{A|c}\Psi^Ap^c+\frac{1}{\ell}\,\overline\epsilon^A\Gamma^{ab}\Psi_A \notag\\
&& -\overline\Theta^{ab}_{A|c}\epsilon^AV^c+\epsilon^{AB}\tilde F^{ab}\overline\Psi_A\epsilon_B +\ii \,\epsilon^{AB}\,{}^*\tilde F^{ab}\overline\Psi_A\Gamma_5\epsilon_B \,,\notag\\
\delta\Psi^A&=&-\frac{1}{4}\,j^{ab}\Gamma_{ab}\Psi^A-\frac{\ii}{2\ell}\,\Gamma_a\Psi^Ap^a+2 \, \tilde\rho^A_{ab}p^aV^b+{\frac{\ii}{2}} \, \Gamma^a\Psi_B p^b\tilde F_{ab}\epsilon^{AB}\notag\\
&& +{\frac{1}{2}}\,\Gamma_5\Gamma^a\Psi_B\,{}^*\tilde F_{ab}\, p^b\epsilon^{AB} +\frac{\hat\lambda}{2\ell}\,\epsilon^{AB}\Psi_B+{\hat{\mathcal D}}\epsilon^A-\frac{1}{2\ell}\,\hat A\epsilon^{AB}\epsilon_B   \notag\\
&& +\frac{\ii}{2\ell}\,\Gamma_a\epsilon^AV^a - {\frac{\ii}{2}} \, \epsilon^{AB}\tilde F_{ab}V^b\Gamma^a\epsilon_B -{\frac{1}{2}} \,\epsilon^{AB}\,{}^*\tilde F_{ab}\,\Gamma_5\Gamma^a\epsilon_BV^b \,, \notag\\
\delta \hat A&=& \diff \hat\lambda +2 \, \overline\epsilon^A\Psi^B\epsilon_{AB} +2 \, \tilde F_{ab}p^aV^b \,.
\label{gauge}
\end{eqnarray}
The latter generalizes to the supersymmetric case the transformation laws \eqref{var}.

In this framework, the supersymmetry invariance of the Lagrangian is expressed by the vanishing of the Lie derivative of the Lagrangian for infinitesimal diffeomorphisms in the fermionic directions, that is, $\delta _{\epsilon }\mathcal{L}=\pounds_{\epsilon }\mathcal{L}=\imath _{\epsilon }\diff
\mathcal{L}+\diff(\imath _{\epsilon }\mathcal{L})=0$. When the spacetime geometry has a non-trivial boundary $\partial \mathcal{M}$ where the superfields  do not vanish, then the condition $\imath _{\epsilon }\mathcal{L}|_{\partial \mathcal{M}}=0$ is non-trivial and determines the precise expression of the boundary contributions to the Lagrangian necessary to preserve supersymmetry invariance.

Let us finally write out the equations of motion of the theory. They can be derived equivalently from the bulk Lagrangian \eqref{Lbulk} or from the full one \eqref{Lfull}, the two expressions differing by the Bianchi relations \eqref{bianchi4d} which are satisfied on-shell.

Using the bulk Lagrangian \eqref{Lbulk} for the variations, one finds
\begin{align}
\delta\hat\omega^{ab}:& \quad V^c\hat\RR^d\epsilon_{abcd}=0 \quad \Rightarrow \quad \hat\RR^a=0 \,,\notag\\
\delta V^{a}:& \quad \frac{1}{2}\,V^b\hat\RR^{cd}\epsilon_{abcd}-\overline\Psi^A\Gamma_a\Gamma_5\hat\rr_A+\,{}^*\tilde F_{ab}\,V^b\hat \FF-\frac{1}{12}\,\tilde F^{ef}\tilde F_{ef}V^bV^cV^d\epsilon_{abcd}=0 \,, \notag \\
\delta\overline\Psi^{A}:& \quad 2\Gamma_aV^a\Gamma_5\hat\rr_A-\epsilon_{AB}\Psi^B{}^*\hat \FF+\ii\epsilon_{AB}	\hat\FF\Gamma_5\Psi^B=0 \,,\label{Psieq} \\
\delta\hat A:& \quad\diff {}^*\hat\FF -2\ii\epsilon^{AB}\overline\Psi_A\Gamma_5\hat\rr_B=0\,.\notag
\end{align}
Considering instead the variation of the Lagrangian \eqref{Lfull}, which includes the boundary contributions, the Euler-Lagrange equations formally read
\begin{eqnarray}\label{elformal}
\frac{\delta \mathcal{L}}{\delta \hat\Phi^\Lambda} \, \delta \hat\Phi^\Lambda=\frac{\partial \mathcal{L}}{\partial F^\Sigma}\cdot \frac{\partial F^\Sigma}{\partial \hat\Phi^\Lambda}\delta \hat\Phi^\Lambda=
\hat{\mathbf{D}}(F^\Sigma \eta_{\Lambda\Sigma}) \delta \hat\Phi^\Lambda + \diff \left(\frac{\partial \mathcal{L}}{\partial F^\Sigma} \delta \hat\Phi^\Sigma\right) \,,
\end{eqnarray}
where $\hat{\mathbf{D}}$ denotes the $\rm{OSp}(2|4)$-covariant derivative (not only Lorentz and gauge one),  that is
\begin{eqnarray} \label{eomwi}
 \delta I &=& \underset{\mathcal{M}^4}{\int}
 \hat{\mathbf{D}}(F^\Sigma \eta_{\Lambda\Sigma}) \, \delta \hat\Phi^\Lambda + \underset{\partial \mathcal{M}}{\int} \frac{\partial \mathcal{L}}{\partial F^\Sigma} \, \delta \hat\Phi^\Sigma=0 \,.
\end{eqnarray}
Invariance of the action  implies,  in all the bulk superspace,  the field equations
\begin{equation}\label{eom}
\hat{\mathbf{D}}\left(\frac{\partial \mathcal{L}}{\partial F^\Lambda} \right)=  \hat{\mathbf{D}}(F^\Sigma \eta_{\Lambda\Sigma})=0\,,
\end{equation}
together with  the boundary conditions
\begin{equation}\label{eombdy}
  \frac{\partial \mathcal{L}}{\partial F^\Sigma} \, \delta \hat\Phi^\Sigma|_{\partial \mathcal{M}}=F^\Lambda \, \eta_{\Lambda\Sigma} \, \delta \hat\Phi^\Sigma|_{\partial \mathcal{M}} = 0 \,.
\end{equation}
Explicitly,  as far as the bulk field equations \eqref{eom} are concerned, we find that the equations of motion for the vielbein and the gauge field have the same expressions given in \eqref{Psieq} as before, while the ones for the spin connection and for the gravitino get replaced by the (equivalent) expressions
\begin{align}
\delta\hat\omega^{ab}:& \quad -\frac{1}{2}\, \hat{\mathcal{D}} \hat\RR^{cd}\epsilon_{abcd}+\ii\, \overline\Psi^A\Gamma_{ab}\Gamma_5\hat\rr_A=0 \,,\\
\delta\overline\Psi^{A}:& \quad \frac{\ell}{4}\, \Gamma^{ab}\Psi_A\hat\RR^{cd}\epsilon_{abcd}-2\ii\ell \Gamma_5 \hat{\mathcal{D}} \hat\rr_A+\ii \Gamma_5\hat A\epsilon_{AB}\hat\rr^B +  \Gamma_aV^a\Gamma_5\hat\rr_A-\epsilon_{AB} \Psi^B{}^*\hat \FF=0 \,.
\end{align}
In our case, on the boundary we have  in general $\delta  \hat\Phi^\Sigma|_{\partial \mathcal{M}}\neq 0$  and the  boundary conditions resulting from \eqref{eombdy}, when expressed in terms of four-dimensional superfields and their derivatives, look like  Neumann boundary conditions on the supercurvatures \eqref{lagsuper},
\begin{equation}
{\bf\hat{{R}}}^{ab}|_{\partial\mathcal M}=0\,,\quad
\boldsymbol{\hat \rho}_A|_{\partial\mathcal M}=0\,,\quad
{\bf \hat F}|_{\partial\mathcal M}=0\,,\quad
{\bf\hat R}^a|_{\partial\mathcal M}=0\,. \label{AdSbound}
\end{equation}
However, let us recall that we have Dirichlet boundary conditions for the holographic fields which, because of spacetime being asymptotically AdS and given the fall-off of other bulk quantities, also implies the vanishing of the supercurvatures.

Thus, to preserve supersymmetry, the $\rm{OSp}(2|4)$ supercurvatures  \eqref{lagsuper}  are constrained on $\partial \mathcal{M}$ to their vacuum values \eqref{AdSbound}, which are indeed the Maurer-Cartan equations  of a rigid $\rm{OSp}(2|4)$ background.
Note that $\rm{OSp}(2|4)$ is also the supergroup of {global} superconformal transformations on $\mathcal{N}=2$ three-dimensional superspace, so that the above relations can be understood from the boundary point of view, in light of the AdS/CFT duality, as the conditions for superconformal invariance of the theory at the asymptotic boundary.

Let us finally mention that, in the geometric approach, in order to obtain the spacetime Lagrangian, one has to project the 4-form Lagrangian from superspace to spacetime (defined by the $\theta = 0$, $\diff \theta = 0$ hypersurface ${\mathcal{M}}^4$), namely, to restrict all the superfields, including the bosonic vielbein $V^a$ and the gravitino $\Psi_{\alpha A}$, to their lowest ($\theta^{\alpha A} = 0$, $\diff\theta^{\alpha A}=0$) components.

In the rest of this paper, we will restrict our analysis to spacetime.

%%%%%%%%%%%%%%%%%%%%%%%%%%%%%%%%%%%%%%%%%%%%%%%%%%%%%%
%%%%%%%%%%%%%%%%%%%%%%%%%%%%%%%%%%%%%%%%%%%%%%%%%%%%%%

\section{Near-boundary analysis of the supergravity fields and local parameters}\label{nba}

In the present section, we are going to apply the holographic techniques combined with the topological terms, outlined in Section \ref{WithoutF}, to the 4D  supergravity theory presented in Section \ref{4d}.

Given the pure, ${\cal N}=2$ supergravity theory, we can deduce the symmetries of its holographically dual QFT in a similar fashion as described in Section \ref{WithoutF} for AdS$_4$ gravity.
The laws (\ref{gauge}) now depend on the local parameters $p^{a},\;j^{ab},\;\hat\lambda$ and$\;\epsilon _{A}$ and we will use this freedom to fix the Lagrange multipliers associated with the radial components of the fields. For the Maxwell field, $A_z$ is not a multiplier, but a non-dynamic variable. Keeping in mind that the  $\mathcal{N}=2$ pure supergravity should, in principle, be able to describe holographically both the standard SCFT and the holographene-like unconventional supersymmetric systems \cite{Alvarez:2011gd, Andrianopoli:2018ymh, Andrianopoli:2019sip}, we will fix the multipliers as generally as possible, focusing on our particular case only starting from Subsection  \ref{FTAEP}.

We have to choose a suitable gauge that generalizes (\ref{holo gauge Vw}). The asymptotic behaviour of the vielbein in the supergravity extensions remains the same as for gravity because it is determined solely by the metric (\ref{FG}). Since the gravitini source the torsion field, we can evaluate the asymptotic behaviour of the spin connection in supergravity from the vanishing supertorsion condition in \eqref{Psieq}, as explicitly worked out in Appendix \ref{spinconnbehaviour}. Similarly, the gravitini also act as a source for the electromagnetic field, which determines the fall-off of the graviphoton connection, that was discussed in Appendix
\ref{A and Psi asymptotics}.

It remains, thus, to analyse the asymptotic behaviour of the gravitini. To this end, it is convenient to express them  in terms of chiral components with respect to the matrix $\Gamma^3$: $\Psi =\Psi _{+}+\Psi _{-}$, where the eigenstates $\Psi _{\pm }$ of the matrix $\Gamma ^{3}$ are defined by eq.~(\ref{Gamma3}). The conventions of gamma matrices are given in Appendix \ref{gammaconventions}.

The asymptotic behaviour of the gravitini is determined by the supertorsion constraints, associated with supersymmetry both in four- and three-dimensional spacetimes.  As a consequence, we are interested in gravitini whose fall-off is $\Psi _{\mu \pm }=\mathcal{O}(z^{\mp 1/2})$ and $\Psi _{z\pm }=\mathcal{O(}z^{\pm 1/2})$, as introduced
in \cite{Andrianopoli:2018ymh}.
From a group theoretical point of view, the same result is obtained from to the request of covariance with respect to the $\rm{OSp}(2|4)$ group (which describes superisometries of the bulk supergravity and superconformal transformations on the boundary), which in particular implies, as we will discuss in general terms in Subsection \ref{SuperA}, a definite scaling ($\pm 1/2$) under the subgroup $\mathrm{SO(1,1)} \subset \mathrm{OSp(2|4)}$ that parametrizes radial rescalings in the bulk and dilations on the boundary. This is better written as
\begin{equation}
\Psi _{A\mu \pm }=\left( \dfrac{z}{\ell }\right) ^{\mp \frac{1}{2}}\varphi_{A\mu \pm }(x,z)\,,\qquad \Psi _{Az\pm }=\left( \dfrac{z}{\ell }\right)
^{\pm \frac{1}{2}}\varphi _{A\pm z}(x,z)\,,  \label{Psi}
\end{equation}
where the Majorana fermions $\varphi _{A\mu \pm }$ and $\varphi _{Az\pm }$ are \textit{regular} functions at the boundary and can be expanded as power series in $z$. The second relation in \eqref{Psi} is consistent with the condition that singles out the spin 3/2 components in the gravitini,
\begin{equation}
    \Gamma^a \Psi_{A\hat\mu}\,V^{\hat\mu}_{\ \ a} =0 \,,
\end{equation}
that in the FG frame \eqref{FG} reads
\begin{equation}\label{prob}
\left( \Gamma^i \Psi_{A\mu} \right)_\pm V^{\mu}_{\ \ i} + \left( \Gamma^3 \Psi_{Az} \right)_\pm V^{z}_{\ \ 3} =0\,.
\end{equation}
 {We do not use {the above equation} in our calculations. {If} we relax it}{, though,} then more general asymptotics for the gravitini components $\Psi_{Az\pm}$ can in principle be considered.
An exploration in this direction could be relevant in view of our interest in unconventional supersymmetry in a holographic SCFT.

Since $\Psi_{A\mu\pm}$ and the transformed field $\Psi_{A\mu\pm }+\delta_{\epsilon }\Psi_{A\mu\pm }$, given by (\ref{gauge}), have to be of the same order in $z$, we have that $\delta _{\epsilon }\Psi _{A\mu \pm
}\sim \mathcal{\hat{D}}_{\mu }\epsilon _{A\pm }\sim \epsilon _{A\pm }$ are of the same order,
\begin{equation}
\epsilon _{A\pm }=\left( \frac{z}{\ell }\right) ^{\mp \frac{1}{2}}H_{A\pm}(x,z)\,,  \label{eta}
\end{equation}
where again the Majorana spinor $H_{A\pm}(x,z)$ is regular on the boundary.

Regarding the bosonic fields,  ${\hat{\omega}}^{ij}$ and {$\hat{A}$} have scaling zero with respect to $\mathrm{SO(1,1)}\subset \mathrm{OSp(2|4)}$, while  $V^i$, ${\hat{\omega}}^{i3}$ do not have a definite scaling. To make this manifest in the supersymmetric theory, it is convenient to define also bosonic quantities with definite $\mathrm{SO(1,1)}$ scaling near the boundary. They are
\begin{equation}
V_{\pm \hat\mu}^{i}=\frac{1}{2}\,\left( \ell \hat{\omega}^{i3}_{\hat\mu}\pm V^{i}_{\ \hat\mu}\right) \,,
\label{Vpm}
\end{equation}
where $V^i_+$ has scaling $+1$ and $V^i_-$ scaling $-1$. They  behave asymptotically as
\begin{equation}
V_{\pm \mu}^{i}=\left( \frac{z}{\ell }\right) ^{\mp 1}E_{\pm \mu }^{i}(x,z)\,,  \label{Epm}
\end{equation}
where the regular functions $E_{\pm }^{i}$ have the following power expansion in $z$,
\begin{eqnarray}
E_{+\mu }^{i} &=&E_{\ \mu }^{i}+\frac{z^{2}}{\ell ^{2}}\,\frac{S_{\ \mu }^{i}-
\tilde{S}_{\ \mu }^{i}}{2}+\frac{z^{3}}{\ell ^{3}}\,\frac{\tau _{\ \mu
}^{i}-2\tilde{\tau}_{\ \mu }^{i}}{2}+\mathcal{O}(z^{4})\,,  \notag \\
E_{-\mu }^{i} &=&-\frac{\ell ^{2}}{2}\,\mathcal{S}_{\ \mu }^{i}-\frac{z}{\ell }\,\frac{\tau _{\ \mu }^{i}+2\tilde{\tau}_{\ \mu }^{i}}{2}+\mathcal{O}(z^{2})\,.
\label{Epm_expand}
\end{eqnarray}

Unless stated differently, all regular functions on the boundary that appear here, $f=\{w^i,\; w^{ij},\; \varphi _{A\mu \pm },\; \varphi _{Az\pm }, \; H_{A\pm }, \ldots \}$, are generically expanded in a power series
\begin{equation}
f(x,z)=\sum_{n=0}^{\infty }\left( \frac{z}{\ell }\right)
^{n}f_{(n)}(x)=f_{(0)}(x)+\frac{z}{\ell }\,f_{(1)}(x)+\frac{z^{2}}{\ell ^{2}}
\,f_{(2)}(x)+\cdots \,.  \label{f}
\end{equation}

Using these conventions, the asymptotic expansion of the spin connection is computed in Appendix \ref{spinconnbehaviour}. It is found (see eqs.~(\ref{w})) that a suitable gauge fixing which includes gravitini has $\hat{\omega}_{z}^{ab}\neq 0$, but it is still subleading on the boundary. We choose arbitrary functions $\hat{\omega}_{z}^{i3}=w^{i}(x,z)$ and $\hat{\omega}_{z}^{ij}=\frac{z}{\ell }\,w^{ij}(x,z)$ in such a way
that they are consistent with the vanishing supertorsion condition, but we treat them off-shell as independent variables in first order formulation of supergravity.

In order to ensure that the gauge fixing of $\hat{A}_{z}$ is consistent with the supergravity dynamics imposed later, it has to satisfy the radial component of the graviphoton equation in \eqref{Psieq}, which is shown in Appendix \ref{A and Psi asymptotics}.
It turns out that having two independent components $\Psi _{Az\pm}$ is too restrictive in the context of holography because it would not allow the components of the gravitini on $\partial\mathcal{M}$, $\varphi^A_{\pm \mu}$, to be the only source of the electromagnetic field,  $\mathcal{F}=\diff A$ on $\partial\mathcal{M}$, which would be a behaviour  similar  to the one in Einstein-Maxwell gravity,
\begin{equation}
\mathbf{\hat{F}}_{\mu \nu }=0\quad \Rightarrow \quad \mathcal{F}_{\mu \nu
}=4\epsilon _{AB}\,\overline{\varphi }_{+[\mu }^{A}\varphi _{-\nu ]}^{B}\,,
\label{Abelian F}
\end{equation}
and have the U(1) gauge parameter finite on $\partial\mathcal{M}$, namely $\hat{\lambda}=\mathcal{O}(1)$. Then, as explained in Appendix \ref{A and Psi asymptotics}, the leading order of the component  $\hat{A}_{z}$, denoted by $\frac{\ell }{z}\,A_{(-1)z}$, is related to the leading order of the component $\Psi _{-Az}$, that is the function $\varphi_{-Az(0)}$. The general solution given by eq.~(\ref{A(-1)z}) requires that either both functions vanish, or $A_{(-1)z}$ to be constant and $\varphi_{(0)-Az}$ determined in terms of it.

If we are interested in a theory consistent with supersymmetry on the boundary, we have two options. The first one is to relax the gauge fixing of $\Psi _{Az-}$ by imposing the stronger condition
\begin{equation}
\Psi _{Az-}=0\,.  \label{Psi_z=0}
\end{equation}
The second one is to change the asymptotic structure of the $\rm{U}(1)$ sector, allowing for a divergent leading contribution in $\hat A_z$.

In sum, the results of Appendix \ref{spinconnbehaviour} and \ref{A and Psi asymptotics} show that the holographic gauge-fixing conditions on the local parameters $p^{a},\;j^{ab},\;\lambda ,\;\epsilon _{A}$ in AdS space have the form
\begin{equation}
\begin{array}[t]{lllll}
V_{\ z}^{3} & =\dfrac{\ell }{z}\,,\quad  & \hat{\omega}_{z}^{i3} &
=w^{i}(x,z)\,, & \Psi _{\pm Az}=\left( \dfrac{z}{\ell }\right) ^{\pm \frac{1}{2}}\varphi _{\pm Az}(x,z)\,,\medskip  \\
V_{\ z}^{i} & =0\,, & \hat{\omega}_{z}^{ij} & =\dfrac{z}{\ell }\,w^{ij}(x,z)\,,
& \hat{A}_{z}=\dfrac{\ell }{z}\,A_{(-1)z}(x)+\dfrac{z}{\ell }\,A_{(1)z}(x)+
\mathcal{O}(z^{3})\,,
\end{array}
\label{holo gauge}
\end{equation}
where we can distinguish particular cases
\begin{equation}
\begin{array}{lllll}
\Psi _{z\pm }\neq 0 & \ \Rightarrow \  & \hat{A}_{z}=\mathcal{O}(1/z)\,, &
w^{i}=\mathcal{O}(1)\,, & w^{ij}=\mathcal{O}(1)\,,\medskip  \\
\Psi _{z-}=0 & \ \Rightarrow  & \hat{A}_{z}=\mathcal{O}(z)\,, & w^{i}=%
\mathcal{O}(z^{2})\,,\quad  & w^{ij}=\mathcal{O}(1)\,,\medskip  \\
\Psi _{z\pm }=0 & \ \Rightarrow  & \hat{A}_{z}=\mathcal{O}(z)\,, & w^{i}=0\,,
& w^{ij}=\mathcal{O}(1)\,.%
\end{array}
\label{cases}
\end{equation}
Because now the gauge-fixing functions also depend on the radial and boundary coordinates,  they can be power-expanded using eq.~(\ref{f}), and for the fermions we use the notation \begin{eqnarray}
\Psi _{+z}^{A} &=&\sqrt{\frac{z}{\ell }}\,\varphi _{+z}^{A}(x,z)=\sqrt{\frac{z}{\ell }}\left[ \binom{\psi _{+z}^{A}}{0}+\frac{z}{\ell }\binom{\zeta
_{+z}^{A}}{0}+\mathcal{O}(z^{2})\right] ,  \notag \\
\Psi _{-z}^{A} &=&\sqrt{\frac{\ell }{z}}\,\varphi _{-z}^{A}(x,z)=\sqrt{\frac{\ell }{z}}\left[ \binom{0}{\psi _{-z}^{A}}+\frac{z}{\ell }\binom{0}{\zeta
_{-z}^{A}}+\mathcal{O}(z^{2})\right] .
\end{eqnarray}
It is important to emphasize that we assume that the gauge-fixing functions $\Psi_z^A(x)$ and $\hat A_z(x)$ do not transform under local transformations. This is equivalent to the statement that their transformation law can always be reabsorbed in higher-order terms of the asymptotic transformations. In contrast, the quantities $w^i(x)$ and $w^{ij}(x)$ introduced in \eqref{holo gauge} do transform, because on-shell they have to allow for the vanishing supertorsion condition. However, in first order formalism we treat them off-shell, thus they enter at the same footing as other gauge-fixing functions, with the only difference that we do not require them to be invariant under residual transformations.
Indeed, using  explicit expressions given by eqs.~\eqref{w_asympt} and \eqref{w} of Appendix \ref{Appasexp}, it is straightforward to check by varying the supertorsion that $\delta w^i$, $\delta w^{ij}\neq 0$ and that we can always set $w^i=0$ consistently (with $\delta w^i=0$), but if $w^i\neq 0$, then $\delta w^i\neq 0$ as well, otherwise imposing it would break all asymptotic symmetries.  The same is true for $w^{ij}$. Nonetheless, $\delta w^i$ and $\delta w^{ij}$ always appear at higher-order and they do not influence the near-boundary expressions.

The conditions (\ref{holo gauge}) produce the
following generic asymptotic behaviour of the boundary fields,
\begin{eqnarray}
V_{\ \mu }^{i} &=&\frac{\ell }{z}\,E_{\ \mu }^{i}+\frac{z}{\ell }\,S_{\ \mu}^{i}+\frac{z^{2}}{\ell^{2}}\,\tau^{i}{}_{\mu}+\mathcal{O}(z^{3})\,,
\nonumber \\
\hat{\omega}_{\mu }^{i3} &=&\frac{1}{z}\,E_{\ \mu }^{i}-\frac{z}{\ell ^{2}}\,
\tilde{S}_{\ \mu }^{i}-\frac{2z^{2}}{\ell ^{3}}\,\tilde{\tau}^{i}{}_{\mu }+
\mathcal{O}(z^{3})\,,  \nonumber \\
\hat{\omega}_{\mu }^{ij} &=&\omega _{\mu }^{ij}(x,z)=\omega _{\mu }^{ij}+
\frac{z}{\ell }\,\omega _{(1)\mu }^{ij}+\frac{z^{2}}{\ell ^{2}}\,\omega_{(2)\mu }^{ij}+\mathcal{O}(z^{3})\,,   \nonumber  \\
\hat{A}_{\mu } &=&A_{\mu }(x,z)=A_{\mu }+\frac{z}{\ell }\,A_{(1)\mu } +\frac{z^{2}}{\ell ^{2}}\,
A_{(2)\mu}+\mathcal{O}(z^{3})\,, \label{asympt_field (1)}\\
\Psi _{\mu +}^{A} &=&\sqrt{\frac{\ell }{z}}\,\varphi _{\mu +}^{A}(x,z)
=\sqrt{\frac{\ell }{z}}\left[ \binom{\psi _{\mu +}^{A}}{0}+\frac{z}{\ell }
\binom{\zeta _{\mu +}^{A}}{0}+\frac{z^{2}}{\ell ^{2}}\binom{\Pi _{\mu +}^{A}}{0}+
\mathcal{O}(z^{3})\right] ,  \nonumber \\
\Psi _{\mu -}^{A} &=&\sqrt{\frac{z}{\ell }}\,\varphi _{\mu -}^{A}(x,z)
=\sqrt{\frac{z}{\ell }}\left[ \binom{0}{\psi _{\mu -}^{A}}+\frac{z}{\ell }
\binom{0}{\zeta _{\mu -}^{A}}+\mathcal{O}(z^{2})\right] ,  \nonumber
\end{eqnarray}
where all functions defined on $\partial \mathcal{M}$ are finite at $z=0$. The fermions acquire a half-integer power expansion in $z$ because their bilinears, which arise from the supersymmetry transformation of the bosons, have integer power expansion in $z$. We also allow for the linear terms in $z$, absent in pure AdS gravity, because in principle they could be switched on by the supersymmetric partners.

Even though the supertorsion is zero, the torsion $\hat{T}^{a}$ does not vanish, so that  $\hat{\omega}_{\mu }^{ab}$ cannot be entirely determined by the bosonic vielbein. In
particular, the relation $\hat{\omega}_{\mu }^{i3}\sim \frac{1}{\ell }\,V_{\ \mu }^{i}$ at the leading order (see Appendix \ref{spinconnbehaviour}) is inherited from the Riemannian geometry ($K_{\mu \nu }\sim \frac{1}{\ell }\,\hat{g}_{\mu \nu }$). The subleading terms in the expansion $\tilde{S}_{\ \mu }^{i}$ and $\tilde{\tau}_{\ \mu }^{i}$ are different from the Riemannian counterparts $S_{\ \mu }^{i}\ $and $\tau _{\ \mu }^{i}$ in the supersymmetric case. The boundary Schouten tensor is now defined as
\begin{equation}
\mathcal{S}_{\ \mu }^{i}=\frac{1}{\ell ^{2}}\,( S_{\ \mu }^{i}+\tilde{S}_{\ \mu }^{i}) \,,  \label{Shouten}
\end{equation}
which is the gauge field associated with special conformal transformations, as we will identify at the end of this section. Similarly, we will later see that $-(\tau _{\ \mu }^{i}+2\tilde{\tau}_{\ \mu }^{i})/\ell$ becomes the holographic stress tensor, up to the fermionic terms.

Notice that now there is an obstruction to symmetrize $\mathcal{S}_{\ \mu }^{i}$ and the holographic stress tensor because the terms $\tilde{S}_{\ \mu }^{i}$ and $\tilde{\tau}_{\ \mu }^{i}$ are not a priori symmetric in the presence of the gravitini.

%%%%%%%%%%%%%%%%%%%%%%%%%%%%%%%%%%%%%%

\subsection{The Schouten tensor in \texorpdfstring{$d=3$}{d=3} and its superconformal extension \label{ST}}

We already saw in previous sections that the Schouten tensor plays an important role in pure AdS gravity, as it describes the first near-boundary correction of the metric given by eq.~(\ref{g(2)}). From the CFT side, it arises as a component of the superconformal connection, as shown at the beginning of Section \ref{SuperWard}. In this paragraph, we will focus on its geometric properties derived in the context of conformal gravity (for a review, see \cite{Fradkin:1985am}).

Consider a $d$-dimensional manifold characterized by a metric $g_{\mu \nu }$ and a torsionful affine connection  $\Gamma_{\phantom{\lambda}\mu \nu }^{\lambda }={\mathring{\Gamma} ^{\lambda }}_{\mu \nu }-{K^\lambda}_{\mu \nu }$, where ${\mathring{\Gamma}}  ^{\lambda }_{\mu \nu }$ is the Levi-Civita connection and ${K^{\lambda }}_{\mu \nu }$ is the contorsion tensor ${{K}^{\lambda }}_{\mu \nu }=g^{\rho \lambda }\left( T_{\rho \mu \nu }+T_{\rho \nu \mu }-T_{\mu \nu \rho }\right) $. Here, ${T_{\mu \nu }}^{\lambda }\equiv \Gamma_{\phantom{\lambda}[\mu \nu ]}^{\lambda }$ is the torsion tensor. Then the Schouten tensor obtained from the conformal constraint equation on the conformal curvature components is defined by \cite{Fradkin:1985am}
\begin{equation}
\mathcal{S}_{\mu \nu }={\mathcal R}_{\mu \nu\ }-\frac{1}{2(d-1)}\,g_{\mu \nu }\,{\mathcal R}\,, \label{Fradkin}
\end{equation}
where ${\mathcal R}_{\mu \nu }$ and ${\mathcal R}$ are, respectively, the Ricci curvature tensor and the Ricci scalar constructed from the torsionful affine connection $\Gamma_{\phantom{\lambda}\mu \nu }^{\lambda }$.  This formula coincides with \eqref{g(2)} in pure AdS gravity: in that case the Ricci tensor is symmetric and this implies that $\mathcal{S}_{\mu \nu }$ is also symmetric. In presence of torsion, the Schouten tensor has both symmetric and antisymmetric parts,
\begin{eqnarray}
\mathcal{S}_{(\mu \nu )} &=&{\mathcal R}_{(\mu \nu )}-\frac{1}{2(d-1)}\,g_{\mu
\nu }{\mathcal R}\,,  \notag \\
\mathcal{S}_{[\mu \nu ]} &=&{\mathcal R}_{[\mu \nu ]}\,. \label{SmunuSymmAndAS}
\end{eqnarray}
In particular, in $d=3$, we can explicitly evaluate its symmetric and antisymmetric parts as
\begin{eqnarray}
\mathcal{S}_{(\mu \nu )} &=&\mathcal{\mathring{R}_{\mu \nu }}-\frac{1}{4}\,g_{\mu \nu }  \mathcal{\mathring{R}} -\frac{1}{2}\,g_{\mu \nu}T_\lambda T^\lambda + T_\mu T_\nu + \tilde{T}_{\lambda \rho \nu} \left(\tilde{T}^{\lambda \rho}_{\phantom{\lambda \rho}\mu} - \tilde{T}_\mu^{\phantom{\mu} \lambda \rho} \right) - \tilde{T}_{\lambda \rho \mu} \tilde{T}_\nu^{\phantom{\nu} \lambda \rho} \notag \\
&& - \frac{1}{2}\,g_{\mu \nu}\tilde{T}_{\lambda \rho \sigma}\left( \frac{1}{2}\,\tilde{T}^{\lambda \rho \sigma} + \tilde{T}^{\lambda \sigma \rho} \right) - \nabla_{(\mu} T_{\nu)} + 2\,\nabla_\lambda \tilde{T}_{(\mu \phantom{\lambda} \nu)}^{\phantom{(\mu}\lambda} \,,
\notag \\
\mathcal{S}_{[\mu \nu ]} &=&T^{\lambda }\left( \tilde{T}_{\mu \lambda \nu }+\tilde{T}_{\mu \nu \lambda }-\tilde{T}_{\nu \lambda \mu
}\right) +2\tilde{T}_{\lambda \rho [\nu}\tilde{T}_{\mu]}^{\,\,\lambda \rho }+\nabla _{\lambda }\tilde{T}_{\mu \nu
}^{\,\,\,\,\,\,\lambda }+\nabla _{[\mu }T_{\nu ]}\,,  \label{Geometric S}
\end{eqnarray}
where we have also exploited the trace decomposition of the torsion tensor ${T_{\lambda \mu }}^{\nu }={\delta _{[ \mu }}^{\nu }T_{\lambda ]}+\tilde{T}_{\lambda \mu }^{\,\,\,\,\,\,\nu }$, with $T_{\lambda }$ and $\tilde{T}_{\lambda \mu }^{\,\,\,\,\,\,\nu }$ its trace and traceless parts, respectively. Here, $\nabla = \nabla(\mathring{\Gamma})$ denotes the  derivative with respect to the Levi-Civita affine connection and $\mathcal{\mathring{R}_{\mu \nu }}$ and $\mathcal{\mathring{R}}$ are the Ricci tensor and curvature scalar of the Levi-Civita connection, respectively.

When the torsion is non-vanishing, such as in presence of fermions, in general we  have $\mathcal{S}_{[\mu \nu ]}\neq 0$ and the symmetric part $\mathcal{S}_{(\mu \nu)}$ acquires the torsionful term.\footnote{The antisymmetric contribution is still vanishing in the special case where the torsion contains only one component, the trace $T_\lambda$, which should be also covariantly constant.} Thus, we expect that, in the context of supergravity, the ``super-Schouten tensor'' (\ref{Shouten}) is not symmetric and that it is a superconformal extension of the expression \eqref{SmunuSymmAndAS}.

The equations written above are general, valid for any Riemann-Cartan manifold. In our particular case, we have the following quantities that arise from the asymptotic expansion,
\begin{eqnarray}
S_{\mu \nu } &=&E_{i\mu }S_{\ \nu }^{i}\,,\qquad \tau _{\mu \nu }=E_{i\mu }\tau _{\ \nu }^{i}\,,  \notag \\ \label{projectingS}
\tilde{S}_{\mu \nu } &=&E_{i\mu }\tilde{S}_{\ \nu }^{i}\,,\qquad \tilde{\tau}_{\mu \nu }=E_{i\mu }\tilde{\tau}_{\ \nu }^{i}\,, \\
\mathcal{S}_{\mu \nu } &=&E_{i\mu }\mathcal{S}_{\ \nu }^{i}\,.  \notag
\end{eqnarray}
It can be shown from eq.~(\ref{w_asympt}) in Appendix \ref{spinconnbehaviour}
that, when $\varphi _{-z}^{A}=0$, the tensors $\tilde{S}_{\mu \nu }$ and $\tilde{\tau}_{\mu \nu }$ acquire the form
\begin{eqnarray}
\tilde{S}_{\mu \nu } &=&S_{\nu \mu }-\ell \,\overline{\varphi }_{(0)A+[\mu }\varphi _{(0)-\nu ]}^{A}+\ii \ell \,\overline{\varphi }_{(0)A+(\nu }\Gamma _{\mu )}\varphi _{(0)+z}^{A}\,,  \notag \\
\tilde{\tau}_{\mu \nu } &=&\frac{\tau _{\mu \nu }+3\tau _{\nu \mu }}{4}\,+\frac{\ell }{2}\,\left( -\overline{\varphi }_{A+[\mu }\varphi _{-\nu ]}^{A}+\ii \overline{\varphi }_{+(\mu }^{A}\Gamma _{\nu )}\varphi _{A+z}\right) _{(1)}\,,
\end{eqnarray}
where the last line is relevant for the holographic stress tensor, whose direct relation to $\tau _{\mu \nu }+2\tilde{\tau}_{\mu \nu }$ will be shown in Section \ref{SuperWard}.

It means that, even if we symmetrize $S_{\mu \nu }$ and $\tau _{\mu \nu } $ by suitable gauge fixing of the residual Lorentz transformations, the
fermions $\psi _{A\pm \mu }$ become an obstruction to make the tensors $\tilde{S}_{\mu \nu }$ and $\tilde{\tau}_{\mu \nu }$ symmetric for arbitrary $\psi_{A+z}$ because of the following form of their antisymmetric parts,
\begin{eqnarray}
\tilde{S}_{[\mu \nu ]} &=&S_{[\nu \mu ]}-\ell \,\overline{\varphi }_{(0)A+[\mu }\varphi _{(0)-\nu ]}^{A}\,,  \notag \\
\tilde{\tau}_{[\mu \nu ]} &=&\frac{1}{2}\,\tau _{\lbrack \nu \mu ]}-\frac{\ell }{2}\,\left( \overline{\varphi }_{(0)A+[\mu }\varphi _{(1)-\nu ]}^{A}+ \overline{\varphi }_{(1)A+[\mu }\varphi _{(0)-\nu ]}^{A}\right) \,.
\end{eqnarray}

Focusing on the Schouten tensor (\ref{Shouten}), we find, for its generalization to the  superconformal case, what we will refer to in the following as ``super-Schouten'',
\begin{equation}
\mathcal{S}_{\mu \nu }=\frac{2}{\ell ^{2}}\,S_{(\mu \nu )}-\frac{1}{\ell }\,
\overline{\varphi }_{(0)A+[\mu }\varphi _{(0)-\nu ]}^{A}+\frac{\ii }{\ell }\,\overline{\varphi }_{(0)A+(\nu }\Gamma _{\mu )}\varphi
_{(0)+z}^{A}\,,
\label{Shouten from T=0}
\end{equation}
which implies
\begin{eqnarray}
\mathcal{S}_{(\mu \nu )} &=&\frac{2}{\ell ^{2}}\,S_{(\mu \nu )}+\frac{\ii }{\ell }\,\overline{\varphi }_{(0)A+(\mu }\Gamma _{\nu )}\varphi
_{(0)+z}^{A}\,,  \notag \\
\mathcal{S}_{[\mu \nu ]} &=&-\frac{1}{\ell }\,\overline{\varphi }_{(0)A+[\mu
}\varphi _{(0)-\nu ]}^{A}\,. \label{schoutenAS}
\end{eqnarray}
This result matches eq.~(\ref{Geometric S}), showing that the symmetric part of the super-Schouten tensor contains not only  the metric term, $S_{(\mu \nu )}$, but also the fermionic terms. In addition, the antisymmetric part does not vanish for arbitrary fermions $\psi _{\pm \mu }$.
Therefore, we are not able to symmetrize the super-Schouten tensor, as this procedure would lead to conditions on the leading terms of the boundary gravitini, which have to remain unconstrained.

Similarly, the term relevant for the holographic stress tensor,
\begin{eqnarray}
\tau _{\mu \nu }+2\tilde{\tau}_{\mu \nu } &=&3\tau _{(\mu \nu )}+\ell \,\left( -\overline{\varphi }_{(0)A+[\mu }\varphi _{(1)-\nu ]}^{A}-\overline{\varphi }_{(1)A+[\mu }\varphi _{(0)-\nu ]}^{A}\right.   \notag \\
&&+\left. \ii \overline{\varphi }_{(0)+(\mu }^{A}E_{\nu )}^{i}\Gamma_{i}\varphi _{(1)A+z} +\ii \overline{\varphi }_{(0)+(\mu }^{A}E_{\nu)}^{i}\Gamma _{i}\varphi _{(1)A+z}\right) \,,
\end{eqnarray}
is not symmetric in general,
\begin{equation}\label{tauantisymm}
\tau _{\lbrack \mu \nu ]}+2\tilde{\tau}_{[\mu \nu ]}=-\ell \,\left(
\overline{\varphi }_{(0)A+[\mu }\varphi _{(1)-\nu ]}^{A}+\overline{\varphi }
_{(1)A+[\mu }\varphi _{(0)-\nu ]}^{A}\right)  \,.
\end{equation}
We will discuss more about symmetry of the holographic stress tensor in Section \ref{SuperWard}.

%%%%%%%%%%%%%%%%%%%%%%%%%%%%%%%%%%%%%%%%%%%%%%%%%%%%%%

\subsection{Field transformations and asymptotic symmetries}  \label{FTAEP}

So far, we have chosen Lagrange multipliers and other non-dynamic variables (\ref{holo gauge}) that generate the asymptotic expansion of the fields (\ref{asympt_field (1)}). In this and in the following section, we will focus on the case with $\Psi _{Az-}=0$. A stronger condition $\Psi _{Az\pm }=0$, referred to as `FG gauge', was considered in \cite{Amsel:2009rr} in the context of $\mathcal{N}=1$ AdS$_4$ supergravity.
An advantage of having $\Psi _{Az+}\neq 0$ is to provide more freedom that could be used to simplify complicated fermionic expressions. We will see, though, that the presence of this particular field would not modify the asymptotic behaviour of the theory.

%%%%%%%%%%%%%%%%%%%%%%%%%%%%%%%%%%%%%%%%%%%%%%%%%%%%%%
\subparagraph{Boundary conditions on the curvatures.}

The $\mathrm{OSp}(2|4)$ supercurvatures vanish at the boundary in asymptotically AdS space, as expressed by the conditions (\ref{AdSbound}). In particular, the supertorsion vanishes exactly and its consequences are discussed in Appendix \ref{spinconnbehaviour}. The other supercurvature conditions at the boundary, whose explicit expressions are given by eq.~(\ref{MCboundary}) in Subsection \ref{SuperA}, boil down to the following constraints on $\partial \mathcal{M}$,
\begin{align}
\mathcal{D}E^{i}-\frac{\ii }{2}\,\overline{\psi}{}_{+}^{A}\wedge \gamma^{i}\psi _{A+}& =0\,,  \notag \\
\mathcal{R}^{ij}-2\,E^{[i}\wedge \mathcal{S}^{j]}-\frac{1}{\ell }\,\overline{\psi }{}_{+}^{A}\wedge \gamma ^{ij}\psi _{A-}& =0\,,  \notag \\
\nabla \psi _{+}^{A}+\frac{\ii }{\ell}\,E^{i}\wedge \gamma _{i}\psi_{-}^{A}& =0\,,
\label{AAdSspace}
\end{align}
where $\mathcal{R}^{ij}$ is the Riemann curvature tensor 2-form at the boundary and $\mathcal{S}^{i}$ is the boundary super-Schouten 1-form defined in (\ref{Shouten}).

The first equation ensures the vanishing boundary supertorsion, by fixing the boundary torsion $T^{i}=\mathcal{D}E^{i}$ in terms of the gravitini. The second equation involves the boundary Weyl tensor $W^{ij}=\mathcal{R}^{ij}-2\,E^{[i}\wedge \mathcal{S}^{j]}$ and it can be interpreted as the super Weyl tensor that vanishes on the boundary.

All three equations can be explicitly solved in the boundary fields $\omega ^{ij}$, $\mathcal{S}^{i}$ and  $\psi _{-}^{A}$. While the spin connection has been solved in Appendix \ref{spinconnbehaviour}, here we focus on the other two composite fields. Using the gamma matrix relation $\gamma_{\mu\nu}= \gamma_\mu \gamma_\nu -g_{\mu\nu}$, from the third  of \eqref{AAdSspace} we get the conformino,
\begin{equation}
\psi_{-A\,\mu}=-\frac{\ell}{2e_{3}}\,\epsilon^{\lambda\nu\rho}\gamma_\lambda\gamma_\mu\nabla_\nu \psi_{+A\,\rho}\,,
\label{conformino}
\end{equation}
while from the second one we solve the super-Schouten tensor,
\begin{equation}
\mathcal{S}_{\mu\nu}=\mathcal{R}_{\mu\nu} - \frac 14 \,g_{\mu\nu}  \mathcal{R} -\frac 1{\ell}
\left(\overline{\psi}_{+A\,\rho}\gamma^\rho_{\ \mu} \psi_{-A\, \nu} -\overline{\psi}_{+A\,\nu}\gamma^\rho_{\ \mu} \psi_{-A\, \rho} -\frac 12 \,g_{\mu\nu}\overline{\psi}_{+A\,\rho}\gamma^{\rho \lambda} \psi_{-A\, \lambda}
\right).
\label{schouten}
\end{equation}
We see that the above tensor is indeed  a superconformal extension of the expression \eqref{SmunuSymmAndAS}.

This result implies that the super-Schouten tensor $\mathcal{S}^i_{\ \mu}$ and its superpartner, the conformino $\psi_{-A\,\mu}$, are not independent sources on $\partial \mathcal{M}$, since they can be expressed in terms of the supervielbein $(E^i_{\ \mu},\psi_{+A\, \mu})$ and their curvatures.

At the end, let us comment that, at first sight, it looks like we are dealing with several different expressions for the Schouten tensor. Its definition (\ref{Shouten}) has a geometric origin, as explained in Subsection \ref{SuperA}, and it is a component of the $d=3$ superconformal field associated with the conformal boosts. From the point of view of the $D=4$ bulk fields, it comes from the vielbein and the spin-connection combined in the negative grading quantity with respect to $\mathrm{O}(1,1) \subset \mathrm{SO}(2,4)$ dilations. The vanishing supertorsion condition leads to the $\mathcal{R}$-independent kinematic relation between the super-Schouten tensor (\ref{Shouten from T=0}) and $S_{(\mu \nu )}$ in the superconformal case. In contrast, the asymptotically AdS condition and the vanishing supercuvatures on the boundary (\ref{AAdSspace}) lead to the $\mathcal{R}$-dependent Schouten tensor (\ref{schouten}). Matching these two formulas expresses $S_{(\mu \nu )}$ in terms of the boundary curvature $\mathcal{R}_{\mu\nu}$ plus the fermion bilinears, that has to be fulfilled on-shell. In pure AdS gravity, for instance, it comes down to the known relation $\mathcal{S}_{\mu \nu }=\frac{2}{\ell^2}\,S_{\mu \nu }=\mathcal{R}_{\mu \nu }-\frac 14 \,g_{\mu \nu }\mathcal{R}$ obtained by solving the Einstein equations near the boundary. Thus, two equations have different origin, but they have to be consistent on-shell.

On the other hand, the definition of the Schouten tensor (\ref{Fradkin}) is the one usually found in the literature \cite{Fradkin:1985am}, obtained from the conformal constraint equation. The superconformal version of this constraint leads to the super-Schouten tensor (\ref{schouten}) found in our case, together with its superpartner \eqref{conformino}.

%%%%%%%%%%%%%%%%%%%%%%%%%%%%%%%%%%%%%%%%%%%%%%%%%%%%%%
\subparagraph{Rheonomic parametrizations.}

The transformation laws \eqref{gauge} depend explicitly on the contractions of the supercurvature. A proper way to account for all contributions requires to know the near-boundary behaviour of the rheonomic parametrizations that appear in eqs.~(\ref{gauge}).

The simplest way to proceed is to project the expressions \eqref{rhopara} for the rheonomic pa\-ra\-me\-tri\-zation of the supercurvatures on the spacetime manifold and identify their asymptotic behaviour  with the  one of the spacetime projections of the supercurvatures (\ref{curv}). One can start  from  the $\rm{U}(1)$ field strength, whose parametrization in \eqref{rhopara} in the case at hand takes value on the 2-vielbein component only. One then proceeds to find $\tilde{\rho}_{ab}^{A}$ from the curvature of the gravitino,  which can be further used to compute $\Theta _{A|c}^{ab}$ and {$\tilde{R}_{\ \ cd}^{ab}$} in the last of \eqref{rhopara}.

Following this procedure, we determine the asymptotic behaviour of all the supercovariant field strengths, whose derivation is fully carried out in Appendix \ref{appC}. The asymptotic expansion of  $\tilde{F}_{ab}$ and $\tilde{\rho}_{ab}^{A}$
leads to%
\begin{equation}
\begin{array}[b]{llll}
\tilde{F}_{ij} & =\mathcal{O}(z^{3})\,, & \tilde{F}_{i3} & =-\dfrac{1}{2\ell
}\left( \dfrac{z}{\ell }\right) ^{2}A_{(1)\mu }E_{i}^{\mu }+\mathcal{O}%
(z^{3})\,,\medskip  \\
\tilde{\rho}_{ij+}^{A} & =\mathcal{O}(z^{5/2})\,, & \tilde{\rho}_{i3+}^{A} &
=-\dfrac{1}{2\ell }\left( \dfrac{z}{\ell }\right) ^{\frac{3}{2}}E_{i}^{\mu
}\zeta _{\mu +}^{A}+\mathcal{O}(z^{5/2})\,,\medskip  \\
\tilde{\rho}_{ij-}^{A} & =\mathcal{O}(z^{5/2})\,,\qquad  & \tilde{\rho}%
_{i3-}^{A} & =\mathcal{O}(z^{5/2})\,.%
\end{array}
\label{gravcurvexp}
\end{equation}
In order to find a radial power expansion of $\tilde{R}^{ab}{}_{cd}$, one needs the $\Theta _{A|c}^{ab}$ coefficients, which are found by inserting (\ref{gravcurvexp}) into the definition (\ref{thetadef}), as shown in Appendix \ref{appC}. After lengthy but straightforward calculation, one obtains
\begin{align}
\tilde{R}^{i3}{}_{jk}& =\frac{\ii}{2\ell }\left( \frac{z}{\ell }\right) ^{2}E_{[j}^{\mu }E_{k]}^{\nu }\overline{\psi }_{\mu +}^{A}\left(
\rule{0pt}{13pt}\gamma ^{i}\zeta _{A\nu +}+\gamma ^{l}\zeta _{A\rho +}E_{l\nu }E^{i\rho }\right) +\mathcal{O}(z^{3})\,,  \notag \\
\tilde{R}^{ij}{}_{k3}& =-\frac{1}{2\ell }\left( \frac{z}{\ell }\right)
^{2}E_{k}^{\mu }\left( \rule{0pt}{13pt}\omega _{(1)\mu }^{ij}-\ii\,
\overline{\psi }_{\mu +}^{A}\gamma ^{\lbrack i}E^{j]\nu }\zeta _{A\nu+}\right) +\mathcal{O}(z^{3})\,,  \notag \\
\tilde{R}^{i3}{}_{j3}& =\mathcal{O}(z^{3})\,,\qquad \tilde{R}^{ij}{}_{kl}=%
\mathcal{O}(z^{3})\,.  \label{Rheo R}
\end{align}

It is worthwhile noticing that all expansions (\ref{gravcurvexp}) and (\ref{Rheo R}) are subleading in $z$ and, when they are slower than $\mathcal{O}(z^{3})$, this is due to the presence of $\omega _{(1)\mu }^{ij}$ and $\zeta _{\mu +}^{A}$.
We will show below that the higher-order residual symmetries can be used to cancel out such linear terms, similarly as in pure AdS gravity.

%%%%%%%%%%%%%%%%%%%%%%%%%%%%%%%%%%%%%%%%%%%%%%%%%%%%%%

\subparagraph{Residual symmetries.}
We look for the residual symmetries of the form (\ref{gauge}) that leave the gauge fixing unaltered on the boundary,
\begin{equation}
\delta V_{\ z}^a=0\,,\quad \delta \hat{\omega}_{z}^{ij}={\mathcal O(z)}\,,\quad \delta\hat\omega^{i3}_z=\mathcal O(z^2)\,,\quad \delta
\hat{A}_{z}=0\,,\quad \delta \Psi _{\pm Az}=0\,.  \label{res}
\end{equation}
The non-dynamic fields in (\ref{holo gauge}) are functions of the boundary coordinates through $w^{i}$, $w^{ij}$, $\varphi _{+Az}$ and ${\hat{A}}_z$. In (\ref{res}), we assume that $\hat{A}_z(x)$ and $\Psi _{\pm Az}(x)$ do not change under general coordinate transformations, even though they depend on $x^{\mu }$. We will show that this assumption will not break the boundary symmetries, but only modify subleading parameters.
On the other hand, the functions $w^i(x)$ and $w^{ij}(x)$ change under the coordinate transformations because, on-shell, they have to satisfy the supertorsion constraint. In fact, it would have been more natural to allow all $x^\mu$-dependent quantities to transform non-trivially under boundary coordinate transformations, but we do not account it for simplicity.
Allowing the fields {$\hat{A}_z$ and $\Psi _{+Az}$} to transform might be related to the unconventional supersymmetry on the boundary discussed in \cite{Alvarez:2011gd,Andrianopoli:2018ymh}, where a spinor $\chi(x^{\mu })$ arises from the gauge fixing of the gravitini \cite{Andrianopoli:2019sqe}.

The corresponding parameters can be expanded as in eq.~(\ref{f}), where we keep the same notation for the leading orders of the bosonic parameters as in (\ref{parameters_0}),
\begin{align}
p^{i}& =\frac{\ell }{z}\,\xi ^{i}+\frac{z}{\ell }\,p_{(1)}^{i}+\frac{z^{2}}{%
\ell ^{2}}\,p_{(2)}^{i}+\mathcal{O}(z^{3}),  \notag \\
p^{3}& =-\ell \sigma +\frac{z}{\ell }\,p_{(1)}^{3}+\frac{z^{2}}{\ell ^{2}}%
\,p_{(2)}^{3}+\frac{z^{3}}{\ell ^{3}}\,p_{(3)}^{3}+\mathcal{O}(z^{4}),
\notag \\
j^{ij}& =\theta ^{ij}+\frac{z}{\ell }\,j_{(1)}^{ij}+\frac{z^{2}}{\ell ^{2}}%
\,j_{(2)}^{ij}+\frac{z^{3}}{\ell ^{3}}\,j_{(3)}^{ij}+\mathcal{O}(z^{4})\,,
\notag \\
j^{i3}& =\frac{1}{z}\,\xi ^{i}+\frac{z}{\ell}\,j_{(1)}^{i3}
+\frac{z^{2}}{\ell ^{2}}\,j_{(2)}^{i3}+\mathcal{O}(z^{3})\,,  \notag \\
\hat \lambda & =\lambda +\frac{z}{\ell }\,\lambda _{(1)}+\mathcal{O}(z^{2})\,,
\notag \\
\epsilon _{+}^{A}& =\sqrt{\frac{\ell }{z}}\,H_{+}(x,z)=\sqrt{\frac{\ell }{z}}%
\binom{\eta _{+}^{A}}{0}+\sqrt{\frac{z}{\ell }}\binom{\eta _{(1)+}^{A}}{0}+%
\mathcal{O}(z^{3/2})\,,  \notag \\
\epsilon _{-}^{A}& =\sqrt{\frac{z}{\ell }}H_{-}(x,z)=\sqrt{\frac{z}{\ell }}%
\binom{0}{\eta _{-}^{A}}+\left( \frac{z}{\ell }\right) ^{\frac{3}{2}}\binom{0}{\eta _{(1)-}^{A}}+\mathcal{O}(z^{5/2})\,.
\end{align}
In the above expansion, the first subleading Lorentz parameter can be consistently set to zero,
\begin{equation}
j_{(1)}^{ij}=0\,.
\end{equation}

As a first step in finding the asymptotic symmetries, we will analyse the linear terms in the transformation laws. The equation $\delta \hat{\omega}_{z}^{ij}=0$ from (\ref{res}) leads to a simple expression
\begin{equation}
\partial _{z}j^{ij}-\frac{1}{\ell }\,\xi ^{\mu }\omega _{(1)\mu }^{ij}-\frac{\ii}{\ell }\,\xi ^{\mu }\overline{\psi }_{\mu +}^{A} E^{\nu [i}\gamma ^{j]}\zeta _{A\nu +}+\frac{\ii}{\ell }\,\overline{\eta }_{+}^{A}E^{\nu [i}\gamma ^{j]}\zeta _{A\nu +}+\mathcal{O}(z)=0\,,
\end{equation}
which, taken at the leading order, amounts to solving the algebraic equation
\begin{equation}
\xi ^{\mu }\omega _{(1)\mu }^{ij}=\ii\left( \overline{\eta }_{+}^{A}-\xi
^{\mu }\overline{\psi }_{\mu +}^{A}\right) E^{\nu [i}\gamma^{j]}\zeta _{A\nu +}\,.
\end{equation}
Since $\xi ^{i}$ and $\eta _{+}^{A}$ are arbitrary and we also know that $\omega _{\mu }^{ij}$ is the composite field (explicitly computed in Appendix \ref{spinconnbehaviour}) that does not contain the linear terms, $\omega_{(1)\mu }^{ij}=0$, we can choose a particular solution for $\zeta _{A\mu +}$ that vanishes, with the result
\begin{equation}
\omega _{(1)\mu }^{ij}=0\,,\qquad \zeta _{A\mu +}=0\,.  \label{omega (1)}
\end{equation}
This choice has also been made in \cite{Amsel:2009rr} in ${\cal N}=1$ supergravity. In our case, when ${\cal N}=2$, it becomes the unique solution both when $\Psi_{-z}=0$ and $\Psi_{-z}\neq 0$ if one imposes the stronger gauge-fixing condition (\ref{strongercond}) (for more detailed discussion, see eq.~(\ref{varphi1pmzero}) in Appendix \ref{A and Psi asymptotics}). It is crucial that these fields  remain zero after a generic local transformation, namely $\delta \omega _{(1)\mu }^{ij}=0$ and $\delta \zeta _{A\mu+}=0$, as we discuss in the next paragraph.

Another constraint on the parameters arises from the fact that the FG coordinate frame (\ref{FG}) does not admit the finite terms in the expansions of $V_{\ \mu }^{i}$ and $\hat{\omega}_{\mu }^{i3}$. Local invariance preserves this frame only if
\begin{equation}
0= \delta V_{(0)\mu }^{i}=-\frac{1}{\ell }\,E_{\ \mu }^{i}\, p_{(1)}^{3}\quad \Rightarrow \quad p_{(1)}^{3}=0\,.  \label{p(1)3}
\end{equation}
Then, using the expansion of the rheonomic parametrizations given in Appendix \ref{appC}, we find that $\delta \hat{\omega}_{(0)\mu }^{i3} =-\frac{1}{\ell ^{2}}\,E_{\ \mu }^{i}\,p_{(1)}^{3}=0$ is satisfied as well.

On the other hand, the invariance of $\Psi _{\pm z}^{A}$ under (\ref{res}) yields at the leading order
\begin{eqnarray}
0 &=&\delta \Psi _{+z}^{A}\quad \overset{\mathrm{order } \sqrt{\frac{\ell}{z}}}{\Longrightarrow}  \quad 0=\frac{1}{\ell }\left( \eta_{(1)+}^{A}-\xi ^{\mu }\zeta _{\mu +}^{A}\right) \,,
 \\
0 &=&\delta \Psi _{-z}^{A}\quad \overset{\mathrm{order } \sqrt{\frac{z}{\ell}}}{\Longrightarrow}  \quad 0=\frac{1}{\ell }\,\left( \eta_{(1)-}^{A}{-}\xi ^{\mu } \zeta _{\mu -}^{A}\right){+\frac{\ii}{{4}\ell} \epsilon^{AB} A_{(1)\mu} \gamma^\mu \left( \eta_{B+}-\xi^\nu \psi_{B\nu+} \right) }\, , \notag
\end{eqnarray}
which can be solved using eq.~(\ref{omega (1)}) as
\begin{equation}\label{eta1+ANDeta1-}
\eta _{(1)+}^{A}=0\,,\qquad \eta _{(1)-}^{A}=\xi ^{\mu }\zeta _{\mu -}^{A}{-\frac{\ii}{{4}}\, \epsilon^{AB} A_{(1)\mu} \gamma^\mu \left( \eta_{B+}-\xi^\nu \psi_{B\nu+} \right) }\,.
\end{equation}

In addition, the transformation law of the radial component of the graviphoton implies
\begin{equation}
0=\delta \hat{A}_{z}=\frac{1}{\ell }\,\lambda _{(1)}-\frac{1}{\ell }
\,A_{(1)\mu }E_{\ i}^{\mu }\xi ^{i}+\mathcal{O}(z)\quad \Rightarrow \quad
\lambda _{(1)}=A_{(1)\mu }\xi ^{\mu }\,.
\end{equation}

Finally, let us require $\delta \hat{\omega}_{z}^{i3}=0$ and $\delta V_{\ z}^{i}=0$ in eqs.~(\ref{res}). At the finite order, they have the form
\begin{equation}
0=\delta V_{(0)z}^{i} =\ell \,\delta \hat{\omega}_{(0)z}^{i3} =j_{(1)}^{i3}+\frac{1}{\ell }\,p_{(1)}^{i}+w_{(0)}^{ij}\,\xi_{j} +\ii \,\bar{\eta}_{+A}\gamma ^{i}\,\psi _{+z}^{A}\,.
\end{equation}
There are two unknown parameters, namely $p_{(1)}^{i}$ and $j_{(1)}^{i3}$, and only one equation, that leads to an arbitrary vector $K^{i}(x)$ in the solution, associated with the special conformal transformations on $\partial \mathcal{M}$, as we will prove later. The solution for the first order parameters is
\begin{align}p_{(1)}^{i}& =\ell m^{i}+\frac{\ell ^{2}}{2}\,K^{i}\equiv b^{i}, \\
\ell j_{(1)}^{i3}& =\ell m^{i}-\frac{\ell ^{2}}{2}\,K^{i}\equiv -\tilde{b}^{i}\,,  \notag
\end{align}
where $m^{i}(x)$ is a function that depends on the gauge fixing,
\begin{equation}
m^{i}(x)=-\frac 12\,\left( w^{ij}_{(0)}\xi _{j}+\ii \,\overline{\eta }
_{+}^{A}\gamma ^{i}\psi _{Az+}\right)\,.  \label{m}
\end{equation}
At the linear order in $z$, we get
\begin{align}
0& =\delta V_{(1)z}^{i}=j_{(2)}^{i3}+\frac{2}{\ell }\,p_{(2)}^{i}+n^{i}\,, \notag \\
0& =\ell \,\delta \hat{\omega}_{(1)z}^{i3} =2j_{(2)}^{i3}+\frac{1}{\ell }\,p_{(2)}^{i}+s^{i}\,,
\end{align}
where we denoted
\begin{eqnarray}
n^i(x) &=&w_{(1)}^{ij}\xi_{j}+\ii\,\bar{\eta}_{+A} \gamma^{i}\,\zeta _{+z}^{A}\,,  \notag \\
s^i(x) &=&-\frac{1}{\ell }\,\xi ^{\mu }(\tau -4\tilde{\tau})^{i}{}_{\mu }+\ii\,\,\bar{\eta}_{+A}\gamma ^{i}\,\zeta _{+z}^{A}-\xi ^{\mu }E^{\nu i
}\bar{\psi}_{+A\mu }\zeta _{-\nu }^{A}-\ii \,\xi ^{\mu }\bar{\psi}_{+A\mu }\gamma ^{i}\zeta _{+z}^{A}  \notag \\
&&-\frac{\ii}{{4}}\xi ^{\mu }E^{\nu i}\epsilon _{AB}\overline{\psi }_{\mu +}^{A}\gamma ^{\rho }\psi _{\nu +}^{B}A_{(1)\rho }+E^{\mu i}\overline{\eta }_{+}^{A}\left( {\frac{\ii}{{4}}\epsilon _{AB}\gamma ^{\rho }\psi_{+\mu }^{B}A_{(1)\rho }}+{\zeta _{A-\mu }}\right) .
\end{eqnarray}
The function $w_{(1)}^{ij}$ can be determined from the vanishing supertorsion equation (\ref{t1muzzero}) in Appendix \ref{SuperCurvatures},
\begin{equation}
{w_{(1)}^{ij}}=-\frac{2}{\ell \,}\,(\tau -\tilde{\tau})^{ij}-\ii\,E^{\mu j}\overline{\psi }_{+A\mu }\gamma ^{i}\,\zeta _{+z}^{A}\,.
\end{equation}
The solution for the second order parameters $p_{(2)}^{i}$ and $j_{(2)}^{i3}$ is unique,
\begin{eqnarray}
p_{(2)}^{i} &=&\frac{\ell }{3}\left( s^{i}-2n^{i}\right) \,,  \notag \\
\ell j_{(2)}^{i3} &=&\frac{\ell }{3}\left( n^{i}-2s^{i}\right) \,.
\end{eqnarray}
In our computations, we will need only the following combination of the parameters,
\begin{eqnarray}\label{pminusj}
\ell j_{(2)}^{i3}-p_{(2)}^{i} &=&\ell \left( n^{i}-s^{i}\right) =-\xi ^{\mu}(\tau +2\tilde{\tau})_{\ \mu }^{i}+\ell \xi ^{\mu }E^{\nu i} \bar{\psi}_{+A\mu }\zeta _{-\nu }^{A}   \\
&&+{\frac{\ii\ell }{{4}}}\,\xi ^{\mu }E^{\nu i}\epsilon _{AB}\overline{\psi }_{\mu +}^{A}\gamma ^{\rho }\psi _{\nu +}^{B}A_{(1)\rho }-\ell E^{\mu i}\overline{\eta }_{+}^{A}\left( {{\frac{\ii}{{4}}}\,\epsilon _{AB}\gamma^{\nu }\psi _{+\mu }^{B}A_{(1)\nu }}+{\zeta _{A-\mu }}\right) . \notag
\end{eqnarray}

After all the above considerations and writing only first few terms, the residual local parameters can be written as
\begin{align}
p^{3}& =-\ell \sigma +\mathcal{O}(z^2)\,,  \notag \\
p^{i}& =\frac{\ell }{z}\,\xi ^{i}+\frac{z}{\ell }\,b^{i}+\frac{z^{2}}{\ell
^{2}}\,p_{(2)}^{i}+\mathcal{O}(z^{3})\,,  \notag \\
j^{i3}& =\frac{1}{z}\,\xi ^{i}-\frac{z}{\ell ^{2}}\,\tilde{b}^{i}+\frac{z^2}{\ell ^2}\, j^{i3}_{(2)}+\mathcal{O}(z^{3})\,,  \notag \\
j^{ij}& =\theta ^{ij}+\mathcal{O}(z^{2})\,,   \label{parameters_b} \\
\hat{\lambda}& =\lambda +\frac{z}{\ell}\,A_{(1)\mu }\xi ^{\mu} +\mathcal{O}(z^{2})\,,  \notag \\
\epsilon _{+}^{A}& =\sqrt{\frac{\ell }{z}}\binom{\eta _{+}^{A}}{0}+\mathcal{O%
}(z^{1/2})\,,  \notag \\
\epsilon _{-}^{A}& =\sqrt{\frac{z}{\ell }}\binom{0}{\eta _{-}^{A}}+\mathcal{O%
}(z^{3/2})\,, \notag
\end{align}%
where the $p_{(2)}^{i}$ and $j_{(2)}^{i3}$ contributions will play a role in cancellation of terms in the next step, but they will not influence the transformation law of the holographic fields. We also expect that the conservation laws do not depend on $m^{i}$ because it is a gauge-fixing function. Without the gravitini, we have $b^{i}=\tilde{b}^{i}=\frac{\ell^2}{2}\,K^{i}$, $w^{ij}=0$,
and the result coincides with the pure AdS case (\ref{parameters_0}).

Therefore, the independent residual parameters in $\mathcal{N}=2$ AdS$_{4}$
supergravity  are
\begin{equation*}
\sigma (x),\;\xi ^{i}(x),\;\theta ^{ij}(x)\,,\;\lambda (x)\,,\;\eta_{\pm}^{A}(x)
\end{equation*}
and they are associated, respectively, with the dilatations, diffeomorphisms, Lorentz, Abelian, and supersymmetry transformations in the holographically dual theory.

The parameters $b^{i}$ and $\tilde{b}^i$ have not been taken into account because $b^i-\tilde{b}^i= 2\ell m^i$ is non-physical and $b^i+\tilde{b}^i= \ell^2 K^i $ is not independent due to the last condition (\ref{z-orthogonality}). Its invariance implies
\begin{equation}\label{delta_V3}
0=\delta V_{\ \mu }^{3}=-\ell \mathcal{\partial }_{\mu }\sigma -
\ell E_{\ \mu }^{i}K_{i} +\ell \xi _{i}\mathcal{S}_{\ \mu }^{i}+\overline{\eta }
_{A+}\psi_{-A\mu }-\overline{\eta }_{A-}\psi_{+A\mu }+\mathcal{O}(z)\,.
\end{equation}
The finite part of the above equation can be solved in $K^i= ( b^{i}+\tilde{b}^{i})/ \ell^2 $ as
\begin{equation}
 K^i=\frac{1}{\ell} \,E^{\mu i}\left( -\ell \mathcal{\partial }_{\mu }\sigma +\ell \xi _{j}\,\mathcal{S}_{\ \mu }^{j}+\overline{\eta }_{A+}\psi^A_{-\mu }-\overline{\eta }_{A-}\psi^A_{+\mu }\right)
\,,  \label{xi_minus}
\end{equation}
confirming that it is not an independent local parameter. This analysis completes the radial expansion of the asymptotic parameters up to the relevant order.

%%%%%%%%%%%%%%%%%%%%%%%%%%%%%%%%%%%%%%%%%%%%%%%%%
\subparagraph{Transformation law of the holographic fields.}

It remains to determine the transformation law of the boundary fields. This is fundamental for their identification  with the
sources in the boundary CFT.

The bulk fields (\ref{asympt_field (1)}) can be cast in the form
\begin{align}
V_{\ \mu }^{i}& =\frac{\ell }{z}\,E_{\ \mu }^{i}+\frac{z}{\ell }\,S_{\ \mu }^{i}+\frac{z^{2}}{\ell ^{2}}\,\tau _{\ \mu }^{i} +\mathcal{O}(z^{3})\,,  \notag \\
\hat{\omega}_{\mu }^{i3}& =\frac{1}{z}\,E_{\ \mu }^{i}-\frac{z}{\ell ^{2}}\,\tilde{S}_{\ \mu }^{i}-\frac{2z^{2}}{\ell ^{3}}\,\tilde{\tau}_{\ \mu }^{i}+
\mathcal{O}(z^{3})\,\,,  \notag \\
\hat{\omega}_{\mu }^{ij}& =\omega _{\mu }^{ij}+\frac{z^{2}}{\ell ^{2}}\,\omega _{(2)\mu }^{ij}+\mathcal{O}(z^3)\,,  \notag \\
\hat{A}_{\mu }& =A_{\mu } +\frac{z}{\ell}\,A_{(1)\mu}
+\frac{z^{2}}{\ell ^{2}}\,A_{(2)\mu} +\mathcal{O}(z^{3})\,,  \label{asympt_field} \\
\Psi _{\mu +}^{A}& =\sqrt{\frac{\ell }{z}}\left[ \left(
\begin{matrix}
\psi _{\mu +}^{A} \\
0
\end{matrix}
\right) +\frac{z^{2}}{\ell ^{2}}\left(
\begin{matrix}
\Pi _{\mu +}^{A} \\
0
\end{matrix}
\right) +\mathcal{O}(z^{3})\right] ,  \notag \\
\Psi _{\mu -}^{A}& =\sqrt{\frac{z}{\ell }}\left[ \left(
\begin{matrix}
0 \\
\psi _{\mu -}^{A}
\end{matrix}
\right) +\frac{z}{\ell }\left(
\begin{matrix}
0 \\
\zeta _{\mu -}^{A}
\end{matrix}
\right) +\mathcal{O}(z^2)\right] .
\notag
\end{align}

Directly from (\ref{gauge}) and writing the boundary 1-forms in the basis \eqref{asympt_field}  on $\partial \mathcal{M}$, we find for the transformation law of the bosonic fields
\begin{eqnarray}
\delta E^{i} &=&\mathcal{D}\xi ^{i}+\sigma E^{i}-\theta ^{ij}E_{j}+\ii  \,\overline{\eta }^A_{+}\gamma ^i  \psi_{+A } \,,  \notag \\
\delta \omega ^{ij} &=&\mathcal{D}\theta ^{ij} +2\xi ^{[i}\mathcal{S}^{j]} +2\,K^{[i}E^{j]}
+\frac{1}{\ell }\,\overline{\eta }_{+}^{A}\gamma ^{ij} \psi _{-A }+\frac{1}{\ell }\,\overline{\eta }_{-}^{A}\gamma ^{ij}  \psi _{+A } \,,  \notag \\
\delta A &=&\diff \lambda +2\epsilon _{AB}\,\overline{\eta }_{+}^{A} \psi _{-}^{B} +2\epsilon _{AB}\,\overline{\eta }_{-}^{A} \psi_{+}^{B}\,,  \label{delta_B}
\end{eqnarray}
and for the gravitino
\begin{eqnarray}
\delta \psi _{+A}  &=&\mathcal{D}\eta _{A+}+\frac{\ii }{\ell }\,E^{i}\gamma_{i}\eta _{A-}-\frac{\ii }{\ell }\,\xi ^{i}\gamma_{i}  \psi _{-A} +\frac{1}{2}\,\sigma   \psi_{+A } \notag \\
&&-\frac{1}{4}\,\theta ^{ij}\gamma_{ij}  \varphi _{A+} +\frac{1}{2\ell }
\,\lambda \epsilon _{AB}\, \psi_{+}^{B}-\frac{1}{2\ell }\,A\,\epsilon
_{AB}\eta _{+}^{B}\,. \label{delta_F}
\end{eqnarray}
The super-Schouten tensor and its superpartner conformino are the composite fields that appear at the subleading order of eqs.~(\ref{gauge}), and they transform as
\begin{eqnarray}
\delta \mathcal{S}^{i} &=&\mathcal{D}K^{i}-\sigma \mathcal{S}^{i}-\theta
^{ij}\mathcal{S}_{j}+\frac{2\ii}{\ell ^{2}}\,\overline{\eta }_{-}^{A}\Gamma^{i}\psi _{-A}+\mathcal{E}^{i}\,,  \notag \\
\delta \psi _{-A} &=&\mathcal{D}\eta _{A-}+\frac{\ii\ell }{2}\,\mathcal{S}^{i}\gamma _{i}\eta_{A+} -\frac{\ii\ell}{2}\,K^{i}\gamma_{i}\psi _{+A}-\frac{1}{2}\,\sigma \psi _{-A}  \notag \\
&&-\frac{1}{4}\,\theta ^{ij}\gamma _{ij}\varphi _{-A}+\frac{1}{2\ell }\,\lambda \epsilon _{AB}\psi _{-}^{B}-\frac{1}{2\ell }\,A\epsilon _{AB}\eta _{-}^{B}+\Sigma_A\,.  \label{delta_S}
\end{eqnarray}
Eqs.~\eqref{delta_B}--\eqref{delta_S},  together with the transformation law of $B\equiv V^3_{\ \mu} \diff x^\mu$ given by  eq.~\eqref{delta_V3},  define the full set of $\mathcal{N}=2$ superconformal transformations of the boundary 1-forms $E^i$, $B$, $\mathcal{S}^i$, $\omega^{ij}$, $A$, $\psi_{\pm A}$. The $\ell $ factors ensure dimensional consistency of the equations with $[V^{i}{}_{\mu }]=L^{0}$, $[\mathcal{S}_{\ \mu }^{i}]=L^{-2}$, $[\psi _{\pm \mu }^{A}]=L^{-1/2}$, $[\xi ^{\mu }]=[\xi ^{i}]=L$ and $[\eta ]=L^{1/2}$.

Similarly as the Cotton tensor appear in the transformation law of the pure AdS gravity arising from the Lie derivative, as discussed at the end of Section \ref{WithoutF}, here we have the tensor ${\cal E}^i={\cal E}_{\ \mu }^{i}\,\diff x^{\mu }$ that comes from the linear in $z$ terms\footnote{In our conventions, the $z$-expansion coefficients of the 4-spinor-tensor $\Theta _A^{ab|c}$ are written as the bispinor-tensors $\Theta _{(n)\pm A}^{ab|c}$. Similarly, the 4-spinors $\tilde\rho^{ab}_A$ have the bispinor coefficients $\tilde\rho^{ab}_{(n) \pm A}$.} and the spinor $\Sigma^A=\Sigma_{\mu }^A\,\diff x^{\mu }$ appearing at the order $z^{1/2}$,
\begin{eqnarray}
\mathcal{E}_{\ \mu }^{i} &=&\frac{2}{\ell }\,\tilde{R}_{(3)}^{i3}{}_{jk}\xi
^{k}E_{\ \mu }^{j}+\frac{1}{\ell }\,\overline{\Theta }_{(5/2)-A|j}^{i3}\left( \eta _{+}^{A}E_{\ \mu }^{j}-\psi _{+\mu }^{A}\xi ^{j}\right) ,  \notag
\\
\Sigma_{\ \mu }^{A} &=&2E_{\ [i}^{\nu }E_{\ j]}^{\lambda }\left( \nabla _{\nu }\psi _{\lambda -}^{A}+\frac{\ii\ell }{2}\,\mathcal{S}_{\ \nu
}^{k}\gamma _{k}\psi _{\lambda +}^{A}\right) \xi ^{i}E_{\ \mu }^{j}\,.
\end{eqnarray}
To explicitly relate them to the Cotton tensor, we recall that in pure gravity, geometrically, the linear term of $\hat{R}_{\mu \nu }^{i3}$ is related to the Cotton tensor through eq.~(\ref{pure R}). Thus, the ${\cal N}=2$ supersymmetric extension of the Cotton tensor ($\mathcal{C}_{\ \mu \nu }^{i}$) and its superpartner, the Cottino ($\Omega^A_{ \mu \nu }$), are the first subleading terms in the corresponding supercurvature expansions, defined by
\begin{eqnarray}
{\mathbf{\hat {R}}^{i3}_{\mu \nu} }&=& -z\,\mathcal{C}_{\ \mu \nu }^{i}+\mathcal{O}(z^{2})\,,\notag \\
{\hat{\rr}^A_{-\mu\nu}}&=&  \sqrt{\frac{z}{\ell }}\,
{\binom{0}{\Omega^A _{\mu \nu }}}+\mathcal{O}(z^{3/2})\,, \label{rho-munu}
\end{eqnarray}
giving rise, {by means of (\ref{lagsuper}),} to the expressions
\begin{eqnarray}
\mathcal{C}_{\ \mu \nu }^{i} &=&2\mathcal{D}_{[\mu }\mathcal{S}_{\ \nu ]}^{i}-\frac{2\ii}{\ell ^{2}}\,\overline{\psi}^A_{-[\mu }\gamma ^{i}\psi _{-A|\nu ]}\,,
\label{supercotton}\\
\Omega^A _{\mu \nu } &=&2\nabla _{[ \mu }\psi^A _{-\nu]}-\ii\ell \,\gamma _{i}\psi^A _{+[\mu }\mathcal{S}_{\ \nu ]}^{i}\,.\label{supersupercotton}
\end{eqnarray}
An easy way to connect the above quantities to the additional terms in the transformation law of the super-Schouten tensor and the conformino is using the rheonomic parametrizations of the supercurvatures $\hat{\mathbf{R}}_{\mu \nu }^{i3}$ and $\hat{\rr}_{-\mu \,\nu }^{A}$ given by the last two equations in \eqref{rhopara}, related but not equal to $\tilde{R}^{i3}{}_{jk}V_{\ \mu }^{j}V_{\ \nu }^{k}$ and $\tilde{\rho}_{-ij}^{A}V_{\ \mu }^{i}V_{\ \nu }^{j}$, as discussed in Section \ref{4d}. Taking all the terms into account, the super-Cotton tensor and the Cottino are evaluated as
\begin{eqnarray}
-\ell \,\mathcal{C}_{\ \mu \nu }^{i} &=&\hat{\mathbf{R}}_{(1)\mu \nu }^{i3}=2\tilde{R}_{(3)jk}^{i3}E_{\ \mu }^{j}E_{\ \nu }^{k}-2\overline{\psi }_{+A[\mu}E_{\ \nu ]}^{j}\Theta _{(5/2)-A|j}^{i3}\,, \\
\Omega _{\mu \nu }^{A} &=&\hat{\rr}_{{(1/2)}-\mu \,\nu }^{A}=2\tilde{\rho}_{(5/2)-ij}^{A}E_{\ \mu }^{i}E_{\ \nu }^{j}=4\Theta_{(5/2)-A|j}^{i3} E_{i[\mu }E^j_{\ \nu] }\,.  \notag
\end{eqnarray}
The last step makes use of the explicit expressions of Appendix \ref{appC} to decompose the spinor-tensor coefficient $\Theta _{(5/2)-A}^{i3|j}$ into its symmetric part, $-2\ii \gamma ^{(i}\tilde{\rho}_{(5/2)A+}^{j)3}$ and the antisymmetric part $\frac 12\,\Omega^{A ij}$.
As a result, the additional terms in the transformation law \eqref{delta_S} are recognized as the contractions of the super-Cotton tensor and Cottino with respect to the boundary superdiffeomorphism parameters $\xi ^{i}$ and $\eta
_{+}^{A}$,
\begin{eqnarray}
\Sigma^{A} &=&i_{\xi }\Omega ^{A}\,,  \notag \\
\mathcal{E}^{i} &=&i_{\xi }\mathcal{C}^{i}+\frac{1}{\ell }\,\left( \overline{\eta }_{+}^{A}-\overline{\psi }_{+A\nu }\xi ^{\nu }\right) \Theta_{(5/2)-A|j}^{i3}\,E^{j}\,. \label{P}
\end{eqnarray}
Finally, we obtain an expected result for $\delta \mathcal{S}^{i}$ and $\delta \psi _{-A}$. The contribution of the symmetric part of the spinor-tensor $\Theta _{(5/2)-A}^{i3|j}$ is non-physical, as it depends on the gauge-fixing functions $\psi _{+zA}$ and $A_{(1)z}$.
We can, in principle, further gauge fix the higher-order parameters such that $\tilde{\rho}_{(5/2)A+}^{i3}$ vanishes as a consequence of $\boldsymbol{\hat{\rho}}_{(1/2)\mu z+}^{A}=0$. However, the result does not have observable consequences near the boundary, thus we will not proceed in this direction.

Notice that not all contractions of the  $\mathrm{OSp}(2|4)$ supercurvatures have appeared in the transformation laws \eqref{delta_B}--\eqref{delta_S} of the ${\cal N}=2$ superconformal algebra $\mathfrak{osp}(2,4)$, but only the ones that have origin in the negative grading supercurvatures. This is because, after imposing eqs.~\eqref{AdSbound}, all the $\mathrm{OSp}(2|4)$ supercurvatures vanish on $\partial \mathcal{M}$, except two, namely {$\mathbf{\hat{R}}^{i3}_{\mu \nu}$} and {$\boldsymbol{\hat{\rho}}^A_{-\mu\,\nu}$}. Indeed, the conditions \eqref{AdSbound} lead to the weaker condition on two supercurvatures,
\begin{equation}
\epsilon _{ijk}\,\mathcal{C}_{\ [\mu \nu }^{i}E_{\ \rho ]}^{k}+2\overline\psi_{+A[\mu }\gamma _{j}\Omega _{A\nu \rho ]}=0\,,
 \label{r-scft}
 \end{equation}
 which implies in particular  $\gamma_{[\mu} \Omega^A_{\nu \rho]}=0$ and, consequently, $\gamma^\nu \Omega^A_{\nu \rho}=0$.

As a matter of fact, non-trivial $\mathcal{C}^{i}$ and  $\Omega _{A}$ on $\partial \mathcal{M}$ mean that a holographic SCFT is not invariant under local $\mathrm{OSp}(2|4)$ transformations, for the same reason as $\mathrm{SO}(2,3)$ is not a local symmetry of the bulk gravity --namely, they are only general coordinate transformations rewritten in a gauge-covariant form. This explains an origin of the contractions of the supercurvatures in transformation laws and structure functions in the algebra, as also pointed out in \cite{Korovin:2017xqu} in the bosonic case.

In the gauge $V^3_{\ \mu}=0$, the boundary supersymmetry reduces to super-Weyl transformations. In the spirit of the analysis in \cite{Howe:1995zm,Butter:2013goa}, such transformations can be obtained from gauging the $\mathcal{N}=2$ superconformal algebra $\mathfrak{osp}(2|4)$ within $\mathcal{N}=2$ superspace in three dimensions, whose supervielbein is given by $(E^i, \psi_{+ A} )$.

Indeed, if we restrict the set of fields to $E^i_{\ \mu}$, $\psi_{+A\mu}$, $\omega^{ij}_\mu$, $A_\mu$ and the parameters to $\xi^i$, $\eta_{+A}$, $\theta^{ij}$, $\lambda$, we see that $E_{\ \mu }^{i}$ transforms as a boundary vielbein,  $\omega _{\mu }^{ij}$ as a boundary spin-connection, and $\psi_{+A\mu}$ as a boundary gravitino, charged with respect to the $\mathrm{SO(2)}$ R-symmetry connection $A_\mu$. Correspondingly, the parameters $\xi^i$, $\eta_{+A}$, $\theta^{ij}$, and $\lambda$ are associated with boundary  diffeomorphisms,  supersymmetry, Lorentz, and $\mathrm{SO(2)}$ gauge transformations,  respectively.

On the other hand, the boundary function $\sigma $, with respect to which all the above fields have definite weight (1 for $E^i_{\ \mu}$, $1/2$ for $\psi_{+ A \mu }$, and 0 for $\omega _{\mu }^{ij}$ an $A_\mu$), is identified with the local parameter associated with Weyl dilatations because it produces rescaling of the vielbein and therefore of the metric.

In the same fashion, the superconformal transformation is characterized by the local parameter $\eta _{-A}$, with the corresponding gauge field $\psi _{A-}$. The parameter $K^{i}$, although not independent within the gauge choice $V^3_{\ \mu}=0$, corresponds to special conformal transformations, whose associated gauge connection is the super-Schouten tensor.

%%%%%%%%%%%%%%%%%%%%%%%%%%%%%%%%%%%%%%%%%%%%%%%%%%%%%

\subparagraph{Consistency of the subleading gauge fixings.}

On top of the previous analysis of the asymptotic parameters, it remains to look for potential inconsistencies in having some linear terms vanishing, in particular $V_{(1)\mu }^{3}=\omega _{(1)\mu }^{ij}=\zeta_{\mu +}^{A}=0$.
Using the transformation law of the gauge fields, it is straightforward to find
\begin{align}\nonumber
\delta V_{(1)\mu }^{3}&=\frac{2}{\ell}\,\xi^\nu(\tau+2\tilde\tau)_{[\nu\mu]}+2\xi^\nu\overline\psi^A_{+[\nu}\zeta_{\mu]-A}=0\,,\\
\delta\zeta^A_{\mu+}&=-\frac{\ii}{\ell}\,\gamma_i\zeta^A_{-\mu}\xi^i-\frac{1}{2\ell}\,\psi^A_{\mu+}p^3_{(1)}+2\tilde\rho^A_{(5/2)+ij}\xi^iE^j_{\ \mu}-\frac{1}{4\ell}\,\xi^\rho A_{(1)\rho}\epsilon^{AB}\psi_{B\mu+}\,,\\ \nonumber
&+\frac{\ii}{4\ell}\,\gamma^i\psi_{B+\mu}\epsilon_{ijk}A_{(1)\rho}E^{\rho k}\xi^j+\frac{\lambda_{(1)}}{2\ell}\,\epsilon^{AB}\psi_{B+\mu}+\frac{\ii}{\ell}\,\gamma_i\eta^A_{(1)-}E^i_{\ \mu}-\frac{1}{2\ell}\,A_{(1)\mu}\epsilon^{AB}\eta_{B+}\\ \nonumber
&+\frac{1}{4\ell}\,A_{(1)\mu}\epsilon^AB\eta_{B+}-\frac{\ii}{4\ell}\,\epsilon^{AB}\epsilon_{ijk}\gamma^i\eta_{B+}E^j_{\ \mu} A_{(1)\rho}E^{\rho k}=0\,,
\end{align}
where the first condition holds by virtue of eq.~\eqref{tauantisymm} and the second one follows from plugging in the expressions of $\tilde \rho^A_{ij}$, $\lambda_{(1)}$ and $\eta^A_{(1)-}$, and by using $p^3_{(1)}=0$.
Finally, a variation of \eqref{R(0)mn} enables to solve
\begin{equation}
\delta \omega _{(1)\mu }^{ij}=\ii\,E^{\nu i}E^{\lambda j}E_{k\mu }\,\delta \overline{\zeta }_{+[\nu }^{A}\gamma ^{k}\psi _{\lambda ]+}^{A}-2\ii\,E^{\nu [ i}\delta \overline{\zeta }_{+A[\mu }\gamma ^{j]}\psi _{\nu ]+}^{A}\,,
\end{equation}
finding that $\delta \zeta _{A\mu +}=0$ implies also $\delta \omega _{(1)\mu }^{ij}=0$.

%%%%%%%%%%%%%%%%%%%%%%%%%%%%%%%%%%%%%%%%%%%%%%%%%%%

\section{Superconformal currents in the holographic quantum theory \label{SuperWard}}

In the previous section we showed that the asymptotic symmetries of pure $\mathcal{N}=2$ AdS$_{4}$ supergravity are given by the three-dimensional superconformal {transformations}. According to the AdS/CFT correspondence, these are also asymptotic symmetries of an underlying superconformal field theory (SCFT).

The superconformal group on a three-dimensional manifold contains Lorentz transformations (with the local parameter $\theta ^{ij}$), coordinate transformations ($\xi ^{i}$), dilatations ($\sigma $), special conformal transformations ($K^{i}$), supersymmetry trasformations ($\eta _{A+}$), special superconformal transformations ($\eta _{A-}$) and the R-symmetry ($\lambda $). Within a gauge theory, the corresponding gauge fields are the spin connection $\omega _{\mu }^{ij}$, the vielbein $E_{\ \mu }^{i} $, the dilatation gauge field $B_{\mu }$, the super-Schouten tensor $\mathcal{S}_{\ \mu }^{i}$, the gravitino $\psi _{+\mu }^{A}$, the conformino $\psi _{-\mu }^{A} $ and the graviphoton $A_{\mu }$.

It is useful to present this \textit{superconformal structure} of the three-dimensional boundary by listing all the transformations, associated local parameters and gauge fields (sources in SCFT), and the conserved currents (quantum operators in SCFT) in the following table:
\begin{equation*}
\fbox{
\begin{tabular}{cccc}
\textbf{Transformation} & \textbf{Local parameter} & \textbf{Source} &
\textbf{Current} \\
Lorentz & $\theta ^{ij}$ & $\omega _{\mu }^{ij}$ & $J_{\ ij}^{\mu }=0$ \\
Translation & $\xi ^{i}$ & $E_{\ \mu }^{i}$ & $J_{\ i}^{\mu }$ \\
Dilatation & $\sigma $ & $B_{\mu }=0$ & $J_{\ (D)}^{\mu }=0$ \\
Special conformal & $K^{i}$ & \quad $\mathcal{S}_{\ \mu }^{i}$
\quad  & \quad $J_{(K)i}^{\mu }=0$\quad  \\
Abelian R-symmetry & $\lambda $ & $A_{\mu }$ & $J^{\mu }$ \\
Supersymmetry & $\eta _{A+}$ & $\psi _{A+\mu}$ & $J_{\ A+}^{\mu }$ \\
Superconformal & $\eta _{A-}$ & $\psi_{A-\mu}$ & $J_{\ A-}^{\mu }=0$
\end{tabular}}
\end{equation*}

When all sources are independent, the currents are also independent. When one imposes the constraints over supercurvatures with a purpose to eliminate non-physical degrees of freedom, some parameters result to be realized non-linearly and the corresponding sources become composite fields, with the associated currents vanishing.

In supergravity, the spin connection is a composite field determined by a constraint on the translation curvature (supertorsion). The gauge field of special conformal transformations (super-Schouten tensor) and its supersymmetric partner (conformino) are also composite, obtained from the constraint on the conformal supercurvatures, equations \eqref{conformino} and \eqref{schouten}. Our particular gauge fixing $B_{\mu }=V_{\ \mu }^3=0$ eliminates the dilatation gauge field and the corresponding dilatation current. The inclusion of $B_\mu$ has been discussed in pure AdS gravity in \cite{Korovin:2017xqu}.

Before moving on to the explicit analysis of quantum symmetries in a three-dimensional field theory holographically dual to  $\mathcal{N} = 2$ AdS$_4$ supergravity, let us first understand more precisely its superalgebra structure.

%%%%%%%%%%%%%%%%%%%%%%%%%%%%%%%%%%%%%%%%%%%%%%%%%%%%%

\subsection{\texorpdfstring{$d=3$}{d=3} superconformal algebra}
\label{SuperA}

The superisometry group $\rm{OSp}(2|4)$ of the vacuum of the bulk theory is encoded in the definition of its curvatures $\hat{{\bf R}}^{\Lambda} =\{ \mathbf{\hat{R}}^{ab}, \mathbf{\hat{R}}^a, \hat{\rr}^A, \mathbf{\hat{F}} \}$,
\begin{equation}
\hat{{\bf R}}^{\Lambda} \equiv  \diff \boldsymbol{\mu}^{\Lambda}+\frac{1}{2}\,C_{\Sigma\Gamma}{}^\Lambda\,\boldsymbol{\mu}^{\Sigma}\wedge\boldsymbol{\mu}^{\Gamma}
\,,\label{bocu}
\end{equation}
where $C_{\Sigma\Gamma}{}^\Lambda$ are the $\mathfrak{osp}(2|4)$ structure constants and $\boldsymbol{\mu}^{\Lambda}=\{ \hat{\omega}^{ab},\,V^a,\,\Psi_A,\,\hat{A} \}$ the Cartan 1-forms.
Asymptotic expansions of the supercurvatures $\hat{{\bf R}}^{\Lambda}$ are given in Appendix \ref{SuperCurvatures}.
Moreover, $\mathfrak{osp}(2|4)$ also {describes} the superconformal {structure of} the boundary. This is made manifest by decomposing the Cartan 1-forms in irreducible representations with respect to the  SO(1,1)$\times$SO(2,1) subgroup of $\rm{OSp}(2|4)$, where SO(2,1) is the (connected component of) the Lorentz group at the boundary and SO(1,1)  is the isometry group which acts as a rescaling on the coordinate $z$ in the FG parametrization: $z\rightarrow \mathrm{e}^\sigma z$.
This decomposition requires splitting the index $a$ into $(i,3)$, where $i=0,1,2$. Moreover, $V^i$ and $\hat{\omega}^{i3}$ naturally combine into $V_\pm^i$ introduced in eq.~\eqref{Vpm}, which have definite scalings with respect to the ${\rm SO}(1,1)$ group. Finally, since the spinorial representation of the generator $T_0$ of the  ${\rm SO}(1,1)$ group is
\begin{equation}
(T_0)^\alpha{}_\beta=-\frac{\ii}{2}\, (\Gamma^3)^\alpha{}_\beta\,,
\label{T0}
\end{equation}
the four-dimensional gravitini naturally split into $\Psi_{\pm A}$ with definite radial chirality. In terms of the ${\rm SO}(1,1)\times{\rm SO}(2,1)$ irreducible forms $\hat{\omega}^{ij},\,V_+^i,\,V_-^i,\,V^3,\,A,\,\Psi_\pm^A$, where we recall the expressions (\ref{Vpm}), the bulk supercurvatures \cite{Andrianopoli:2019sip}  given by eq.~(\ref{lagsuper}) become
\begin{align}
\hat{{\bf R}}^{ij}&= \hat{\mathcal{R}}^{ij} +\frac{4}{{\ell}^2}\,V_+^{[i}\wedge V_-^{j]} -\frac{1}{{\ell}}\,\overline\Psi^A_+\wedge\Gamma^{ij}\Psi_{A-}\notag \,,
\\
\hat{{\bf R}}_\pm^{i}&= \hat{\mathcal{D}}V^i_\pm \mp\frac{1}{{\ell}}\,V^i_\pm\wedge V^3\mp\frac{\ii}{2}\, \overline\Psi^A_\pm\wedge\Gamma^i\Psi_{A\,\pm}\,, \notag
\\
\hat{{\bf R}}^{3}&=\diff V^3+\frac{2}{\ell}\,V^i_+\wedge V_{-i}+\overline\Psi^A_-\wedge\Psi_{A+}\,, \label{MCboundary}
\\
\hat{{\bf F}}&=\diff \hat{A} -2\epsilon_{AB}\,\overline\Psi{}^{A}_+\wedge\Psi^{B}_-\,,
\notag
\\
\hat{\boldsymbol{\rho}}^{A}&=
\hat{\mathcal{D}}\Psi^{A}_\pm \pm
\frac{\ii}{{\ell}}\,V_\pm^i\wedge\Gamma_i\Psi^{A}_\mp \pm\frac{1}{2{\ell}}\,V^3\wedge\Psi^{A}_\pm -\frac{1}{2\ell}\,\epsilon_{AB}\hat{A}\wedge\Psi^{B}_\pm\,.
\notag
\end{align}
The right-hand sides of the above equations encode the algebraic structure of the superconformal algebra in $d=3$, where $V^3$ is the 1-form associated with the Weyl transformations, $V_+^i$ the ones associated with the spacetime translations, $V_-^i$  with the conformal boosts, $\Psi^{A}_+$ with the supersymmetries, $\Psi^{A}_-$ with the superconformal transformations \cite{Castellani:1998nz,DallAgata:1998lrx}. The connection components $\hat{\omega}^{ij}$  correspond to the Lorentz algebra at the boundary.
The precise connection to the Cartan 1-forms of the superconformal algebra in $d=3$ is that the leading order 1-form in the $z$-expansion of the above bulk quantities are identified with the Cartan 1-forms dual to the corresponding superconformal generators. Let us summarize below the correspondence between the $D=4$ gauge field and $d=3$ superconformal field:
\begin{equation*}
\begin{array}{llll}
\hat\omega^{ij} & \rightarrow  & \omega^{ij}\quad  & \text{Lorentz symmetry}\,, \\
V^3 & \rightarrow  & B & \text{Weyl symmetry}\,, \\
V_+^i & \rightarrow  & E^{i} & \text{spacetime translations}\,,\\
V_-^i & \rightarrow  & \mathcal{S}^i & \text{conformal boosts}\,, \\
\Psi_+^A & \rightarrow  & \psi_+^A & \text{supersymmetry}\,, \\
\Psi_-^A & \rightarrow  & \psi_-^A & \text{superconformal symmetry}\,, \\
\hat A & \rightarrow  & A & \text{SO(2) R-symmetry}\,.
\end{array}
\end{equation*}
This can also be understood as the boundary conditions set imposed on the bulk fields in an asymptotically AdS space.

Let us make this connection more precise. To this end, we perform the redefinitions (\ref{Psi}) and (\ref{Epm}) and define the gauge vector associated with the Weyl rescalings as follows,
\begin{equation}
B=\frac{1}{\ell}\,\left(V^3 -\ell \frac{\diff z}{z}\right) =B_\mu(x)\,\diff x^\mu\,.
\end{equation}
Note that, in order for $B$ to be non-vanishing, we have to generalize the FG parametrization \eqref{FG} to allow for a non-trivial component $V^3_{\ \mu}$ for the vielbein. After rescaling the various fields by $z/\ell$ factors according to their ${\rm O}(1,1)$ grading, the $\diff z/z$ term in $V^3$, within the definitions of the curvature/field strengths, cancel.
Next we recall the relation between the $d=3$ super-Schouten tensor and $E^i_-$ given by the second of eqs.~(\ref{Epm_expand}),
\begin{equation}
    \mathcal{S}^i=-\frac{2}{\ell^2}\,\left.E_-^i\right\vert_{z=0}\,.
\end{equation}
Rescaling the field strengths associated with $\Psi_\pm$ and $V_\pm^i$, in eqs.~(\ref{MCboundary}), correspondingly, we can evaluate the right-hand side at $z=0,\,\diff z=0$ and find the following supercurvatures in the dual field theory (see Appendix \ref{SuperCurvatures}),
\begin{align}
{\bf R}^{ij}&= \mathcal{R}^{ij}-2\,E^{[i}\wedge \mathcal{S}^{j]}-\frac{1}{{\ell}}\,\overline\psi^A_+\wedge\gamma^{ij}\psi_{A-}\,, \notag
\\
{\bf R}_+^i&=\mathcal{D}E^i+B\wedge E^i-\frac{\ii}{2}\, \overline\psi^A_+\wedge\gamma^i\psi_{A\,+}\,,
\notag
\\
\mathcal{C}^i&\equiv -\frac{2}{\ell^2}\, {\bf R}_{-}^{i}=\mathcal{D} \mathcal{S}^i-B\wedge \mathcal{S}^i-\frac{\ii}{\ell^2}\, \overline\psi^A_-\wedge\gamma^i\psi_{A\,-}\,,  \notag
\\
{\bf R}&=\diff B-E^i\wedge \mathcal{S}_i+\frac{1}{\ell}\,\overline\psi^A_-\wedge\psi_{A+}\,, \notag
\\
{\bf F}&=\diff A -2\epsilon_{AB}\, \overline\psi{}^{A}_+\wedge\psi^{B}_-\,, \label{Mcboundary}
\\
\boldsymbol{\rho}^{A}_{+}&=\mathcal{D}\psi^{A}_+ +\frac 12\, B\wedge\psi^{A}_+ + \frac{\ii}{\ell}\,E^i\wedge\gamma_i\psi^{A}_-  -\frac{1}{2\ell}\,\epsilon_{AB}{A}\wedge\psi^{B}_+\,, \notag \\
\Omega^A&\equiv \boldsymbol{\rho}_{-}^{A} =\mathcal{D} \psi^{A}_- -\frac 12\,B\wedge\psi^{A}_- + \frac{\ii\ell}{2}\,\mathcal{S}^i\wedge\gamma_i\psi^{A}_+  -\frac{1}{2\ell}\,\epsilon_{AB} {A}\wedge\psi^{B}_-\,, \notag
\end{align}
where $\mathcal{D}$ is the Lorentz-covariant derivative.
Each $\mathcal{D}$ always appears in the combination  $\mathcal D + \Delta B$ of the Weyl-covariant derivative, as naturally expected from a theory with local Weyl symmetry. The  Weyl weight $\Delta$ of the corresponding field is equal to its scaling dimension, namely $\Delta(E^i_\pm)=\pm 1$, $\Delta(\psi^A_\pm)=\pm \frac 12$, $\Delta(\mathcal S^i)=-1$ and $\Delta(\omega^{ij})=\Delta(A)=\Delta(B)=0$. This feature can be used to reconstruct the $B$-terms in the transformations laws \eqref{delta_B}--\eqref{delta_S}, similarly as it was done in the pure AdS gravity case given by eqs.~(\ref{full}).

Note that, for $B=0$, the third and the last of eqs.~(\ref{Mcboundary}) yield the definitions of $\mathcal{C}^i$ and $\Omega^A$ in eqs.~(\ref{supercotton}) and (\ref{supersupercotton}), respectively.

Finally, let us recall once again that, while the boundary theory possesses global $\mathrm{OSp}(2|4)$ isometry, it is not also locally $\mathrm{OSp}(2|4)$ invariant, but the transformation law of the gauge fields is put in an $\mathrm{OSp}(2|4)$-covariant form thanks to the superdiffeomorphisms written in a suitable way through a field-dependent gauge transformation.

%%%%%%%%%%%%%%%%%%%%%%%%%%%%%%%%%%%%%%%%%%%%%%%%%%%%%

\subsection{Superconformal currents}

To explore the quantum symmetries in a SCFT dual to supergravity with $\Psi^A_{z-}=0$, we apply the AdS/CFT correspondence summarized in Section \ref{AdS/CFT} to the case when the boundary fields are $\mathcal{J}^{\Lambda }(x)=\{E_{\ \mu }^{i}(x)$, $\omega_{\mu }^{ij}(x)$, $\psi _{+ A\,\mu }(x)$, $A_{\mu }(x)\}$.
They become sources for the corresponding operators in the dual SCFT. Generalizing eq.~(\ref{Z}) to the supergravity case, the bulk action in the classical supergravity approximation is identified with the effective action of the dual boundary theory as
\begin{equation}
I_{\mathrm{on-shell}}[E^{i},\,\omega ^{ij},\,\psi _{+}^{A},\,A]=W[E^{i},\,\omega ^{ij},\,\psi _+^{A},\,A]=-\ii \,\ln
(Z[E^{i},\,\omega ^{ij},\,\psi _+^{A},\,A])\,.
\end{equation}
The sources $\mathcal{J}^{\Lambda }$ couple to the operators in quantum field theory $J_{\Lambda }^{\mu }=\{J_{\ i}^{\mu }$, $J_{\ ij}^{\mu }$, $J_{A+ }^{\mu }$, $J^{\mu }\}$, which are the energy-momentum tensor, spin current, supercurrent, and U$(1)$-current, respectively. The latter are identified with the 1-point functions of the Noether currents in the presence of arbitrary sources, associated with the residual symmetries of the boundary action, see Section \ref{AdS/CFT} and the above table. However, we shall refrain from writing explicitly the symbol $\left\langle \cdots \right\rangle _{\mathrm{CFT}}$. We will also express the currents in terms of their Hodge-dual 2-forms in the boundary theory, to be denoted by the same  symbol, as defined by eq.~(\ref{Hodgestar}).

The explicit expression of these currents is inferred from the variation of the effective action with respect to the sources (eq.~(\ref{varW}) generalized to supergravity),
\begin{equation}
    \delta W={\int\limits_{\partial \mathcal{M}}} \delta\mathcal{J}^\Lambda\wedge J_\Lambda ={\int\limits_{\partial \mathcal{M}}} \left(\delta E^i\wedge J_i+\frac{1}{2}\,\delta \omega^{ij}\wedge J_{ij}+\overline{J}{}^A_+\wedge\delta\psi_{A+} +J\wedge \delta A\right)\,.\label{deltaWgen}
\end{equation}
Invariance of the boundary effective action with respect to the residual symmetries of the boundary theory implies conservation laws to be satisfied by the currents. As we shall prove, they are satisfied by virtue of the ``constraint''  equations of motion in the bulk. Namely, in the radial foliation of spacetime, the bulk equations of motion are divided into the ones describing the radial ``evolution'' (that were used to determine radial expansions of the bulk fields) and the ``constraints'', which do not contain radial derivatives $\partial _z$ and that should give rise to conservation laws in the holographic QFT.

In the following{,} we shall first derive the expressions of the currents and the corresponding conservation laws. Eventually, using the bulk equations of motion, we shall show that these conditions are indeed satisfied at the quantum level and they represent the Ward identities in the SCFT.

%%%%%%%%%%%%%%%%%%%%%%%%%%%%%%%%%%%%%%%%%%%%%%%%%%%%%

\paragraph{SCFT currents.}
In this derivation it is somewhat convenient to retain, in the computation of $\delta W$, a four-dimensional notation, writing it in terms of the bulk fields and their curvatures, keeping in mind that, in the boundary integral, they are meant to be functions of the corresponding boundary values through the supergravity solution. So when we write $\delta \hat{\omega}^{ab},\,\delta\Psi^A,\,\delta\hat{A}${,} we mean the variations of the bulk fields in a supergravity solution, originating from a variation of the corresponding boundary conditions. Using the compact form (\ref{Lfull}) of the full supergravity action and using the field equations, we find
\begin{align}\label{Weff}
\delta W=\delta I_{\rm on-shell}={\int\limits_{\partial \mathcal{M}}} \left(-\frac{\ell^2}{4}\,\delta\hat\omega^{ab}\hat\RR^{cd}\epsilon_{abcd}-2\ii\ell\delta\overline\Psi^A\Gamma_5\hat\rr_A+\frac{1}{2}\,\delta\hat A\ {}^*\hat\FF\right)\bigg |^{\text{on-shell}}_{z=\diff z=0},
\end{align}
where we have explicitly indicated that the quantities in the integral are to be computed on the boundary $\partial\mathcal M$, namely at $z=\diff z=0$. Using the boundary expansion of the four-dimensional fields in (\ref{asympt_field}), we can write the above variation in the form (\ref{deltaWgen}) (recall that we have set ${\omega_{(1)}}$ and $\zeta_+^A$ to zero) and read off the explicit form of the external current 2-forms on $\partial {\cal M}$,
\begin{align}
J_i&=\frac{1}{2}\,\epsilon_{ijk}\bigg[\frac{2}{\ell}\,E^j{\wedge}(\tau^k+2\tilde\tau^k)+\overline\psi^{A}_+{\wedge}\gamma^{jk}\zeta_{A-}\bigg],\nonumber \\
J_{ij}&=0\,,\nonumber\\
J&=\left.\frac{1}{2}\,\epsilon_{ijk}\,\tilde{F}^{i3}\,V^j{\wedge}V^k\right\vert_{z=0}, \nonumber \\
J^A_+&=-2\,\ii \,E^i{\wedge}\gamma_i\zeta^{A}_-
+ A_{(1)}\wedge \epsilon_{AB}\,\psi_{+}^B\,,
\label{currents}
\end{align}
where $\tilde{F}_{ab}$ are the components of the supercovariant field strength associated with the graviphoton, see eq.~\eqref{rhopara}. The current associated with the Lorentz transformation ($J_{ij}$) is zero because it corresponds to the field that is composite ($\omega _{\mu }^{ij}$), but it has been treated as independent in first order formulation of gravity. The other composite fields (${\cal S}^i _{\ \mu }$ and $\psi _{A-\mu }$) have not been taken into account as sources.

From the above expressions for the conserved
current 2-forms, $J_{\Lambda }$, we can obtain the Noether currents $J_{\Lambda }^{\mu }$ as the Hodge-dual 3-vectors $^{\ast }J_{\Lambda}=J_{\Lambda \mu }\,\diff x^{\mu }$ defined by eq.~(\ref{Hodgestar}). The non-vanishing currents are
\begin{align}
J_{\ i}^{\mu }& =-\frac{1}{\ell }\,\left( (\tau _{\ i}^{\mu }+2\tilde{\tau}_{\ i}^{\mu })-E_{\ i}^{\mu }(\tau _{\ k}^{k}+2\tilde{\tau}_{\ k}^{k})\right) +
\frac{\ii }{e_{3}}\,\epsilon ^{\mu \nu \rho }\,\bar{\psi}_{+\nu }^{A}\gamma _{i}\zeta _{A-\rho }\,,  \notag \\
J_{A+}^{\mu }& =-\frac{2\ii }{e_{3}}\,\epsilon ^{\mu \nu \rho }\gamma _{\nu }\zeta _{A-\rho } +\frac{1}{e_3}\,\epsilon^{\mu\nu\rho}\,A_{(1)\nu}\epsilon_{AB}\,\psi^B_{+\rho}\,,  \notag \\
J^{\mu }& =-g_{(0)}^{\mu\nu}\,\tilde{F}_{\nu z}=\frac{1}{2\ell}\,g_{(0)}^{\mu\nu}\,A_{(1)\nu}\,, \label{Jimutt}
\end{align}
where in the first equation the traces $\tau^k{}_k,\,\tilde{\tau}^k{}_k$ are defined using the vielbein tensor (e.g. $\tau^k{}_k\equiv \tau^k{}_\mu\,E^\mu_{\ k}$). In the last equation we have used the fact that the contribution of $A_z$ to $\tilde{F}_{\mu z}$ is subleading in $z$, while the fermion bilinears do not contribute at $z=0$ having set $\varphi_{-Az}=0$.

In particular, the holographic stress tensor is $J_{\mu \nu }=J_{\mu i}E_{\ \nu }^{i}$. Recall that, in the CFT$_{d}$ dual to pure AdS$_{d+1}$ gravity, this tensor is proportional to the (symmetric) metric coefficient $g_{(d)\mu \nu }\propto \tau _{\mu \nu }$ whose trace is zero. Indeed, the above result in pure gravity with the traceless $\tau _{\ i}^{\mu }=\tilde{\tau}_{\ i}^{\mu }$ reduces to $J_{\mu \nu }^{\mathrm{pure \, GR}}=-\frac{3}{\ell
}\,\tau _{\mu \nu }$. In the SCFT$_{3}$, the relevant bosonic coefficient is $\tau _{\mu \nu }+2\tilde{\tau}_{\mu \nu }$ and generally it is not symmetric any longer because of $\tilde{\tau}_{\mu \nu }$. Furthermore, the trace of $\tau _{\mu \nu }+2\tilde{\tau}_{\mu \nu }$ is not necessarily zero --it has to be computed from the conservation law of the local Weyl symmetry.

In supergravity, the holographic stress tensor contains the fermionic contribution. Which particular fermionic coefficient becomes holographic can be determined by simple power counting in the variation of the action. Since the on-shell action is always a boundary term, the Jacobian $e$ given by (\ref{e}) expressed in terms of the boundary Jacobian $e_{3}$ has the factor $1/z^{4}$, but on the boundary $z=const$ it becomes $1/z^{3}$. Thus, the holographic order --the one that contributes to the holographic current-- is always the \textit{third} order in $z$ of the variation of the Lagrangian density on-shell on the three-dimensional boundary. For the metric, it means the third coefficient in the expansion ($\tau _{\mu \nu }$). For fermions, it means $\Psi_{(3/2)-\mu }=\zeta_{-\mu}$. Similarly, the third coefficient on the boundary of the Maxwell Lagrangian comes from $(\partial _{z}\hat{A}_{\mu })^2$,
implying that the finite part of $\partial _z\hat{A}_{\mu }$, that is $\hat{A}_{(1)\mu }$, enters the holographic current. In $d$ dimensions, the respective holographic orders are $\tau ^i_{\ \nu }=\hat{E}^i_{(d) \nu }$, $\Psi_{(d/2)-\mu }$, and $\hat{A}_{((d-1)/2)\mu }$. They are the last terms in the near-boundary power expansion of the variation of the action which do not vanish when $z=0$.

%%%%%%%%%%%%%%%%%%%%%%%%%%%%%%%%%%%%%%%%%%%%%%%%%%%%%

\paragraph{Conservation laws in SCFT.}

We observe that, in the boundary expansion of the integrand form in (\ref{Weff}), the divergent terms vanish by virtue of the conditions (\ref{AdSbound}) that, in components, are given by eqs.~(\ref{AAdSspace}). These conditions therefore guarantee consistency of the holographic construction. Namely, both the currents and the conservation laws become finite, confirming that the bulk supergravity has been properly regularized in the asymptotic region.

Being the leading terms in the boundary expansion of the bulk curvatures zero by (\ref{AAdSspace}), from eq.~(\ref{Weff}) it follows that the currents in (\ref{currents}) are expressed in terms of the subleading terms in the same expansions. The reader can check, for instance, that
\begin{equation}\label{JR1}
    J_{ij}=-\ell^2\,\epsilon_{ijk}\,\hat{{\bf R}}^{k3}_{(0)}
    \,\,,\,\,\,J_i=-\frac{\ell}{2}\,\epsilon_{ijk}\,\hat{{\bf R}}^{jk}_{(1)}\,\,,\,\,\,J_{A+}=-2\ell\,
    \hat{\boldsymbol{\rho}}_{(1/2)\,A+}\,.
\end{equation}

Next we seek for the form of conservation laws associated with the residual symmetry discussed in Section \ref{FTAEP}, in case when the quantum effective action is invariant (after that  we will have to check whether the obtained supercurrents indeed  satisfy these conservation laws and since they are quantum, in fact they will give the Ward identities.) The corresponding transformations are parametrized by $\xi ^{i}$, $\theta ^{ij}$, $\lambda $, $\eta _{\pm }^{A}$. This means that $\delta W$ evaluated on the corresponding symmetry transformations of the fields must vanish and amounts to the following conservation laws for the Noether currents which are the generalization of the pure gravity laws (\ref{Wardclassical}) (we omit the wedge symbol),
\begin{align}
    \mathcal{D}J_i&=\mathcal{S}^j\,J_{ij}-\frac{\ii }{\ell}\overline{J}{}^A_+\gamma_i\psi_{A-}+\mathcal{S}^k{}_i
    \,J_{kj}\, E^j-
    \frac{\ii \ell}{2}\,\mathcal{S}^j{}_i\,\overline{J}{}^A_-\gamma_j\psi_{A+}\,,\nonumber\\
     \mathcal{D}J_{ij}&=2\,E_{[i}\,J_{j]}-\frac{\ii }{2}\,\overline{J}{}^A_+\gamma_{ij}\psi_{A+}-\frac{\ii }{2}\,\overline{J}{}^A_-\gamma_{ij}\psi_{A-}\,,\nonumber\\
   0&= \partial_\mu\left[E^{\mu i}\left(J_{ij}\, E^j-
    \frac{\ii \ell}{2}\,\overline{J}{}^A_-\gamma_i\psi_{A+}\right)\right]+
    E^i\,J_i+\frac{1}{2}\,\overline{J}{}^A_+\,\psi_{A+}-
    \frac 12 \,\overline{J}{}^A_-\,\psi_{A-}\,,\nonumber\\
    \diff J&=\frac{1}{2\ell}\,\epsilon_{AB}\left(\overline{J}{}^A_+\,\psi_{B+}+\overline{J}{}^A_-\,\psi_{B-}\right)\,,\nonumber\\
    \nabla J_{A+}&=\frac{1}{2\ell}\,\gamma^{ij}\psi_{A-}\,J_{ij}+\ii \,\gamma^i\,\psi_{A+}\,J_i-\frac{\ii \ell}{2}\,\mathcal{S}^i\,\gamma_i\,J_{A-}+2 \,\epsilon_{AB}\,\psi_{B-}\,J+\frac{1}{\ell}\,\psi_{A-}^i\,J_{ij}\,E^j\nonumber\\
    &\phantom{=}-
    \frac{\ii }{2}\,\psi_{A-}^i\,\overline{J}{}_{B-}
\gamma_i \psi_{B+}\,,\label{WID}\\
   \nabla J_{A-}&=\frac{1}{2\ell}\,\gamma^{ij}\psi_{A+}\,J_{ij}+2 \,\epsilon_{AB}\,\psi_{B+}\,J-\frac{\ii }{\ell}\,E^i\,\gamma_i\,J_{A+}-\frac{1}{\ell}\,\psi_{A+}^i\,J_{ij}\,E^j
   +\frac{\ii }{2}\,\psi_{A+}^i\,\overline{J}{}_{B-}
\gamma_i \psi_{B+}\,. \nonumber
\end{align}
We use the boundary vielbein $E^i_{\ \mu}$ and its inverse  $E_{\ i}^\mu$ to project the boundary spacetime indices ($\mu,\nu,\ldots$) to the boundary Lorentz ones ($i,j,\ldots$) and vice versa.
Note that the above conservation laws reduce to those in  (\ref{Wardclassical}) in the pure gravity case, namely in the absence of the fermionic superpartners and of the ${\rm U}(1)$ gauge field.
This is best seen from the pure gravity laws \eqref{full} when the dilatation gauge field is $B=0$ and the conformal current is $J_{(K)i}=0$. Then the dilatation current $J_{(D)}$ is not independent and can be solved from the last (algebraic) equation in \eqref{full}, leading to the identities $\ell \mathcal{S}_{i}J_{(D)}=\mathcal{S}_{i}^{\ k}J_{kj}E^{j}$ and $\ell \diff J_{(D)}=\partial _{\mu}\left( E^{\mu i}J_{ij}E^{j}\right) $. The obtained set of equations matches \eqref{WID} when all spinors are zero and $\mathcal{S}_{ij}$ is symmetric. In addition, it is explicit from \eqref{WID} that the fermions are sources of the electromagnetic current $J$.

As a final comment we observe that, in supergravity, invariance of the boundary action under Weyl transformations is guaranteed by the third of eqs.~(\ref{WID}) which, taking into account
 eqs.~(\ref{currents}), amounts to the  condition
 \begin{equation}
    E^i\wedge J_i=-\frac{1}{2}\,\overline{J}{}^A_+\,\wedge\psi_{A+}+
    \frac{1}{2}\,\overline{J}{}^A_-\,\wedge\psi_{A-}=-\frac{1}{2}\,\overline{J}{}^A_+\,\wedge\psi_{A+} \,.\label{TheTrace}
 \end{equation}
Let us now use the explicit form of the currents, given in eqs.~(\ref{currents}), to write eq.~(\ref{TheTrace}) in components. Using eq.~(\ref{Jimutt}) we find the trace of the bosonic part of the holographic stress tensor, namely
\begin{equation}
    (2\tilde{\tau}+\tau)^l_{\ l} =-\ii \,\ell\,\epsilon^{ijk}\bar{\psi}^A_{+j}\gamma_i\zeta_{A-\,k}\,.
\end{equation}
Using the properties of the gamma matrices, the reader can verify that the above relation is consistent with eq.~(3.34) of \cite{Amsel:2009rr}.

Notice that neither the holographic stress tensor $J_{\mu \nu }$ nor its bosonic part $\tau _{\mu \nu }+2\tilde{\tau}_{\mu \nu }$ have vanishing trace as in pure gravity. This does not mean that we have the trace anomaly because the value of the trace $J^{i}\wedge E_{i}$, given in (\ref{TheTrace}), is  fixed by the structure of the superalgebra. This is consistent with the result in ${\cal N}=1$ supergravity \cite{Amsel:2009rr}. Furthermore, the trace anomaly is a local expression even though
all currents, in particular $\tau^\mu_{\; i}$, $\tilde{\tau}^\mu_{\; i}$ and $J_{ij}$ are, in general, non-local tensors. Having a quantum anomaly would mean that $J^{i}\wedge E_{i}$ is a different expression than the one given in eq.~(\ref{TheTrace}).

Similarly, $J_{\mu \nu }$ and $\tau _{\mu \nu }+2\tilde{\tau}_{\mu \nu }$ are not symmetric: the second conservation law in (\ref{WID}) with $J_{ij}=0$ and $J_{-}=0$ gives the antisymmetric part as  $E_{[i}\wedge J_{j]}= \frac \ii 4 \overline{J}_+\gamma _{ij}\wedge \psi_+$. A reason is that, with our gauge fixing choice, $J_{\mu \nu }$ is not, as in pure gravity, the traceless Belinfante-Rosenfeld stress tensor. However, we know that, in principle, it is possible to use an ambiguity in definitions of Noether currents to construct a so-called `improved' stress tensor which would be symmetric and traceless.

%%%%%%%%%%%%%%%%%%%%%%%%%%%%%%%%%%%%%%%%%%%%%%%%%%%%%

\subsection{The Ward identities}

We now prove that the Ward identities are indeed satisfied by using the explicit form of the currents and showing that $\delta W=0$. We remind the reader that, although all expressions are evaluated on-shell in the bulk supergravity, they present off-shell identities in CFT computed on the curved background.
We start by integrating \eqref{Weff} by parts,
\begin{align}\nonumber
\delta W=\int\limits_{\partial\mathcal M}& \bigg[\frac{\ell^2}{4}j^{ab}\mathcal D\hat\RR^{cd}\epsilon_{abcd}-\frac{\ell^2}{4}\left(\frac{2}{\ell^2}p^aV^b+\frac{1}{\ell}\overline\epsilon^A\Gamma^{ab}\Psi_A\right)\hat\RR^{cd}\epsilon_{abcd}+2\ii\ell\overline\epsilon^A\Gamma_5\mathcal D\hat \rr_A\\ \nonumber
&-2\ii\ell\left(\frac{1}{4}j^{ab}\overline\Psi^A\Gamma_{ab}+\frac{\ii}{2\ell}p^a\overline\Psi^A\Gamma_a+\frac{1}{2\ell}\lambda\epsilon^{AB}\overline\Psi_B-\frac{1}{2\ell}\hat A\epsilon^{AB}\overline\epsilon_B-\frac{\ii}{2\ell}\overline\epsilon^A\Gamma_aV^a\right)\Gamma_5\hat\rr_A\\
&-\frac{1}{2}\,\lambda \ \diff \, {}^*\hat\FF+\overline\epsilon^A\Psi^B\epsilon_{AB}\, {}^*\hat\FF\bigg]\bigg|^{\text{on-shell}}_{z=\diff z=0}.
\end{align}
We now make use of the Bianchi identities (\ref{bianchi4d}), to obtain
\begin{align}\nonumber
\delta W=\int\limits_{\partial\mathcal M}& \bigg[\frac{\ell}{4}j^{ab}\overline\Psi^A\Gamma^{cd}\hat\rr_A\epsilon_{abcd}-\frac{\ell^2}{4}\left(\frac{2}{\ell^2}p^aV^b+\frac{1}{\ell}\overline\epsilon^A\Gamma^{ab}\Psi_A\right)\hat\RR^{cd}\epsilon_{abcd}\\ \nonumber
&+2\ii\ell\left(\frac{1}{2\ell}\hat A\epsilon^{AB}\overline\epsilon_A\Gamma_5\hat\rr_B-\frac{\ii}{2\ell}\overline\epsilon^A\Gamma_5\Gamma_a\hat\rr_AV^a+\frac{1}{4}\hat\RR^{ab}\overline\epsilon^A\Gamma_5\Gamma_{ab}\Psi_A-\frac{1}{2\ell}\epsilon^{AB}\hat\FF\overline\epsilon_A\Gamma_5\Psi_B\right)\\ \nonumber
&-2\ii\ell\left(\frac{1}{4}j^{ab}  \overline\Psi^A\Gamma_{ab}+\frac{\ii}{2\ell}p^a \overline\Psi^A\Gamma_a+\frac{1}{2\ell}\lambda\epsilon^{AB}\overline\Psi_B-\frac{1}{2\ell}\hat A\epsilon^{AB} \overline\epsilon_B-\frac{\ii}{2\ell}\overline\epsilon^A\Gamma_aV^a\right) \Gamma_5\hat\rr_A\\
&-\frac{1}{2}\lambda \ \diff {}^*\hat\FF+ \overline\epsilon^A\Psi^B\epsilon_{AB}\,{}^*\hat\FF\bigg]\bigg|^{\text{on-shell}}_{z=\diff z=0}.
\end{align}
We are now able to write the Ward identities, in the four-dimensional notation, which have to hold on-shell. They originate from requiring the vanishing of the  coefficient of the independent symmetry parameters in $\delta W$.
Let us denote the independent asymptotic parameters by $\Lambda (x)=\left\{\theta ^{ij},\xi ^{i},\sigma ,\eta _{\pm }^{A},\lambda \right\} $, computed
in Subection \ref{FTAEP} as the radial expansion of the bulk parameters $\hat{\Lambda}(x,z)=\{j^{ab},p^{a},\epsilon _{\pm }^{A},\hat{\lambda}\}$.
Since in the quantum effective action all divergences cancel out and the subleading terms vanish on the boundary, we can identify the bulk gauge transformations with the boundary ones,
\begin{equation}
\delta W\equiv \delta _{\Lambda }W=\left. \delta _{\hat{\Lambda}}W\right| _{z=\diff z=0}^{\text{on-shell}}.
\end{equation}
This method makes use of the fact that the quantum effective action has already been renormalized and enables to prove the invariance of the action (and therefore the validity of the Ward identities) by looking directly at the bulk parameters $\hat{\Lambda}$.

%%%%%%%%%%%%%%%%%%%%%%%%%%%%%%%%%%%%%%%%%%%%%%%%%%%%%

\paragraph{Lorentz transformations.}
We can easily verify that the coefficient of the four-dimensional Lorentz parameters $j^{ab}$ vanishes identically due to the identity (\ref{G5}) for four-dimensional gamma matrices whose properties are given in Appendix \ref{gammaconventions},
\begin{align}
\frac{\ell}{4}\,j^{ab}\overline\Psi^A\Gamma^{cd}\hat\rr_A\,\epsilon_{abcd}-\frac{\ii\ell}{2}\,j^{ab}\overline\Psi^A\Gamma_{ab}\Gamma_5\hat\rr_A=0 \,.
\end{align}

%%%%%%%%%%%%%%%%%%%%%%%%%%%%%%%%%%%%%%%%%%%%%%%%%%%%%

\paragraph{Translations.}
As for the terms containing $p^a$ one finds, up to terms which vanish  in the $z\rightarrow 0$ limit,
\begin{align}
-\frac{1}{2}\,p^aV^b\hat\RR^{cd}\epsilon_{abcd}+p^a\overline\Psi^A\Gamma_a\Gamma_5\hat\rr_A\,.
\end{align}
The above expression vanishes at the boundary by effect of the Einstein equations in the bulk (see the second of eqs.~\eqref{Psieq}),
\begin{align}
-\frac{1}{2}\,p^aV^b\hat\RR^{cd}\epsilon_{abcd}+p^a\overline\Psi^A\Gamma_a\Gamma_5\hat\rr_A=\frac{1}{2}\,p^a\epsilon_{abcd}V^b \left(\hat F^{cd}\hat\FF-\frac{1}{6}\hat F_{ef}\hat F^{ef}V^cV^d \right) \,,\label{pawa}
\end{align}
since the two terms on the right-hand side are zero at $z=0$.

%%%%%%%%%%%%%%%%%%%%%%%%%%%%%%%%%%%%%%%%%%%%%%%%%%%%%

\paragraph{Supersymmetry.}
The terms involving the parameter $\epsilon_A$ are given by
\begin{align}\nonumber
&\ii\hat A\epsilon^{AB}\overline\epsilon_A\Gamma_5\hat\rr_B+\overline\epsilon^A\Gamma_5\Gamma_a\hat\rr_AV^a+\frac{\ii\ell}{2}\hat\RR^{ab}\overline\epsilon^A\Gamma_5\Gamma_{ab}\hat\rr_A-\ii\epsilon^{AB}\hat\FF \overline\epsilon_A\Gamma_5\Psi_B\\ \nonumber
&-\frac{\ell}{4}\overline\epsilon^A\Gamma^{ab}\Psi_A\hat\RR^{cd}\epsilon_{abcd}+\ii\hat A\epsilon^{AB}\overline\epsilon_B\Gamma_5\hat\rr_A-\overline\epsilon^A\Gamma_a\Gamma_5\hat\rr_AV^a+\overline\epsilon^A\Psi^B\epsilon_{AB}\, {}^*\hat\FF \\
&=\overline\epsilon^A(-2\Gamma_aV^a\Gamma_5\hat\rr_A+\epsilon_{AB}\Psi^B\,{}^*\hat\FF-\ii\epsilon_{AB}	\hat\FF\Gamma_5\Psi^B) \,.
\end{align}
They vanish as a consequence of the equations of motion of the gravitini (\ref{Psieq}).

%%%%%%%%%%%%%%%%%%%%%%%%%%%%%%%%%%%%%%%%%%%%%%%%%%%%%

\paragraph{Abelian transformations.}
Finally, we evaluate the terms depending on  $\hat\lambda$ and find
\begin{align}
\hat\lambda\left(-\frac{1}{2}\,\diff {}^*\hat\FF-\ii\epsilon^{AB}\overline\Psi_B\Gamma_5\hat\rr_A\right)\,,
\end{align}
which vanishes by virtue of the gauge field equation of motion in \eqref{Psieq}.\bigskip

This proves that, on-shell, $ \delta W=0$, namely that the equations (\ref{WID}), which were derived from $ \delta W=0$ in the three-dimensional notation, are indeed satisfied. This can be seen as a consequence of the absence of any anomaly, in particular conformal anomaly, in $d=3$.
Note that the term in \eqref{pawa} which is proportional to $p^3$ and which, as we have shown above, vanishes once the $a=3$ component of the Einstein equations in the bulk (the second of eqs.~\eqref{Psieq}) is implemented, coincides, once integrated over the boundary, with the variation of the generating functional under a dilatation, being $p^3=-\ell\sigma$ at $z=0$. Its vanishing provides the trace Ward identity \eqref{TheTrace}.

The above explicit proof can also be seen as following from the general form of the field equations (\ref{eom}) derived from the $\mathcal{N}=2$ bulk Lagrangian (\ref{Lfull}), which is of MacDowell–Mansouri type. Indeed, being the currents identified with subleading terms in the boundary expansions of the curvatures, see eq.~(\ref{JR1}), one can view the Ward identities as following from eq.~(\ref{eom}), computed at the boundary.

Note that, in the above derivation, we have neglected the curvature-contraction terms occurring in the general expression of the symmetry variations of the fields (\ref{gauge}),\footnote{These are the terms in the symmetry transformation of the fields which, according to the general formula given in footnote \ref{footnote1}, are expressed in terms of the superspace components of the curvatures along the anholomic basis $(V^a,\Psi_A)$, whose expressions can be found in eqs.~(\ref{rhopara}).} which one can check to give vanishing contributions at the boundary.

%%%%%%%%%%%%%%%%%%%%%%%%%%%%%%%%%%%%%%%%%%%%%%%%%%%%%

\section{Discussion}\label{concl}

In the present paper we have developed in detail
the holographic framework for an $\mathcal{N}=2$ pure AdS$_4$ supergravity in the first order formalism, including all the contributions in the fermionic fields.  This analysis, which generalizes the one of \cite{Amsel:2009rr,Papadimitriou:2017kzw}, includes a general discussion of the  gauge-fixing conditions on the bulk fields which yield the asymptotic symmetries at the boundary. The corresponding currents of the boundary theory are constructed and shown to satisfy the associated Ward identities, once the field equations of the bulk theory are imposed.

Consistency of the holographic setup, in particular the finiteness of the quantum generating functional of the boundary theory, is shown to require the vanishing of the super-AdS curvatures computed at the boundary, which was proven in \cite{Andrianopoli:2014aqa} to be a necessary condition for a consistent definition of the bulk supergravity. In particular, the vanishing of $\hat{{\bf R}}^{ij}\vert_{\partial \mathcal{M}}$ determines the general expression of the super-Schouten tensor $\mathcal{S}^i$ of the boundary theory, which generalizes the more familiar bosonic expression of standard gravity by the inclusion of gravitini bilinears, see eq.~(\ref{schouten}). The same applies to the superpartner of $\mathcal{S}^i$, namely the conformino.
Working in the first order formalism, we are able to keep the full superconformal structure of the theory manifest in principle, even if only a part of it is realized as a symmetry of the theory on $\partial\mathcal{M}$, as the rest appears as a non-linear realization on $\partial\mathcal{M}$.
Furthermore, an important role in our analysis is played by the supertorsion constraint $\hat{\bf R}^a=0$, where $\hat{\bf R}^a$ was defined in eq.~(\ref{lagsuper}), which determines the bulk spin connection.
In particular, the radial component, $\hat{\bf R}^3=0$, of this condition poses general constraints on the sources of the boundary CFT. In the FG parametrization of the bulk background, that condition implies a non-vanishing antisymmetric component of the super-Schouten tensor, proportional to the gravitini bilinear $\overline\psi_{A+[\mu}\psi_{A-\nu]}$, see eq.~(\ref{schoutenAS}). This shows that in general the superconformal structure and  the conformino field $\psi_{A-\mu}$  pose an obstruction to the symmetrization of $\mathcal{S}_{\mu\nu}$. For a special choice of background, for which $\psi_{A-\mu}\propto \psi_{A+\mu}$, $\overline\psi_{A+[\mu}\psi_{A-\nu]}=0$ and the super-Schouten tensor becomes symmetric, i.e. ${\mathcal S}^i\wedge E_{i}=0$. This latter property restricts ${\mathcal S}^i$ to be proportional to $E^i$. The manifest SCFT symmetry is then broken to  the symmetry of the chosen background which, in this case, is a maximally symmetric spacetime: AdS$_3$  ($\mathcal{S}^i\neq 0$, $\psi_{\pm\mu}\neq 0$),  dS$_3$ ($\mathcal{S}^i\neq 0$, $\psi_{\pm\mu}=0$) or Mink$_3$ ($\psi_{-\mu}={\mathcal S}^i=0$),  and provides the vacuum of the boundary theory.\footnote{The AdS$_3$ and dS$_3$ cases are distinguished by the sign of the proportionality factor between $\mathcal{S}^i$ and $E^i$.} The three (super)algebras associated with the symmetries of these backgrounds are defined by suitable projections on the ${\rm OSp}(2|4)$ asymptotic symmetry group.

As far as the gauge fixing conditions are concerned, we refrain from imposing $\gamma^\mu\psi_{\pm \mu}=0$ in SCFT, having in mind  generalizations of standard holography where this condition is relaxed in the boundary theory. This has a bearing on the radial gauge fixing condition on the gauge field. This generalization is needed in particular to apply the holographic analysis to the AVZ model \cite{Alvarez:2011gd} as boundary field theory,
where the only propagating degrees of freedom are associated with a spin-$1/2$ field $\chi$, which is identified with the contraction $\gamma^\mu\psi_\mu$ itself. This theory is naturally defined on an AdS$_3$ background. In \cite{Andrianopoli:2019sqe} it was shown that the spinor $\chi$ is actually the Nakanishi-Lautrup field associated with the covariant gauge fixing of the odd local symmetries in a three-dimensional Chern-Simons theory with gauge supergroup ${\rm OSp}(2|2)\times {\rm SO}(2,1)$. This opens a window on the definition of the dual field theory of which the AVZ model provides an effective description. We shall pursue this objective in a  future investigation. Other future directions of research would be an extension of the present analysis to $\mathcal{N}>2$ bulk supergravity, along the lines of \cite{Andrianopoli:2019sip}, or the $D>4$ bulk dimensions where, for odd $D$, quantum anomalies would arise in a boundary SCFT. Furthermore, a  generalization of the present work to the case where the  FG  choice of parametrization is relaxed, which would allow the full superconformal  symmetry of the boundary theory to be linearly reali{z}ed, will also be object of our investigation.

%%%%%%%%%%%%%%%%%%%%%%%%%%%%%%%%%%%%%%%%%%%%%%%%%%%%%%

\section*{Acknowledgments}

We are grateful to Andrés Anabal\'on, Riccardo D'Auria, Stefan Theisen and Jorge Zanelli for many stimulating discussions.
This work was funded in part by FONDECYT Grants N$^\circ$1190533 (O.M.) and N$^\circ$1170765 (R.O.), as well as VRIEA-PUCV Grant N$^\circ$123.764.
L.R. would like to thank the Department of Applied Science and Technology of the Polytechnic University of Turin and in particular Fabrizio Dolcini and Andrea Gamba, for financial support.

%%%%%%%%%%%%%%%%%%%%%%%%%%%%%%%%%%%%%%%%%%%%%%%%%%%%%%

\appendix

\section{Conventions
\label{conventions}}

\subsection{Curvature conventions
\label{Rconventions}}

In our conventions, the bulk local coordinates are denoted by $x^{\hat{\mu}}=(x^{\mu },z)$ and the boundary coordinates by $x^{\mu }$ ($\mu =0,\ldots 3$). In general, the hatted quantities always refer to the bulk and the non-hatted ones to the boundary placed at $z=0$.

As respect to the connection and curvature conventions, apart from the hatted (bulk) ones $\{\hat{\omega},\hat{\Gamma},\hat{\mathcal{R}},\hat{R}, \hat{\rho}\}$ and the non-hatted (boundary) ones $\{\omega ,\Gamma,\mathcal{R}, R, \rho\}$, the circle above the quantity, $\{\mathring{\omega},\mathring{\Gamma}, \mathcal{\mathring{R}}\}$, denotes that it is torsion-free and the bold symbol, $\{\mathbf{\hat{R}},\mathbf{R},\boldsymbol {\hat{\rho}},\boldsymbol {\rho}\}$, denotes that it is super-covariant. Here
$\{ \hat{\rho},\boldsymbol {\hat{\rho}},\rho,\boldsymbol {\rho} \}$ correspond to the fermionic components of the supercurvatures. Similar notation applies for the Abelian supercurvatures  $\{ \hat{F},\mathbf{\hat{F}},F,\mathbf{F}\}$ where, furthermore, the Maxwell field strength on the boundary is denoted by $\mathcal{F}$.

Explicitly, we have in the bulk the Lorentz curvature 2-form $\hat{\mathcal{R}}^{ab}=\frac{1}{2}\,\hat{\mathcal{R}}_{\ \ \hat{\mu}\hat{\nu}}^{ab}$ $\diff x^{\hat{\mu}}\wedge \diff x^{\hat{\nu}}$ defined in terms of the bulk spin connection $\hat{\omega}_{\hat{\mu}}^{ab}$. Using the first vielbein
postulate,
\begin{equation}
\mathcal{\partial }_{\hat{\mu}}V_{\ \hat{\nu}}^{a}+\hat{\omega}_{\hat{\mu}}^{ab}V_{b\hat{\nu}}=\hat{\Gamma}_{\hat{\nu}\hat{\mu}}^{\hat{\lambda}}V_{\
\hat{\lambda}}^{a}\,, \label{FVpostulate}
\end{equation}
it is mapped to the bulk curvature tensor,
\begin{equation}
\hat{\mathcal{R}}_{\ \hat{\sigma}\hat{\mu}\hat{\nu}}^{\hat{\lambda}}(\hat{\Gamma})=\hat{\mathcal{R}}_{\ \ \hat{\mu}\hat{\nu}}^{ab}(\hat{\omega})V_{\
a}^{\hat{\lambda}}V_{\hat{\sigma}b}\,,
\end{equation}
expressed in terms of the bulk affine connection $\hat{\Gamma}_{\hat{\nu}\hat{\mu}}^{\hat{\lambda}}$.
The bulk AdS curvature 2-form is denoted by $\hat{R}^{ab}$ and the super AdS curvature by $\mathbf{\hat{R}}^{ab}$.

On the other hand, on the boundary, the Lorentz curvature 2-form is $\mathcal{R}^{ij}=\frac{1}{2}\,\mathcal{R}_{\ \ \mu \nu }^{ij}$ $\diff x^{\mu
}\wedge \diff x^{\nu }$, from which we can obtain $\mathcal{R}_{\ \sigma \mu \nu }^{\lambda }(\Gamma )=\mathcal{R}_{\ \ \mu \nu }^{ij}(\omega )E_{\ \
i}^{\lambda }E_{\sigma j}$, where $\Gamma _{\nu \mu }^{\lambda }$ and $\omega _{\mu }^{ij}$ are the (torsionful) affine and spin connection,
respectively. The boundary AdS curvature 2-form is $R^{ij}$ and the super AdS curvarure $\mathbf{R}^{ij}$. Similarly, the torsionless quantities on
the boundary are $\mathcal{\mathring{R}}_{\ \ \mu \nu }^{\lambda \sigma }=\mathcal{\mathring{R}}_{\ \ \mu \nu }^{ij}E_{\ \ i}^{\lambda }E_{\ \
j}^{\sigma }$, where the corresponding Levi-Civita connections are $\mathring{\Gamma}_{\nu \mu }^{\lambda }$ and $\mathring{\omega}_{\mu }^{ij}$.\medskip

In the following list, we summarize different Lorentz and AdS (super)curvatures and the places where they appear for the first time in the text.\medskip

\textbf{Pure gravity}

$d=3$

In (\ref{g(2)}): $\mathring{\mathcal{R}}_{\ \nu \lambda \sigma }^{\mu }$ torsionless Lorentz curvature of $\mathring{\Gamma}_{\nu \lambda }^{\mu }=\mathring{\Gamma}_{\nu \lambda }^{\mu }(g_{(0)})$

In (\ref{tildeOmega}): $\mathring{\Gamma}^\mu_{\nu\lambda}(g)$ $z$-dependent affine Levi-Civita

In (\ref{W=0}): $\mathring{\mathcal{R}}^{ij}$ torsionless Lorentz curvature of $\mathring{\omega}^{ij}$ \medskip

$D=4$

In (\ref{var}): $\hat R^{ab}$ AdS curvature of $\hat \omega^{ab}$

After (\ref{var}): $\hat{\mathcal{R}}^{ab}$ Lorentz curvature of $\hat \omega^{ab}$

In (\ref{omega pure}): $\hat{\Gamma}_{\hat{\nu}\hat{\mu}}^{\hat{\lambda}}$ affine connection \medskip

\textbf{Supergravity}

$d=3$

In (\ref{Fradkin}): $\mathcal{R}_{\ \nu \lambda \sigma }^{\mu }$ torsionful Lorentz curvature of $\Gamma _{\phantom{\lambda}\mu \nu }^{\lambda }$

In (\ref{Geometric S}): $\mathring{\mathcal{R}}_{\ \nu \lambda \sigma }^{\mu }$ torsionless Lorentz curvature of $\mathring{\Gamma}_{\phantom{\lambda}\mu
\nu }^{\lambda }$

In (\ref{AAdSspace}): $\mathcal{R}^{ij}$  torsionful Lorentz curvature of $\omega ^{ij}$

In (\ref{schouten}): $\mathcal{R}_{\ \nu \lambda \sigma }^{\mu }$ torsionful Lorentz curvature of $\Gamma _{\phantom{\lambda}\mu \nu }^{\lambda }$

In (\ref{Mcboundary}): $\mathbf{R}^{ij}$ boundary AdS supercurvature
\medskip

$D=4$

In (\ref{curv}): $\hat{\mathcal{R}}^{ab}$ Lorentz curvature of $\hat \omega^{ab}$

In (\ref{lagsuper}): $\mathbf{\hat{R}}^{ab}$ super-AdS curvature

In (\ref{Rheo R}): $\tilde{R}^{ab}{}_{cd}$ rheonomic parametrization of the supercurvature

%%%%%%%%%%%%%%%%%%%%%%%%%%%%%%%%%%%%%%%%%%%%%%%%%%%%%%

\subsection{Gamma matrices and spinor conventions \label{gammaconventions}}

In the present paper we follow the notation of \cite{Andrianopoli:2019sip}.
The four-dimensional $4\times 4$ gamma matrices $\Gamma ^{a}$ ($a=0,1,2,3$) satisfy the Clifford algebra
\begin{equation}
\{\Gamma ^{a},\Gamma ^{b}\}=2\kappa ^{ab}\,,\quad \kappa ^{ab}=\text{diag}(+,-,-,-)\,,
\end{equation}
and the fifth matrix is defined by
\begin{equation}
\Gamma _{5}=\ii\,\Gamma ^{0}\Gamma ^{1}\Gamma ^{2}\Gamma ^{3}\,.
\end{equation}
They have the properties
\begin{equation}
(\Gamma ^{i})^{\dagger }=\Gamma ^{0}\Gamma ^{i}\Gamma ^{0}\,,\qquad (\Gamma_{5})^{\dagger }=\Gamma _{5},
\end{equation}
and they satisfy the identity
\begin{gather}
\frac 12 \,\epsilon _{abcd}\,\Gamma ^{cd} =\ii
\Gamma _{ab}\Gamma_{5}\,,  \label{G5}
\end{gather}
where
\begin{gather}
\Gamma^{a_1 \cdots a_n}= \Gamma^{[a_1 \cdots a_n]} \equiv
\begin{cases}
\frac{1}{2} \bigl[ \Gamma^{a_1}, \Gamma^{a_2 \cdots a_n} \bigr]\,, \quad \text{for even } n\,,\medskip\\
\frac{1}{2} \bigl \{ \Gamma^{a_1}, \Gamma^{a_2 \cdots a_n} \bigr \}\,, \quad \text{for odd } n\,.
\end{cases}
\end{gather}

We can also define the charge conjugation matrix $C$ that determines the symmetry properties of the gamma matrices,
\begin{equation}
C=\Gamma ^{0}\,, \qquad C\Gamma ^{a}C^{-1}=-(\Gamma ^{a})^{T}\,.
\label{symmetry}
\end{equation}
From this condition, we can derive a general property of the antisymmetric product of $k$ gamma matrices as
\begin{equation}
\left( C\Gamma ^{a_{1}\dots a_{k}}\right) ^{T}=-(-1)^{\frac{k(k+1)}{2}}\,C\Gamma ^{a_{1}\dots a_{k}} \,.
\end{equation}
Furthermore, the following identity holds for the gamma matrices in any $D$ dimensions \cite{Castellani:1991et}
\begin{equation}
\Gamma ^{a_{1}\dotsc a_{n}c_{1}\dotsc c_{q}}\Gamma _{c_{1}\dotsc
c_{q}b_{1}\dotsc b_{m}}=\sum_{k=0}^{\mathrm{inf(}n,m\mathrm{)}%
}c_{k}(q,n,m)\,\delta _{\lbrack b_{1}}^{[a_{1}}\dotsc \delta
_{b_{k}}^{a_{k}}\,\Gamma _{\ \ \ \ \ \ \ \ \ b_{k+1}\dotsc
b_{m}]}^{a_{k+1}\dotsc a_{n}]}\,,  \label{For_App}
\end{equation}
where the coefficients reads
\begin{equation}
c_{k}(q,n,m)=(-1)^{\frac{1}{2}q(q-1)+\frac{k}{2}[k-(-1)^{n-1}]}\binom{n}{k}\,
\binom{m}{k}\,q!\,k!\,\binom{D-n-m+k}{q} \,.
\end{equation}

It is convenient to introduce the $2\times 2$ gamma matrices $\gamma ^{i}$ ($i=0,1,2$) that are the elements of the $d=3$ Clifford algebra
\begin{equation}
\{\gamma ^{i},\gamma ^{j}\}=2\kappa ^{ij}\,,\qquad \kappa ^{ij}=\text{diag}(+,-,-)\,.
\end{equation}
The $D=4$ gamma matrices can be represented in terms of these $d=3$ gamma matrices as
\begin{eqnarray}
\Gamma ^{i} &=&\sigma _{1}\otimes \gamma ^{i}\,,\quad \gamma ^{0}=\sigma_{2}\,,\quad \gamma ^{1}=\ii\sigma _{1}\,,\quad \gamma ^{2}=\ii\sigma _{3}\,,
\notag \\
\Gamma ^{3} &=&\ii\sigma _{3}\otimes \mathbf{1}\,,\quad \Gamma _{5}=\ii \,\Gamma ^{0}\Gamma ^{1}\Gamma ^{2}\Gamma ^{3}=-\sigma _{2}\otimes \mathbf{1}=
\begin{pmatrix}
\mathbb{0} & \ii\mathbb{1}_{2} \\
-\ii\mathbb{1}_{2} & \mathbb{0}
\end{pmatrix}
\,.
\end{eqnarray}
An identity often used in the text is
\begin{equation}
\gamma^i\gamma^j=\kappa^{ij}+\ii \,\epsilon^{ijk}\,\gamma_k\,,\qquad \epsilon ^{012}=1\,,
\end{equation}
that implies
\begin{equation}
\gamma ^{ij}=\ii\,\epsilon ^{ijk}\,\gamma_k \,, \qquad \gamma^{ij} \equiv \frac 12 \, [ \gamma^i, \gamma^j ]\,.
\end{equation}

Let us now focus on the spinor conventions. The Majorana 4-spinor 1-form $\Psi = \Psi_{\hat{\mu}}\,\diff x^{\hat{\mu}}$ has Grassmannian components $\Psi_{\hat{\mu}}$. Using the symmetry properties of the gamma matrices (\ref{symmetry}), we obtain the following ones for the fermionic bilinears,
\begin{equation}
\begin{array}{llllll}
\overline{\Psi }{}_{A\hat{\mu}}\Psi _{B\hat{\nu}} & = & \overline{\Psi }{}_{B\hat{\nu}}\Psi _{A\hat{\mu}}\,,\smallskip  & \overline{\Psi }{}_{A\hat{\mu}%
}\Gamma _{5}\Psi _{B\hat{\nu}} & = & \overline{\Psi }{}_{B\hat{\nu}}\Gamma
_{5}\Psi _{A\hat{\mu}}\,, \\
\overline{\Psi }{}_{A\hat{\mu}}\Gamma ^{a}\Psi _{B\hat{\nu}} & = & -
\overline{\Psi }_{B\hat{\nu}}\Gamma ^{a}\Psi _{A\hat{\mu}}\,,\smallskip  &
\overline{\Psi }{}_{A\hat{\mu}}\Gamma ^{a}\Gamma _{5}\Psi _{B\hat{\nu}} & =
& \overline{\Psi }{}_{B\hat{\nu}}\Gamma ^{a}\Gamma _{5}\Psi _{A\hat{\mu}}\,,
\\
\overline{\Psi }{}_{A\hat{\mu}}\Gamma ^{ab}\Psi _{B\hat{\nu}} & = & -\overline{\Psi }{}_{B\hat{\nu}}\Gamma ^{ab}\Psi _{A\hat{\mu}}\,,\quad  &
\overline{\Psi }{}_{A\hat{\mu}}\Gamma ^{ab}\Gamma _{5}\Psi _{B\hat{\nu}} & =
& -\overline{\Psi }{}_{B\hat{\nu}}\Gamma ^{ab}\Gamma _{5}\Psi _{A\hat{\mu}%
}\,.
\end{array}
\end{equation}
In view of the application to the holographic duality, it is convenient to choose a gamma matrix basis   where only Lorentz invariance in $d=3$ dimensions is manifest, where the radial matrix $\Gamma^3$ is associated with the generator $T_0$ of the $\mathrm{SO}(1,1)$ group given by eq.~\eqref{T0}.
Then, for our purposes, it is useful to decompose the four-spinor $\Psi$ in eigenmodes $\Psi_\pm$ of the matrix $\Gamma^3$,
\begin{equation}
\Gamma^3 \Psi_\pm=\pm \ii\Psi_\pm \label{Gamma3}\,,
\end{equation}
where the projectors and the corresponding  projections are given by
\begin{equation}
\mathbb{P}_\pm = \frac{\mathbb{1}  \mp \ii \Gamma^3}{2} \quad \Rightarrow \quad \mathbb{P}_\pm \Psi_\pm= \Psi_\pm\,,
\quad
\overline \Psi_\pm=\overline \Psi_\pm \mathbb{P}_{\mp}\,.
\label{Ppm}
\end{equation}

Furthermore, in order to find chiral components of the fermionic expressions, we list the following useful identities,
\begin{equation}
\begin{array}{llll}
\mathbb{P}_{\pm }\Gamma ^{3} & =\pm \ii \mathbb{P}_{\pm }\,, & \mathbb{P}_{\pm
}\Gamma _{ij} & =\Gamma _{ij}\mathbb{P}_{\pm }\,, \\
\mathbb{P}_{\pm }\Gamma _{i} & =\Gamma _{i}\mathbb{P}_{\mp }\,,\qquad  &
\mathbb{P}_{\pm }\Gamma _{i3} & =\pm \ii \Gamma _{i}\mathbb{P}_{\mp
}\,,
\end{array}
\label{PPsi}
\end{equation}
as well as
\begin{equation}
\mathbb{P}_{\pm }\Gamma _{5}=\Gamma _{5}\mathbb{P}_{\mp }\,.  \label{P5}
\end{equation}
When the chiral spinors are involved, the fermionic bilinears have only the following non-vanishing terms
\begin{eqnarray}
\overline{\Psi }{}_{\hat{\mu}}\Psi _{\hat{\nu}} &=&\overline{\Psi }{}_{\hat{\mu}+}\Psi _{\hat{\nu}-}+\overline{\Psi }{}_{\hat{\mu}-}\Psi _{\hat{\nu}+}\,,
\notag \\
\overline{\Psi }{}_{\hat{\mu}}\Gamma ^{3}\Psi _{\hat{\nu}} &=&\ii\overline{\Psi }{}_{\hat{\mu}-}\Psi _{\hat{\nu}+}-\ii\overline{\Psi }{}_{\hat{\mu}%
+}\Psi _{\hat{\nu}-}\,,  \notag \\
\overline{\Psi }{}_{\hat{\mu}}\Gamma ^{i}\Psi _{\hat{\nu}} &=&\overline{\Psi}{}_{\hat{\mu}+}\Gamma ^{i}\Psi _{\hat{\nu}+}+\overline{\Psi }{}_{\hat{\mu}%
-}\Gamma ^{i}\Psi _{\hat{\nu}-}\,,  \notag \\
\overline{\Psi }_{\hat{\mu}}\Gamma ^{i} \Gamma^3\Psi _{\hat{\nu}} &=&\ii\overline{\Psi }_{\hat{\mu}+}\Gamma ^{i}\Psi _{\hat{\nu}+}-\ii\overline{\Psi
}_{\hat{\mu}-}\Gamma ^{i}\Psi _{\hat{\nu}-}\,,  \notag \\
\overline{\Psi }{}_{\hat{\mu}}\Gamma ^{ij}\Psi _{\hat{\nu}} &=&\overline{\Psi }{}_{\hat{\mu}+}\Gamma ^{ij}\Psi _{\hat{\nu}-}+\overline{\Psi }{}_{\hat{%
\mu}-}\Gamma ^{ij}\Psi _{\hat{\nu}+}\,,  \notag \\
\overline{\Psi }{}_{\hat{\mu}}\Gamma ^{ij} \Gamma^3\Psi _{\hat{\nu}} &=&\ii \overline{\Psi }{}_{\hat{%
\mu}-}\Gamma ^{ij}\Psi _{\hat{\nu}+}-\ii \overline{\Psi }{}_{\hat{\mu}+}\Gamma ^{ij}\Psi _{\hat{\nu}-} \,,  \notag \\
\overline{\Psi }{}_{\hat{\mu}}\Gamma_{5}\Psi _{\hat{\nu}} &=&\overline{\Psi }{}_{\hat{\mu}+}\Gamma_{5} \Psi _{\hat{\nu}+}+\overline{\Psi }{}_{\hat{\mu}%
-}\Gamma_{5} \Psi _{\hat{\nu}-}\,,  \notag \\
\overline{\Psi }{}_{\hat{\mu}}\Gamma_{5} \Gamma^3 \Psi _{\hat{\nu}} &=&\ii \overline{\Psi }{}_{\hat{\mu}+}\Gamma_{5} \Psi _{\hat{\nu}+}-\ii\overline{\Psi }{}_{\hat{\mu}%
-}\Gamma_{5} \Psi _{\hat{\nu}-}\,,  \notag \\
\overline{\Psi }{}_{\hat{\mu}}\Gamma ^{i}\Gamma_{5}\Psi _{\hat{\nu}} &=&\overline{\Psi }{}_{\hat{\mu}+}\Gamma ^{i}\Gamma_{5}\Psi _{\hat{\nu}-}+\overline{%
\Psi }{}_{\hat{\mu}-}\Gamma ^{i}\Gamma_{5}\Psi _{\hat{\nu}+}\,,  \notag \\
\overline{\Psi }{}_{\hat{\mu}}\Gamma ^{i}\Gamma _{5}\Gamma^3 \Psi _{\hat{\nu}} &=&\ii \overline{%
\Psi }{}_{\hat{\mu}-}\Gamma ^{i}\Gamma_{5}\Psi _{\hat{\nu}+}-\ii \overline{\Psi }{}_{\hat{\mu}+}\Gamma ^{i}\Gamma_{5}\Psi _{\hat{\nu}-}\,,  \notag \\
\overline{\Psi }{}_{\hat{\mu}}\Gamma ^{ij}\Gamma_{5}\Psi _{\hat{\nu}} &=&\overline{\Psi }{}_{\hat{\mu}+}\Gamma ^{ij} \Gamma_5\Psi _{\hat{\nu}+}+\overline{\Psi }{}_{\hat{\mu}-}\Gamma ^{ij}\Gamma_5\Psi _{\hat{\nu}-}\,,  \notag \\
\overline{\Psi }{}_{\hat{\mu}}\Gamma ^{ij}\Gamma_{5}\Gamma^3\Psi _{\hat{\nu}} &=&\ii \overline{\Psi }{}_{\hat{\mu}+}\Gamma ^{ij} \Gamma_5\Psi _{\hat{\nu}+}-\ii \overline{\Psi }{}_{\hat{\mu}-}\Gamma ^{ij}\Gamma_5\Psi _{\hat{\nu}-}\,.
\end{eqnarray}

In the context of holography, only the radial decomposition (with respect to $\Gamma^3$) is relevant and used to define the chiral componets. We do not use the Weyl decomposition of the four-spinor with respect to $\Gamma_5$.

Finally, let us list the three-dimensional Fierz identities used in the main text,
\begin{align}
\psi _{A+}\overline{\zeta }_{B+}=& -\frac{1}{4}\,\delta _{AB}\,\left(\overline{\psi }_{+}^{C}\zeta _{C+}\right) -\frac{1}{4}\,\epsilon
_{AB}\epsilon ^{CD}\,\left( \overline{\psi }_{C+}\zeta _{D+}\right)   \notag
\label{fierzidy} \\
& +\frac{1}{4}\,\delta _{AB}\,\gamma _{i}\left( \overline{\psi }{}%
_{+}^{C}\gamma ^{i}\zeta _{C+}\right) +\frac{1}{4}\,\epsilon _{AB}\epsilon^{CD}\,\gamma _{i}\left( \overline{\psi }{}_{C+}\gamma ^{i}\zeta
_{D+}\right) \,,  \notag \\
\psi _{A+}\overline{\psi }_{B+}=& -\frac{1}{4}\,\epsilon _{AB}\epsilon^{CD}\,\left( \overline{\psi }_{C+}\psi _{D+}\right) +\frac{1}{4}\,\delta
_{AB}\,\gamma _{i}\left( \overline{\psi }{}_{+}^{C}\gamma ^{i}\psi
_{C+}\right) \,,
\end{align}
with the following convention for the SO(2) invariant tensor
\begin{equation}
\epsilon _{AB}=\epsilon ^{AB}=\left(
\begin{array}{cc}
0 & 1 \\
-1 & 0%
\end{array}%
\right) ,\qquad A,B,\dotsc =1,2\,.
\end{equation}

%%%%%%%%%%%%%%%%%%%%%%%%%%%%%%%%%%%%%%%%%%%%%%%%%%%%%

\section{Asymptotic expansions}\label{Appasexp}

%%%%%%%%%%%%%%%%%%%%%%%%%%%%%%%%%%%%%%%%%%%%%%%%%%%%%%%%%%%%%
\subsection{Spin connection}\label{spinconnbehaviour}

In pure AdS$_4$ gravity, a spin connection $\mathring{\omega}_{\hat{\mu}}^{ab}(x,z)$ satisfies the torsion constraint $\hat{T}_{\hat{\mu}\hat{\nu}}^{a}=\mathcal{\mathring{D}}_{\hat{\mu}}V_{\hat{\ \nu}}^{a}- \mathcal{\mathring{D}}_{\hat{\nu}}V_{\hat{\ \mu}}^a=0$, see eq.~(\ref{tildeOmega}). If we use $\mathring{\omega}_{\hat{\mu}}^{ab}$ as a reference spin connection on spacetime also in the supersymmetric case,  where the vielbein satisfies instead the supertorsion constraint $\mathbf{\hat{{R}}}_{\hat{\mu}\hat{\nu}}^{a}=\mathcal{\hat{D}}_{\hat{\mu}}V_{\ \hat{\nu}}^{a}- \mathcal{\hat{D}}_{\hat{\nu}}V_{\ \hat{\mu}}^{a}-\ii\overline\Psi_{A[\hat\mu}\Gamma^a \Psi_{A\hat\nu]}=0$ given by  eq.~(\ref{lagsuper}), then the contribution of the fermions (gravitini and conformini) in the supertorsion can be taken into account as contorsion on spacetime,
\begin{equation}
\hat{\omega}^{ab}=\mathring{\omega}^{ab}+C^{ab}\,,\quad C^{ab}=C_{\ \ \hat{\mu}}^{ab}\,dx^{\hat{\mu}}\,.
\end{equation}
We now evaluate how the fermions contribute to the contorsion using the condition of vanishing supertorsion. From the decomposition
$\mathcal{\hat{D}}_{\hat{\mu}}V_{\ \hat{\nu}}^{a}= \mathcal{\mathring{D}}_{\hat{\mu}}V_{\ \hat{\nu}}^{a}+C_{\ \hat{\nu}\hat{\mu}}^{a}$, we find
\begin{equation}
\mathbf{\hat{R}}_{\hat{\mu}\hat{\nu}}^{a}=0\quad \Leftrightarrow \quad C_{\hat{\lambda}[\hat{\mu}\hat{\nu}]}=-\frac{\ii}{2}\,\overline{\Psi}^A_{\hat{\mu}}\Gamma_{\hat{\lambda}}\Psi_{A\hat{\nu}}\,.
\end{equation}
The solution is
\begin{equation}
C_{\hat{\lambda}\hat{\mu}\hat{\nu}}=\frac{\ii}{2}\, \overline{\Psi}^A_{\hat{\lambda}}\Gamma _{\hat{\mu}}\Psi_{A\hat{\nu}}- \frac{\ii}{2}\, \overline{\Psi}^A_{\hat{\mu}}\Gamma_{\hat{\lambda}} \Psi _{A\hat{\nu}}+ \frac{\ii}{2}\,\overline{\Psi}^A_{\hat{%
\lambda}}\Gamma_{\hat{\nu}}\Psi _{A\hat{\mu}}\,,
\end{equation}
which can be restated in the following way
\begin{equation}
C_{\ \ \hat{\mu}}^{ab}=\frac{\ii}{2}\,V^{\hat{\nu} a}\overline{\Psi}^A_{\hat{\nu}}\Gamma ^{b}\Psi_{A\hat{\mu}}- \frac{\ii}{2}\,V^{\hat{\nu}b}\overline{\Psi}^A_{\hat{\nu}}\Gamma ^{a}\Psi_{A\hat{\mu}}+\frac{\ii}{2} \,V^{\hat{\nu}a}V^{\hat{\lambda}b}V_{c\hat{\mu}}\,\overline{\Psi}^A_{\hat{\nu}}\Gamma ^{c}\Psi_{A\hat{\lambda}}\,.
\end{equation}
Note that, since $\overline{\Psi}^A_{\hat{\nu}}\Gamma ^{c}\Psi _{A\hat{\lambda}}=-\overline{\Psi}^A_{\hat{\lambda}}\Gamma ^{c}\Psi _{A\hat{\nu}}$, the tensor $C_{\ \ \hat{\mu}}^{ab}$ is explicitly antisymmetric in $[ab]$.

To determine a radial dependence of the spin connection as it approaches to the boundary, we express each component of the contorsion in terms of the fermionic fields regular on $\partial \mathcal M^4$ and obtain
\begin{align} \nonumber
C_{\ \ z}^{i3} &=\hat{E}^{\mu i}\left( \overline{\varphi}^A_{+\mu }\varphi _{A-z}- \frac{z^{2}}{\ell ^{2}}\,\overline{\varphi}^A_{-\mu }\varphi _{A+z}\right) +\frac{\ii}{2}\left( \overline{\varphi}^A_{-z}\Gamma ^{i}\varphi _{A-z}+\frac{z^{2}}{\ell ^{2}}\,
\overline{\varphi}^A_{+z}\Gamma ^{i}\varphi _{A+z}\right) \,, \\ \nonumber
C_{\ \ z}^{ij} &=\frac{\ii z}{\ell }\,\hat{E}^{\mu [i}\left( \overline{\varphi}^A_{+\mu }\Gamma ^{j]}\varphi _{A+z}+\overline{\varphi}^A_{-\mu }\Gamma^{j]}\varphi _{A-z}\right) +\frac{z}{2\ell }\,\hat{E}^{\mu i}\,\hat{E}^{\nu j}\,\left( \overline{\varphi}^A_{-\mu }\varphi _{A+\nu }-\overline{\varphi}^A_{+\mu }\varphi
_{A-\nu }\right) \,, \\ \nonumber
C_{\ \ \mu }^{i3} &=\frac{z}{2\ell }\,\hat{E}^{\nu i}\left( \overline{\varphi}^A_{+\nu } \varphi_{A-\mu }- \overline{\varphi}^A_{-\nu}\varphi_{A+\mu}\right) +\frac{\ii z}{2\ell }\,\left( \overline{\varphi}^A_{+z}\Gamma ^{i}\varphi_{A+\mu}+\overline{\varphi}^A_{-z}\Gamma^{i} \varphi_{A-\mu } \right)  \\
&-\frac{\ii z}{2\ell}\,\hat{E}^{\nu i} \hat{E}_{j\mu}\,\left(\overline{\varphi}^A_{+\nu }\Gamma ^{j}\varphi_{A+z}+\overline{\varphi}^A_{-\nu }\Gamma ^{j}\varphi_{A-z}\right) \,, \\
C_{\ \ \mu }^{ij} &=\ii \hat{E}^{\nu [i}\left( \overline{\varphi}^A_{+\nu}\Gamma ^{j]}\varphi_{A+\mu}+\frac{z^{2}}{\ell ^{2}}\,\overline{\varphi}^A_{-\nu}\Gamma ^{j]}\varphi_{A-\mu }\right) +\frac{\ii}{2}\,\hat{E}^{\nu i}\hat{E}^{\lambda j }\hat{E}_{k\mu }\,\left( \overline{\varphi}^A_{+\nu }\Gamma ^{k}\varphi
_{A+\lambda }+\frac{z^{2}}{\ell ^{2}}\,\overline{\varphi}^A_{-\nu }\Gamma^{k}\varphi_{A-\lambda }\right) \,. \nonumber
\end{align}
From eq.~(\ref{tildeOmega}), we find for the full spin-connection
\begin{eqnarray}
\hat{\omega}_{z}^{i3} &=&\left( \overline{\varphi }_{+}^{Ai}+\frac{\ii}{2}\,%
\overline{\varphi }_{-z}^{A}\Gamma ^{i}\right) \varphi _{A-z}+\frac{z^{2}}{%
\ell ^{2}}\,\left( -\overline{\varphi }_{-}^{Ai}+\frac{\ii}{2}\overline{%
\varphi }_{+z}^{A}\Gamma ^{i}\right) \varphi _{A+z}\,,  \notag \\
\hat{\omega}_{z}^{ij} &=&\frac{z}{\ell }\,\left( \ii\overline{\varphi }%
_{+}^{A[i}\Gamma ^{j]}\varphi _{A+z}+\ii\overline{\varphi }_{-}^{A[i}\Gamma
^{j]}\varphi _{A-z}+\overline{\varphi }_{-}^{A[i}\varphi _{A+}^{j]}\right) ,
\notag \\
\hat{\omega}_{\mu }^{i3} &=&\frac{1}{z}\,\hat{E}_{\ \mu
}^{i}-\frac 12\,k_{\mu \nu }\hat{E}^{\nu i}+\frac{z}{2\ell }\left( \rule{0pt}{14pt}\overline{\varphi }_{+}^{Ai}\varphi _{A-\mu }-\overline{\varphi }%
_{-}^{Ai}\varphi _{A+\mu }+\ii\overline{\varphi }_{+z}^{A}\Gamma ^{i}\varphi
_{A+\mu }\right.  \nonumber\\
&&\left. \rule{0pt}{14pt}-\ii\overline{\varphi }_{+}^{Ai}\Gamma _{\mu
}\varphi _{A+z}+\ii\overline{\varphi }_{-z}^{A}\Gamma ^{i}\varphi _{A-\mu }-%
\ii\overline{\varphi }_{-}^{Ai}\Gamma _{\mu }\varphi _{A-z}\right) , \label{w_asympt}
\\
\hat{\omega}_{\mu }^{ij} &=&\mathring{\omega}_{\mu }^{ij}+\ii\overline{\varphi }_{+}^{A[i}\Gamma ^{j]}\varphi _{A+\mu }+\frac{\ii}{2}\,\overline{%
\varphi }_{+}^{Ai}\Gamma _{\mu }\varphi _{A+}^{j}+\frac{z^{2}}{\ell ^{2}}%
\,\left( \ii\overline{\varphi }_{-}^{A[i}\Gamma ^{j]}\varphi _{A-\mu }+\frac{%
\ii}{2}\,\overline{\varphi }{}_{-}^{Ai}\Gamma _{\mu }\varphi
_{A-}^{j}\right) .  \notag
\end{eqnarray}

Therefore the $\mathcal{O}(1/z)$ term of the connection is not
modified by the fermions. This is consistent with the asymptotically AdS behaviour of the extrinsic curvature, being proportional to the induced metric thanks to this fact.

The most general gauge fixing, with $\Psi
_{\pm z}\neq 0$, is
\begin{eqnarray}
\hat{\omega}_{z}^{i3} &=&w^{i}(x,z)\,,  \notag \\
\hat{\omega}_{z}^{ij} &=&\frac{z}{\ell }\,w^{ij}(x,z)\,,  \label{w}
\end{eqnarray}
where $w^{i},w^{ij}=\mathcal{O}(1)$ and the boundary fields are
\begin{eqnarray}
\hat{\omega}_{\mu }^{i3} &=&\frac 1z\, E_{\ \mu }^{i} -\frac{z}{\ell ^{2}}\,
\tilde{S}_{\ \mu }^{i}-\frac{2z^{2}}{\ell ^{3}}\,\tilde{\tau}_{\ \mu }^{i} + \mathcal{O}(z^{3})\,,  \notag \\
\hat{\omega}_{\mu }^{ij} &=&\omega _{\mu }^{ij}+\frac{z}{\ell}\omega^{ij}_{\mu(1)}+\frac{z^{2}}{\ell ^{2}}\,\omega _{(2)\mu}^{ij}+\frac{z^{3}}{\ell ^{3}}\,\omega _{(3)\mu }^{ij}+\mathcal{O}(z^{4})\,,
\end{eqnarray}
where now $S_{\ \mu }^{i}\neq \tilde{S}_{\ \mu }^{i}$, $\tau _{\ \mu }^{i}\neq \tilde{\tau}_{\ \mu }^{i}$ and $\omega _{\mu }^{ij}\neq \mathring{\omega}_{\mu }^{ij}$.

As particular cases, let us notice that when $\Psi ^A_{-z}=0$ and $\Psi^A _{+z}\neq 0$, the behaviour (\ref{w_asympt})
yields $w^{i}=\mathcal{O}(z^{2})$ and all other components remain the same. Furthermore, if we set to zero both components $\Psi ^A_{\pm z}=0$, we have $w^{i}=0$ exactly.

This behaviour of $w^{i}$ and $w^{ij}$ that we just described is summarized in the table (\ref{cases}).

%%%%%%%%%%%%%%%%%%%%%%%%%%%%%%%%%%%%%%%%%%%%%%%%%%%%%

\subsection{The supercurvatures \label{SuperCurvatures}}

In this subsection we evaluate, for the most general gauge fixings, the first contributions in the asymptotic expansion of the super field strengths, decomposing them with respect to a world-volume basis on the four-dimensional spacetime. Let us generically denote by $\hat{\mathbf{R}}^\Lambda=\{ \mathbf{\hat{R}}^{ab}, \mathbf{\hat{R}}^a, \hat{\rr}^A, \mathbf{\hat{F}} \}$ the supercurvature 2-form field strengths given by eq.~\eqref{lagsuper} and further discussed in eq.~\eqref{bocu} of Subsection \ref{SuperA},
\begin{equation}\label{curv2formgendef}
\hat{\mathbf{R}}^\Lambda= \frac 12 \, \hat{\mathbf{R}}^\Lambda_{\hat\mu\hat\nu}\,\diff x^{\hat\mu}\wedge \diff x^{\hat\nu} = \frac 12 \,  \hat{\mathbf{R}}^\Lambda_{\mu\nu}\,\diff x^{\mu}\wedge \diff x^{\nu}+\hat{\mathbf{R}}^\Lambda_{\mu z}\,\diff x^{\mu}\wedge \diff z\,.
\end{equation}
We use the following notation for the supercurvature expansion,
\begin{equation}
\hat{\mathbf{R}}^\Lambda_{\hat{\mu} \hat{\nu}} =\sum_{n=n_{\min }}^{\infty }\left( \frac{z}{\ell }\right) ^{n}\hat{\mathbf{R}}^\Lambda_{(n)\hat{\mu} \hat{\nu} } \, ,
\end{equation}
where $n_{\min }$ denotes the minimal power of $\frac{z}{\ell}$ in the expansion, that is the order of the most divergent term.
Our covariant derivatives $\mathcal{\hat{D}}$ and $\mathcal{D}$, acting as exterior covariant derivatives, contain only the spin-connection.

From the supertorsion constraint $\mathbf{\hat{R}}_{\hat{\mu}\hat{\nu}}^{a}= 2\, \mathcal{\hat{D}}_{[\hat{\mu}}V_{\ \hat{\nu}]}^{a}-\ii \, \overline{\Psi}^A_{\hat{\mu}}\Gamma ^{a}\Psi^A _{\hat{\nu}}=0$, we get
\begin{equation}
{\hat{\mathbf{R}}}_{{\hat{\mu}}{\hat{\nu}}}^a =\sum_{n=n_{\min}}^{\infty}\left(\frac{z}{\ell}\right)^n  \hat{\mathbf{R}}^a_{(n)\hat{\mu}\hat{\nu}}(x) = 0  \,,
\end{equation}
and find the following expansion coefficients in terms of the boundary quantities,
\begin{eqnarray}
\hat{\mathbf{R}}^i_{(-1){\mu}{\nu}}&=& \mathbf{R}^i_{{\mu}{\nu}} \,= \,\,\, 2 \,\mathcal{D} _{[\mu}E^i_{\phantom{i}\nu]}- \ii \, \overline{\psi}^{A}_{+[\mu} \gamma^i \psi^{A}_{\nu]+}=0 \,, \notag \\
\hat{\mathbf{R}}^i_{(0){\mu}{\nu}}&=& 2 \, \omega^{ij}_{(1)[\mu}E_{j|\nu]} - 2 \, \ii \, \overline{\zeta}^{A}_{+[\mu} \gamma^i\psi^{A}_{\nu]+}=0 \,, \label{R(0)mn} \\
\hat{\mathbf{R}}^i_{(1){\mu}{\nu}}&=& 2 \, \mathcal{D}_{[\mu}S^i_{\phantom{i}\nu]}+ 2 \, \omega^{ij}_{(2)[\mu}E_{j |\nu]} \nonumber \\
&& - \ii \, \left(\overline{\zeta}^{A}_{+[\mu} \gamma^i\zeta^{A}_{\nu]+}+ 2 \, \overline{\Pi}^{A}_{+[\mu} \gamma^i \psi^{A}_{\nu]+}+\overline{\psi}^{A}_{-[\mu} \gamma^i\psi^{A}_{\nu]-}\right)=0 \,, \notag  \\
\hat{\mathbf{R}}^i_{(2){\mu}{\nu}}&=& 2 \, \mathcal{D}_{[\mu}\tau^i_{\phantom{i}\nu]}+ 2 \, \omega_{(1)[\mu}^{ij}S_{j|\nu]} + 2 \, {\omega_{(3)[\mu}^{ij}}E_{j|\nu]} \nonumber \\
&& - 2 \, \ii \, \left(
\overline{\zeta}^{A}_{+[\mu} \gamma^i\Pi^{A}_{\nu]+}+  \overline{\mho}^{A}_{+[\mu} \gamma^i\psi^{A}_{\nu]+}+\overline{\zeta}^{A}_{-[\mu} \gamma^i\psi^{A}_{\nu]-}\right) = 0 \,, \notag
\end{eqnarray}
where we identified $\mho^A_{+\mu} = {\psi}^{A}_{(3)+\mu}$. Note that the last equation gives the expression for $\omega_{(3)\mu}^{ij}$ in the supersymmetric case. The next supertorsion components to be expanded are ${\hat{\mathbf{R}}}_{{\mu}z}^i$, for which we obtain
\begin{eqnarray}
\hat{\mathbf{R}}^i_{(0){\mu}{z}}&=& \frac 1{2\ell} \, \left(\tilde{S}^i_{\phantom{i} \mu} -{S}^i_{\phantom{i} \mu}\right) - \frac  12 \, w^{ij}_{(0)} E_{j\mu} \nonumber \\
&& -\frac{\ii}{2} \, \left(\overline{\psi}_{A+\mu}\gamma^i\psi_{A+z} + \overline{\psi}_{A-\mu}\gamma^i\psi_{A-z}\right)=0 \,, \label{SuperT=0} \\
\hat{\mathbf{R}}^i_{(1)\mu z}&=&\frac 1\ell \, \left(\tilde{\tau}^i_{\phantom{i} \mu} - {\tau}^i_{\phantom{i} \mu}\right) - \frac  12 \, w^{ij}_{(1)} E_{j\mu} -\frac{\ii}{2} \, \left(\overline{\psi}_{A+\mu}\gamma^i\zeta_{A+z} \right. \nonumber \\
&& + \left. \overline{\psi}_{A-\mu}\gamma^i\zeta_{A-z} +\overline{\zeta}_{A+\mu}\gamma^i\psi_{A+z} +\overline{\zeta}_{A-\mu}\gamma^i\psi_{A-z}\right)=0 \,. \label{t1muzzero}
\end{eqnarray}
On the other hand, for the $\mathbf{\hat{R}}^3$ components restricted to $\partial \mathcal{M}^4$ we find
\begin{eqnarray}
\hat{\mathbf{R}}^3_{(0){\mu}{\nu}}&=& -\frac{2}{\ell} \, \left(S_{[\mu \nu]} - \tilde{S}_{[\nu \mu]} \right) - 2 \, \ii \, \overline{\psi}_{A+[\mu} \psi_{A-\nu]}  =0 \,, \notag  \\
\hat{\mathbf{R}}^3_{(1){\mu}{\nu}}&=&  -\frac{2}{\ell} \, \left(\tau_{[\mu \nu]} - 2 \, \tilde{\tau}_{[\nu \mu]} \right) - 2 \, \ii \, \left( \overline{\psi}_{A+[\mu} \zeta_{A-\nu]} +  \overline{\zeta}_{A+[\mu} \psi_{A-\nu]}\right) = 0 \,,
\label{tautildetaufromtorsion}
\end{eqnarray}
and projected to $\diff x^\mu \wedge \diff z$ we have
\begin{eqnarray}
\hat{\mathbf{R}}^3_{(-1){\mu}{z}}&=& \frac{1}{2} \, w^i_{(0)} E_{i\mu} - \frac{\ii}{2} \, \overline{\psi}_{A+\mu}\psi_{A-z} = 0 \,, \notag \\
\hat{\mathbf{R}}^3_{(0){\mu}{z}}&=& \frac{1}{2} \, w^i_{(1)} E_{i\mu} - \frac{\ii}{2} \, \left(  \overline{\psi}_{A+\mu}\zeta_{A-z} + \overline{\zeta}_{A+\mu}\psi_{A-z}  \right) = 0 \,, \notag \\
\hat{\mathbf{R}}^3_{(1){\mu}{z}}&=& \frac{1}{2} \, w^i_{(0)} S_{i\mu} + \frac{1}{2} \, {w^i_{(2)}} E_{i \mu} - \frac{\ii}{2} \, \overline{\psi}_{A-\mu} \psi_{A+z}  \nonumber \\
&& - \frac{\ii}{2} \, \left( \overline{\psi}_{A+\mu} {\Pi_{A-z}} + \overline{\zeta}_{A+\mu}\zeta_{A-z}  + \overline{\Pi}_{A+\mu}\psi_{A-z}\right) = 0 \,,
\end{eqnarray}
where $\Pi^A_{-z}=\psi^A_{(2)-z}$. The last equation gives the expression for $w^i_{(2)}$.

Focusing now on the AdS supersurvature, from $\hat {\bf{R}}^{ij}=   {{\mathcal{R}}}^{ij} + \frac 4{\ell^2} \, V^{[i}_+ V^{j]}_- -\frac 1\ell \, \left(\overline{\Psi}^A_+\Gamma^{ij}\Psi^A_- +\overline{\Psi}^A_-\Gamma^{ij}\Psi^A_+\right)$ we get
\begin{eqnarray}
\hat {\bf{R}}^{ij}_{(0)\mu\nu} &=& {\bf R}^{ij}_{\mu\nu} \, = \,\,\,  2 \, {\mathcal{R}}^{ij}_{\mu\nu}-4 \, E^{[i}_{\phantom{[i}[\mu}\mathcal{S}^{j]}_{\phantom{j]}\nu]}-\frac 2\ell \, \overline\psi^{A}_{-\mu} \gamma^{ij}\psi^{A}_{+\nu} =0 \,, \label{rijscft} \\
\hat {\bf{R}}^{ij}_{(1)\mu\nu} &=& 2 \, \mathcal{D}_{[\mu}\omega_{(1)|\nu]}^{ij}-\frac 4{\ell^2} \, E^{[i}_{\phantom{[i} [\mu}(\tau^{j]}_{\phantom{j]} \nu]} + 2 \tilde\tau^{j]}_{\phantom{j]} \nu]}) \nonumber \\
&& -\frac 2\ell \, \left( \overline\psi^{A}_{-[\mu} \gamma^{ij}\zeta^{A}_{+\nu]}+\overline\psi^{A}_{+[\mu} \gamma^{ij}\zeta^{A}_{-\nu]}\right) \,, \notag \\
\hat {\bf{R}}^{ij}_{(-1)\mu z} &=& E^{[i}_{\phantom{[i}\mu} w_{(0)}^{j]}-\frac 1{2\ell} \,\overline{\psi}^A_{+\mu}\gamma^{ij}\psi^A_{-z} \,, \notag \\
\hat {\bf{R}}^{ij}_{(0)\mu z} &=& - \frac 1{2\ell} \left(-2 \ell \, E^{[i}_{\phantom{[i}\mu} w_{(1)}^{j]}+ \omega_{(1)\mu}^{ij}+\overline{\psi}^A_{+\mu}\gamma^{ij}\zeta^A_{-z} \right) \,, \notag \\
\hat {\bf{R}}^{ij}_{(1)\mu z} &=& \frac 12 \, \mathcal{D}_\mu w_{(0)}^{ij}-\frac 1\ell \, \omega_{(2)\mu}^{ij} -\tilde{ S}^{[i}_{\phantom{[i}\mu} w_{(0)}^{j]} \nonumber \\
&& - \frac 1{2\ell}\, \left(\overline{\psi}^A_{-\mu}\gamma^{ij}\psi^A_{+z} +\overline{\psi}^A_{+\mu}\gamma^{ij} {\Pi^A_{-z}} +\overline{\Pi}^A_{+\mu}\gamma^{ij}\psi^A_{-z}\right) \,. \notag
\end{eqnarray}
Next, from $\hat {\bf{R}}^{i3}=   \hat{\mathcal{D}}\hat\omega^{i3} - \frac 1{\ell^2} \, V^{i}V^{3} -\frac \ii{2\ell} \, \left(\overline{\Psi}^A_+\Gamma^{i}\Psi^A_+ -\overline{\Psi}^A_-\Gamma^{i}\Psi^A_-\right)$, we find
\begin{eqnarray}
\hat {\bf{R}}^{i3}_{(-1) \mu \nu} &=& \hat {\bf{R}}^{i}_{(-1) \mu \nu}\,=\,\,\, 0 \,, \notag \\
\hat {\bf{R}}^{i3}_{(0) \mu \nu} &=& \hat {\bf{R}}^{i}_{(0) \mu \nu}\,=\,\,\, 0 \,, \notag \\
\hat {\bf{R}}^{i3}_{(1) \mu \nu} &=& -\ell \, \mathcal{C}^i_{\mu \nu} \,=\,\,\, -\frac{2}{\ell} \, \mathcal{D}_{[\mu}\tilde{S}^i_{\phantom{i}\nu]} + \frac{2}{\ell} \, \omega^{ij}_{(2)[\mu} E_{j|\nu]} +\frac \ii\ell \, \left( \overline\psi^A_{-[\mu}\gamma^i\psi^A_{-\nu]} -\overline\zeta^A_{+[\mu}\gamma^i\zeta^A_{+\nu]} \right.  \notag \\
&&\left. - 2\, \overline\Pi^A_{+[\mu}\gamma^i\psi^A_{+\nu]}  \right) \,, \notag \\
\hat {\bf{R}}^{i3}_{(0)\mu z} &=& \frac 1{2} \, \mathcal{D}_\mu w^i_{(0)}
    +\frac \ii{\ell} \, \overline{\psi}^A_{-\mu}\gamma^{i}\psi^A_{-z} \,, \notag \\
\hat {\bf{R}}^{i3}_{(1)\mu z} &=& \frac{1}{2} \, \mathcal{D}_\mu w^i_{(1)} + \omega^{(1)|ij}_\mu w_{(0)|j} +  \frac 1{2\,\ell^2} \, (2\tilde{\tau}^i_{\phantom{i}\mu} +\tau^i_{\phantom{i}\mu}) + \frac{1}{2\ell} \, w^{ij}_{(0)}S_{j|\mu} \nonumber\\
  && +\frac \ii{\ell} \, \left( \overline{\psi}^A_{-\mu}\gamma^{i}\zeta^A_{-z} + \overline{\zeta}^A_{-\mu}\gamma^{i}\psi^A_{-z} \right) \,,
\end{eqnarray}
where we have also exploited the vanishing supertorsion equations (\ref{SuperT=0}) and (\ref{t1muzzero}).

As regards to the graviphoton super field strength $\hat{\bf{F}}=\diff\hat{A} - 2 \, \epsilon^{AB} \overline\Psi_{+A}\Psi_{-B}$, we obtain
\begin{eqnarray}
\hat{\bf{F}}_{(0)\mu \nu} &=& {\bf F}_{\mu \nu} \,=\,\,\,  2 \, \partial_{[\mu}A_{\nu]} -4 \, \epsilon_{AB} \overline\psi^{A}_{+[\mu} \psi^{B}_{-\nu]} =0 \,, \label{fscft} \\
\hat{\bf{F}}_{(1)\mu \nu} &=& 2 \, \partial_{[\mu}A_{(1)\nu]}-4 \, \left(\overline\psi^A_{+[\mu}\zeta^B_{-\nu]}+\overline\zeta^A_{+[\mu}\psi^B_{-\nu]}\right)\epsilon_{AB} \,, \notag \\
\hat{\bf{F}}_{(-1)\mu z} &=& \frac 12 \, \partial_\mu A_{(-1)z} -\overline\psi^A_{+\mu}\psi^B_{-z}\epsilon_{AB} \,, \notag \\
\hat{\bf{F}}_{(0)\mu z} &=& \frac 12 \, \partial_\mu  {A_{(0)z}} - \frac{1}{2\ell} \, A_{(1)\mu} -\overline\psi^A_{+\mu}\zeta^B_{-z}\epsilon_{AB} \,, \notag \\
\hat{\bf{F}}_{(1)\mu z} &=& \frac 12 \, \partial_\mu  {A_{(1)z}} -\frac 1\ell \,  {A_{(2)\mu}} -\left(\overline\psi^A_{-\mu}\psi^B_{+z}+\overline\psi^A_{+A\mu} {\Pi^B_{-z}}\right)\epsilon_{AB} \,. \notag
\end{eqnarray}

Furthermore, the gravitini supercurvature ${\hat{\boldsymbol{\rho}}}_{+A}= \diff\Psi_{+A} + \frac 14 \, \hat \omega^{ij}\Gamma_{ij} \Psi_{+A}  -\frac 1{2\ell} \, \epsilon_{AB}\hat A\Psi_{+B}$  $+\frac \ii\ell \, V^i_+ \Gamma_i \Psi_{-A} -\frac 1{2\ell} \, \Psi_{+A} V^3$ leads to
\begin{eqnarray}
\hat{\boldsymbol{\rho}}_{(-1/2)+A\mu \nu} &=& \boldsymbol{\rho}_{+A\mu \nu} \,=\,\,\,  2 \, \nabla_{[\mu}\psi_{+\nu]}^{A}+\frac{2\ii}{\ell} \, \gamma_{[\mu}\psi_{-\nu]}^{A}=0  \,, \label{rho+scft} \\
\hat{\boldsymbol{\rho}}_{(1/2)+A\mu \nu}&=&  2 \, \nabla_{[\mu}\zeta_{+A\nu]}+ {\frac 2\ell \, \gamma_{[\mu}\zeta_{-A\nu]}} +  \frac 12 \, \gamma_{ij}\omega^{ij}_{(1)[\mu}\psi_{+A\nu]}
-\frac 1{\ell} \, A_{(1)[\mu}\psi_{+B\nu]}\epsilon_{AB} \,, \notag \\
\hat{\boldsymbol{\rho}}_{(-3/2)+A\mu z} &=&  {\frac \ii{2\ell} \, \left(\gamma_\mu \psi_{-Az} -\frac{\ii}{2} A_{(-1)z} \psi_{+B\mu}\epsilon_{AB}\right)} \,, \notag \\
\hat{\boldsymbol{\rho}}_{(-1/2)+A\mu z} &=&  {\frac{\ii}{2\ell }\gamma_{\mu }\zeta _{-Az}+\frac{1}{4\ell }}\,\left(  \rule{0pt}{15pt} A_{(0)z}\psi _{+B\mu}+A_{(-1)z}\zeta _{+B\mu }\right) \epsilon _{AB}-\frac{1}{2\ell } \,\zeta _{+A\mu }\,,  \notag \\
\hat{\boldsymbol{\rho}}_{(1/2)+A\mu z} &=& \frac 12 \, \nabla_\mu \psi_{+Az}-\frac 18 \, w^{ij}_{(0)}\gamma_{ij}\psi_{+A\mu}+ \frac \ii{4\ell} \, \left(S^i_{\phantom{i}\mu}-\tilde{S}^i_{\phantom{i}\mu}\right)  \gamma_i \psi_{-Az} -\frac{1}{\ell} \, \Pi_{+A\mu}  \nonumber \\
&& -\frac{\ii}{4}\,w^{i}_{(0)}\gamma _{i}\psi _{-A\mu }+{\frac{\ii}{2\ell }\,\gamma _{\mu }\Pi _{-Az}+}\frac{1}{4\ell }\,\left( \rule{0pt}{15pt}{A_{(1)z}\psi _{+B\mu }}+A_{(-1)z}\Pi _{+B\mu }+{A_{(0)z}}\zeta _{+B\mu }\right) \epsilon _{AB}\,.  \notag
\end{eqnarray}

Finally, using the negatively graded fermionic supercuvature ${\hat{\boldsymbol{\rho}}}_{-A}= \diff\Psi_{-A} + \frac 14 \, \hat \omega^{ij}\Gamma_{ij} \Psi_{-A} $ $ -\frac 1{2\ell} \, \epsilon_{AB}\hat A\Psi_{-B} -\frac{\ii}{\ell} V^i_- \Gamma_i \Psi_{+A} +\frac 1{2\ell} \, \Psi_{-A} V^3$, we are left with
\begin{eqnarray}
\hat{\boldsymbol{\rho}}_{(1/2)-A\mu \nu}&=& \Omega_{A\mu \nu} \,=\,\,\, 2 \, \nabla_{[\mu}\psi_{-A\nu]}- \ii \, \ell \, \gamma_i \psi_{+A[\mu}\mathcal{S}^i_{\phantom{i} \nu]} \,, \notag \\
\hat{\boldsymbol{\rho}}_{(3/2)-A\mu \nu}&=& \nabla_{[\mu}\zeta_{-A\nu]}- \frac{\ii}{2} \,\ell \, \gamma_i \zeta_{+A[\mu}\mathcal{S}^i_{\phantom{i}\nu]} +\frac 14 \, \omega_{(1)|[\mu}^{ij}\gamma_{ij}\psi_{-A\nu]} \nonumber\\
&& -\frac 1{2\ell} \, A_{(1)[\mu}\psi_{-B\nu]}\epsilon_{AB} + \frac{\ii}{2\ell} \, \left( \tau^i_{\phantom{i}[\mu} + 2 \, \tilde{\tau}^i_{\phantom{i}[\mu} \right) \gamma_i \psi_{+A\nu]}  \,, \notag \\
\hat{\boldsymbol{\rho}}_{(-1/2)-A\mu z} &=& \frac 12 \, \nabla_\mu \psi_{-Az} + \frac 1{4\ell} \, A_{(-1)z}\, \epsilon_{AB}\psi_{-B\mu}+ \frac{\ii}{4} \gamma_i w^i_{(0)}\psi_{+A\mu} \,, \notag \\
\hat{\boldsymbol{\rho}}_{(1/2)-A\mu z} &=& \frac 12 \, \nabla_\mu \zeta_{-Az}  + \frac 1{4\ell} \,  {A_{(0)z}} \, \epsilon_{AB}\psi_{-B\mu}+ \frac{\ii}{4} \gamma_i w^i_{(1)} \psi_{+A\mu} \nonumber\\
&& -\frac 1{2\ell} \, \zeta_{-A\mu} +\frac 1{4\ell} \, A_{(-1)z} \zeta_{-B\mu}\epsilon_{AB} \,.
\end{eqnarray}

We observe that the $\hat{\mathbf{R}}^\Lambda_{(n_{\min})\mu\nu}$ components of $\hat{\mathbf{R}}^\Lambda=\{ \mathbf{\hat{R}}^{ab}, \mathbf{\hat{R}}^a, \hat{\rr}^A, \mathbf{\hat{F}} \}$ define the curvatures $\{ \mathbf{R}^{ij},  \mathbf{R}^i, {\rr}^A, \mathbf{F}, \mathcal{C}^i, \Omega^A \}$ of the $\mathcal{N}=2$ superconformal group $\rm{OSp}(2|4)$ discussed in Subsection \ref{SuperA} and given by eqs.~(\ref{Mcboundary}). We expect naively that they all vanish in the vacuum with the $\rm{OSp}(2|4)$ isometries in a superconformal theory on the three-dimensional boundary. However, we obtain $\hat{\mathbf{R}}^\Lambda_{(n_{\min})\mu\nu}=0$ for all the curvatures except the ones with the negative grading, $\hat{\bf{R}}^{i3}_{\mu \nu}$ and $\hat{\boldsymbol{\rho}}_{-A\mu\nu}$, where we find instead that the equations \eqref{AdSbound} lead to the weaker condition \eqref{r-scft}.

%%%%%%%%%%%%%%%%%%%%%%%%%%%%%%%%%%%%%%%%%%%%%%%%%%%%%

\subsection{Equations of motion of the graviphoton \label{A and Psi asymptotics}}

Here we analyse a relation between the gauge fixing and the asymptotic behaviour of the fields, using radial field equations. In Appendix \ref{spinconnbehaviour}, a similar problem was discussed for the spin connection using the vanishing supertorsion.

The radial evolution of the graviphoton is given by the respective field equation in \eqref{Psieq} that, in components, with the Hodge star dual (\ref{Hodge}), has the form
\begin{equation}
\mathcal{\hat{D}}_{\hat{\nu}}\mathbf{\hat{F}}^{\hat{\nu}\hat{\mu}}=\frac{\ii}{e}\,\epsilon ^{\hat{\mu}\hat{\nu}\hat{\lambda}\hat{\rho}}\, \overline{\Psi }_{\hat{\nu}}^{A}\Gamma^{5}\hat\rr_{\hat{\lambda}\hat{\rho}}^{B}\,\epsilon _{AB}\,.
\end{equation}
Using the conventions (\ref{3D epsilon}) and \eqref{e}, the component $\hat\mu = \mu$  acquires the form
\begin{equation}
\mathcal{\hat{D}}_{\nu }\mathbf{\hat{F}}^{\nu \mu }+\mathcal{\hat{D}}_{z} \mathbf{\hat{F}}^{z\mu }=-\frac{\ii }{e}\,\epsilon ^{\mu \nu \lambda}\left( 2\overline{\Psi }_{\nu }^{A}\Gamma _{5}\hat\rr_{\lambda
z}^{B}+\overline{\Psi }_{z}^{A}\Gamma _{5}\hat\rr_{\nu \lambda
}^{B}\right) \,\epsilon _{AB}\,.
\end{equation}
For convenience, we factorize  the relevant field strength components as
\begin{equation}
\begin{array}{llll}
 \mathbf{\hat{F}}^{z\mu} & =-\left( \frac{z}{\ell }\right) ^{4}g^{\mu \nu}  \mathbf{\hat{F}}_{ z \nu} \,,\qquad  & \hat\rr_{\mu z\pm }^{A} & =\left( \frac{z}{\ell }\right) ^{\pm \frac{1}{2}}\Xi _{\mu \pm }^{A}\,,\medskip  \\
\mathbf{\hat{F}}^{\mu \nu } & =\left( \dfrac{z}{\ell }\right) ^{4}F^{\mu \nu}\,, & \hat\rr_{\mu \nu \pm }^{A} & =\left( \frac{z}{\ell }\right) ^{\mp \frac{1}{2}}\Xi _{\mu \nu \pm }^{A}\,,
\end{array}  \label{F,Rho}
\end{equation}
where $\mathbf{\hat{F}}_{\mu\nu} =F_{\mu\nu}$ and the tensors $\mathbf{\hat{F}}_{z \mu}$, $F_{\mu\nu}$, $\Xi^A_{\mu\pm}$ and $\Xi^{A}_{\mu \nu \pm}$ have to be expanded in power series in $z$. The metric $g_{\mu\nu}(x,z)$ and its inverse $g^{\mu\nu}$ rise and lower the spacetime indices on $\partial{\cal M}$. Recalling the FG metric (\ref{FG}) and the tensor $k_{\mu\nu}=\partial _z g_{\mu\nu}$ introduced by eq.~(\ref{k}),
as well as using the Christoffel symbols
\begin{equation}
\begin{array}{ll}
\hat{\Gamma}_{\nu z}^{\mu }=-\dfrac{1}{z}\,\delta _{\nu }^{\mu }+\dfrac{1}{2}\,k_{\nu }^{\mu }\,,\qquad \medskip  & \hat{\Gamma}_{zz}^{\mu }=0=\hat{\Gamma}^z_{z\mu}\,, \\
\hat{\Gamma}_{\mu \nu }^{z}=- \dfrac{1}{z}\,g_{\mu \nu }+\dfrac{1}{2}\,k_{\mu\nu } \,, & \hat{\Gamma}_{zz}^{z}=-\dfrac{1}{z}\,,
\end{array}
\label{Gamma}
\end{equation}
the radial graviphoton equation becomes
\begin{eqnarray}
&&\mathcal{D}_{\nu }F^{\nu \mu }-\left( k^{\mu \nu }-\frac{k}{2}\,g^{\mu \nu}\right)\mathbf{\hat{F}}_{\nu z} +g^{\mu \nu }\partial _{z}\mathbf{\hat{F}}_{\nu z}  \label{Maxwell_mu} \\
&=&-\frac{\ii }{\hat{e}_3}\,\epsilon ^{\mu \nu \lambda }\left( 2\,\overline{\varphi }_{+\nu }^{A}\Gamma _{5}\Xi _{\lambda +}^{B} +2\,\overline{\varphi }_{-\nu }^{A}\Gamma _{5}\Xi _{\lambda-}^{B}+\overline{\varphi }_{+z}^{A}\Gamma _{5}\Xi _{\nu \lambda +}^{B}+\overline{\varphi }_{-z}^{A}\Gamma _{5}\Xi _{\nu \lambda -}^{B}\right) \epsilon _{AB}\,.  \notag
\end{eqnarray}
Now we calculate $\mathbf{\hat{F}}_{\mu z}$, $\Xi^A_{\mu\pm}$ and $\Xi^{A}_{\mu \nu \pm}$ defined in (\ref{F,Rho}). Evaluation of the
components
\begin{eqnarray}
\mathbf{\hat{F}}^{\hat{\mu}\hat{\nu}} &=&\hat{g}^{\hat{\mu}\hat{\alpha}}\hat{g}^{\hat{\nu}\hat{\beta}}\left( \partial _{\hat{\alpha}}\hat{A}_{\hat{\beta}%
}-\partial _{\hat{\beta}}\hat{A}_{\hat{\alpha}}-2\epsilon _{AB}\,\overline{\Psi }_{\hat{\alpha}}^{A}\Psi _{\hat{\beta}}^{B}\right) ,  \notag \\
\hat\rr_{\hat{\mu}\hat{\nu}}^{A} &=&2\mathcal{\hat{D}}_{[\hat{\mu%
}}\Psi _{\hat{\nu}]}^{A}-\frac{1}{\ell }\,\epsilon _{AB}\hat{A}_{[\hat{\mu}}\Psi _{\hat{\nu}]}^{B} -\frac{\ii}{\ell }\, \Gamma _{a}\Psi _{[\hat{\mu}}^{A}V_{\ \hat{\nu}]}^{a}
\end{eqnarray}
leads to
\begin{eqnarray}
\mathbf{\hat{F}}_{\mu z} &=&\partial _{\mu }\hat{A}_{z}-\partial _{z}A_{\mu }-\dfrac{2\ell }{%
z}\,\epsilon _{AB}\,\overline{\varphi }_{+\mu }^{A}\varphi _{z-}^{B}\,-\dfrac{2z}{\ell }\,\epsilon _{AB}\,\overline{\varphi }_{-\mu }^{A}\varphi_{z+}^{B}\,,  \notag \\
F^{\mu \nu } &=&g^{\mu \alpha }g^{\nu \beta }\left( \mathcal{F}_{\alpha
\beta }-4\epsilon _{AB}\,\overline{\varphi }_{+\alpha }^{A}\varphi _{-\beta }^{B}\right)=0
\end{eqnarray}
and, by means of the rescalings (\ref{Epm}), we get
\begin{eqnarray}
\Xi _{\mu \pm }^{A} &=&\mathcal{D}_{\mu }\varphi _{\pm z}^{A}-\frac{1}{4}\,\left( \dfrac{z}{\ell }\right) ^{1\mp 1}w^{ij}\Gamma _{ij}\varphi _{\pm\mu }^{A}-\left( \dfrac{z}{\ell }\right) ^{\mp 1}\partial _{z}\varphi _{\pm
\mu }^{A}-\frac{1}{2\ell }\,\epsilon _{AB}\,A_{\mu }\varphi _{\pm z}^{B}
\notag \\
&&\mp \frac{\ii }{2}\,w^{i}\Gamma _{i}\varphi _{\mp \mu }^{A}+\frac{1}{%
2\ell }\left( \dfrac{z}{\ell }\right) ^{\mp 1}\epsilon _{AB}\,\hat{A}%
_{z}\varphi _{\pm \mu }^{B}\pm \frac{\ii }{\ell }\,\left( \dfrac{z}{%
\ell }\right) ^{\mp 2}E_{\pm \mu }^{i}\Gamma _{i}\varphi _{\mp z}^{A}\,,
\notag \\
\Xi _{\mu \nu \pm }^{A} &=&2\mathcal{D}_{[\mu }\varphi _{\nu ]\pm }^{A}\pm
\frac{2\ii }{\ell }\,E_{\pm \lbrack \mu }^{i}\Gamma _{i}\varphi _{\nu
]\mp }^{A}-\frac{1}{\ell }\,\epsilon _{AB}\,A_{[\mu }\varphi _{\nu ]\pm
}^{B}\,.
\end{eqnarray}
We also assume that the gauge-fixing functions are
\begin{eqnarray}
\hat{A}_{z} &=&\frac{\ell}{z}\,A_{(-1)z} +A_{(0)z}
+\frac{z}{\ell }\,A_{(1)z}+
\mathcal{O}(z^{3})\,,  \notag \\
\hat{A}_{\mu } &=&\frac{\ell }{z}\,A_{(-1)\mu }+A_{\mu }+\frac{z}{\ell }\,A_{(1)\mu }+\frac{z^{2}}{\ell ^{2}}\,A_{(2)\mu }+\mathcal{O}(z^{3})\,,
\notag \\
\varphi _{+\mu }^{A} &=&\varphi _{(0)+\mu }^{A}+\frac{z}{\ell }\,\varphi_{(1)+\mu }^{A}+\mathcal{O}(z^{2})\,, \label{extra terms}
\end{eqnarray}
in general allowing for the linear terms (in contrast to eq.~(\ref{asympt_field (1)}) valid in pure gravity), where  $E_\pm^i$ expand as (\ref{Epm_expand}), and we find
\begin{eqnarray}
\mathbf{\hat{F}}_{\mu z} &=&\frac{\ell }{z^{2}}\,A_{(-1)\mu }+\frac{\ell }{z}\,\left(
\partial _{\mu }A_{(-1)z}-2\epsilon _{AB}\,\overline{\varphi }_{(0)+\mu }^{A}\varphi _{(0)z-}^{B}\right) +\mathcal{O}(1)\,,  \notag \\
\Xi _{\mu +}^{A} & = & \frac{\ell}{2z^{2}}\, \left(A_{(-1)z}\,\epsilon^{AB}\,\varphi _{B+\mu }+2\ii E_{\ \mu }^{i}\Gamma _{i}\varphi_{(0)-z}^{A}\right) +\frac{1}{z}\,\left( \frac{1}{2}\,\epsilon _{AB}\,A_{(0)z}\varphi_{(0)+\mu }^{B}-\varphi _{(1)+\mu }^{A}\right) +\mathcal{O}(1)\,,  \notag \\
&&\left. F^{\mu \nu },\;\Xi _{\mu -}^{A},\;\Xi _{\mu \nu \pm }^{A}=\mathcal{O}(1)\,.\right.
\end{eqnarray}
 Remembering that $k_{\mu \nu }= \mathcal{O}(z)$, the graviphoton equation (\ref{Maxwell_mu}) then yields
\begin{eqnarray}
\frac{\ell }{z^{3}} &:&\quad A_{(-1)\mu }=0\,,  \notag \\
\frac{\ell }{z^{2}} &:&\quad \partial _{\mu }A_{(-1)z}=\left( 2\,\overline{\varphi }_{(0)+\mu }^{A}-\frac{1}{e_{3}}\,g_{(0)\mu \sigma }\,\epsilon^{\sigma \nu \lambda }E_{\lambda }^{i}\overline{\varphi }_{(0)+\nu}^{A}\Gamma _{5}\Gamma _{i}\right) \epsilon _{AB}\varphi _{(0)-z}^{B}\,,\notag \\
\frac{1}{z} &:& \quad 0=\epsilon ^{\mu \nu
\lambda }\,\overline{\varphi }_{(0)+\nu }^{A}\Gamma _{5}\left( \dfrac{1}{2}\,A_{(0)z}\varphi _{(0)+\mu }^{A}+\varphi _{(1)+\mu }^{B}\epsilon_{AB}\right)\,,  \label{dA}
\end{eqnarray}
and all other terms are finite. We used the fact that the term $\overline{\varphi }_{+\nu }^{A}\Gamma _{5}\varphi _{A+\lambda }$ is symmetric in $(\nu \lambda )$ so it vanishes when contracted with $\epsilon
^{\sigma \nu \lambda }$.
From the last equation in (\ref{dA}), when $\varphi _{(0)+\mu }^{A}\neq 0$ (and otherwise), we can choose a particular solution $A_{(0)z}=0$, $\varphi _{(1)+\mu }^{A}\ \equiv\binom{\zeta _{\mu +}^{A}}{0}=0$, which is in agreement with eq.~(\ref{omega (1)}) obtained in Subsection  \ref{FTAEP}. This choice was also taken in \cite{Amsel:2009rr} in the context of ${\cal N}=1$ supergravity. We will show below (see (\ref{varphi1pmzero})) that, in fact, this is the only solution only if we assume the stronger condition \eqref{strongercondnew} to hold. Then (\ref{extra terms}) implies
\begin{eqnarray}
\hat{A}_{z} &=&\frac{\ell }{z}\,A_{(-1)z}+\frac{z}{\ell }\,A_{(1)z}+
\mathcal{O}(z^{3})\,,  \notag \\
\hat{A}_{\mu } &=&A_{\mu }+\frac{z}{\ell }\,A_{(1)\mu }+\frac{z^{2}}{\ell
^{2}}\,A_{(2)\mu }+\mathcal{O}(z^{3})\,,  \notag \\
\varphi _{+\mu }^{A} &=&\varphi _{(0)+\mu }^{A}+\mathcal{O}(z^{2})\,.
\end{eqnarray}
We also conclude that the gauge-fixing functions $A_{(-1)z}$ and $\varphi_{(0)-z}^{B}$ are correlated, which is consistent with the table (\ref{cases}). In addition, the boundary graviphoton does not acquire divergent terms of the form $1/z$ even when $\varphi^A_{(0)z-}\neq 0$. We have not considered the logarithmic terms here.

The graviphoton curvature behaves in the following way on the boundary,
\begin{eqnarray}
\mathbf{\hat{F}}_{\mu z} &=&\frac{\ell }{z}\,\left( \partial _{\mu
}A_{(-1)z}-2\epsilon _{AB}\,\overline{\varphi }_{(0)+\mu }^{A}\varphi
_{(0)-z}^{B}\right)-\frac{1}{\ell }\,A_{(1)\mu } +\mathcal{O}(z)\,,  \notag \\
\mathbf{\hat{F}}_{\mu \nu } &=&\mathcal{F}_{\mu \nu }-4\epsilon _{AB}\,
\overline{\varphi }_{+[\mu }^{A}\varphi _{-\nu]}^{B}=0\,.
\end{eqnarray}
This shows that it is possible to have the components $\mathbf{\hat{F}}_{\mu
z}\neq 0$ on the boundary $z=0,\diff z=0$, with a suitable gauge choice which changes the asymptotics.

%%%%%%%%%%%%%%%%%%%%%%%%%%%%%%%%%%%%%%%%%%%%%%%%%%%%%%
\subsection{Equations of motion of the gravitini}

The equation of motion that describes the dynamics of gravitini (\ref{Psieq}) in components has the form
\begin{equation}\label{gravitinieom4d}
0=\epsilon ^{\hat{\mu}\hat{\nu}\hat{\lambda}\hat{\tau}}\left( V_{\ \ \hat{\mu}
}^{a}\Gamma _{a}\Gamma _{5}\hat\rr_{A\hat{\nu}\hat{\lambda}}+\frac{\ii }{2}\,\epsilon _{AB}\mathbf{\hat{F}}_{\hat{\mu}\hat{\nu}}\Gamma _{5}\Psi _{\hat{\lambda}}^{B}\right) +e\,\epsilon _{AB}\Psi _{\hat{\lambda}}^{B}\,\mathbf{\hat{F}}^{\hat{\lambda}\hat{\tau}}\,,
\end{equation}
where the formula (\ref{Hodge}) was applied. The radial expansion of the gravitini is given by the components $\hat{\tau}=\mu $ which, with the conventions (\ref{3D epsilon}) and (\ref{e}), leads to
\begin{eqnarray}
0 &=&\epsilon ^{\mu \nu \lambda }\left( -V_{\ \ z}^{3}\Gamma ^{3}\Gamma _{5}\hat\rr_{A\nu \lambda } -2V_{\ \ \nu }^{i}\Gamma _{i}\Gamma _{5}\hat\rr_{Az\lambda }+\frac{\ii }{2}\,\epsilon _{AB}\mathbf{\hat{F}}_{\nu \lambda }\Gamma _{5}\Psi _{z}^{B}+\ii\,\epsilon _{AB}\mathbf{\hat{F}}_{z\nu }\Gamma _{5}\Psi _{\lambda }^{B}\right)   \notag \\
&&+e\,\epsilon _{AB}\left( \Psi _{z}^{B}\,\mathbf{\hat{F}}^{z\mu }+\Psi
_{\nu }^{B}\,\mathbf{\hat{F}}^{\nu \mu }\right) \,.
\end{eqnarray}
Projecting it by $\mathbb{P}_{\pm }$ defined by (\ref{Ppm}) and applying the identities (\ref{PPsi}) and (\ref{P5}) from  Appendix \ref{gammaconventions}, we find
\begin{eqnarray}
0 &=&\epsilon ^{\mu \nu \lambda }\left( \mp \ii V_{\ \ z}^{3}\Gamma _{5}\hat\rr_{\mp A\nu \lambda }-2V_{\ \ \nu }^{i}\Gamma _{i}\Gamma _{5}\hat\rr_{\pm Az\lambda }+\frac{\ii }{2}\,\epsilon _{AB}\mathbf{\hat{F}}_{\nu \lambda }\Gamma _{5}\Psi _{\mp z}^{B}+\ii\,\epsilon _{AB}\mathbf{\hat{F}}_{z\nu }\Gamma _{5}\Psi _{\mp \lambda }^{B}\right)   \notag \\
&&+ e\,\epsilon _{AB}\left( \Psi _{\pm z}^{B}\,\mathbf{\hat{F}}^{z\mu }+\Psi
_{\pm \nu }^{B}\,\mathbf{\hat{F}}^{\nu \mu }\right) \,.
\end{eqnarray}
Now we can use eqs.~(\ref{F,Rho}), (\ref{e}),
(\ref{holo gauge}) and (\ref{asympt_field}), to obtain the equation
expressed in terms of the auxiliary quantities with known asymptotic
behaviour,
\begin{eqnarray}
0 &=&\left( \dfrac{z}{\ell }\right) ^{\pm \frac{1}{2}-1}\,\epsilon ^{\mu \nu
\lambda }\left( \mp \ii\,\Gamma _{5}\Xi _{\nu \lambda \mp }^{A}+2\hat{E}_{\ \ \nu }^{i}\Gamma _{i}\Gamma _{5}\Xi _{\lambda \pm }^{A}\right) \nonumber\\
&&+\left(\dfrac{z}{\ell }\right) ^{\pm \frac{1}{2}}\epsilon _{AB}\,\left( -\ii \,\epsilon ^{\mu \nu \lambda }\Gamma _{5}\varphi _{\mp \lambda
}^{B}+ e_{3}\,g^{\mu \nu }\varphi _{\pm z}^{B}\right) \mathbf{\hat{F}}_{ \nu z}  \nonumber \\
&&+\left( \dfrac{z}{\ell }\right) ^{\mp \frac{1}{2}}\epsilon _{AB}\left(
\frac{\ii }{2}\,\epsilon ^{\mu \nu \lambda }F_{\nu \lambda }\Gamma
_{5}\varphi _{\mp z}^{B}+e_{3}\,F^{\nu \mu }\varphi _{\pm \nu }^{B}\right)
\,.
\end{eqnarray}
All tensors appearing above are finite, except $\mathbf{\hat{F}}_{\mu z}$ and $\Xi _{\mu +}^{A}$. With this at hand, we identify the leading orders of the $z$-component of the gravitini equations of motion (looking at the two projections separately). By requiring the most divergent terms to vanish (that are $\left( \ell /z\right) ^{5/2}$
and $\left( \ell /z\right) ^{3/2}$ in the two chiralities), we get
\begin{eqnarray}
0 &=&\epsilon ^{ijk}\left( A_{(-1)z}\,\epsilon _{AB}\,\Gamma _{i}\varphi
_{(0) +\mu }^{B}E_{\ j}^{\mu}+2\ii\Gamma_{ij}\varphi_{A(0)-z}\right) \,,  \nonumber \\
0 &=&\epsilon ^{\mu \nu \lambda }\left( \ii\,\Xi _{(0)\nu \lambda+}^{A}-2E_{ \ \nu }^{i}\Gamma _{i}\Xi _{(0)\lambda -}^{A}\right) \label{Gravi_leading} \\
&&+\epsilon _{AB}\,\left( -\ii\,\epsilon ^{\mu \nu \lambda }\varphi_{(0)+\lambda }^{B}+ e_{3(0)}\,g_{(0)}^{\mu \nu }\Gamma _{5}\varphi_{(0)-z}^{B}\right) \left( \partial _{\nu}A_{(-1)z}-2\epsilon _{AC}\,\overline{\varphi }_{(0)+\nu }^{A}\varphi _{(0)-z}^{C}\right) \nonumber \,,
\end{eqnarray}
where we multiplied the equations by $\Gamma _{5}$. Since $\partial _{\nu
}A_{(-1)z}$ is correlated with $\varphi _{(0)-z}^{A}$ through the condition (\ref{dA}), it can be used in the second equation.

It turns out that we can solve the gauge-fixing functions from the first equation in (\ref{Gravi_leading}), in terms of the dynamic fields. Contracting it by $\epsilon _{ki^{\prime }j^{\prime }}$, it acquires an equivalent form
\begin{equation}
0=-A_{(-1)z}\,\epsilon _{AB}\,
E_{\ [i}^{\mu }\Gamma _{j]}\varphi _{(0)+\mu }^{B} + 2 \ii \, \Gamma _{ij}\varphi _{A(0)-z}\,.
\end{equation}
We can contract the above equation by $\Gamma ^{ij}$ and use the contractions of the gamma matrices (\ref{For_App}), which in this case become
$\Gamma _{i}\Gamma ^{i}=3$, $\Gamma ^{ij}\Gamma _{j}=2\Gamma ^{i}$ and $\Gamma ^{ij}\Gamma _{ij}=-6$. As a result, we obtain a solution which relates the gauge fixing $\varphi _{(0)-z}^{A}$ with the gauge fixing $A_{(-1)z}$,
\begin{equation}
\varphi _{(0)-z}^A=\frac{\ii }{6}\,A_{(-1)z}\epsilon ^{AB}\,\Gamma ^{i}\varphi _{B(0)+\mu }E_{\ i}^{\mu }\,. \label{solution}
\end{equation}
Then second equation in (\ref{dA}) becomes a linear differential equation in $A_{(-1)z}$.
One possible solution is $A_{(-1)z}=0$ that, from eq.~(\ref{solution}), yields $\varphi _{(0)-z}^{A}=0$. On the other hand, when $A_{(-1)z}\neq 0$, we can solve $\varphi^A_{(0)+\mu}$ from the first equation in (\ref{Gravi_leading}) as
\begin{equation}\label{varphi0pmexpr}
A_{(-1)z}\varphi_{(0)+\mu}^A=2\ii E_{\ \mu }^{i}\Gamma _{i}\varphi_{(0)-z}^{B}\epsilon _{AB} \,,
\end{equation}
and the differential equation becomes
\begin{equation}
A_{(-1)z}\partial _{\mu }A_{(-1)z}=2\mathrm{i\,}E_{k\mu }\overline{\varphi }_{(0)-z}^{A}\left( 2\Gamma ^{k}+\epsilon ^{ijk}\Gamma _{5}\Gamma _{ij}\right) \varphi _{(0)-z}^{A}=0\,,
\end{equation}
where the last zero is due to antisymmetry of the fermionic bilinears, namely $\overline{\varphi }_{(0)-z}^{A}\Gamma ^{k}\varphi _{(0)-z}^{A}\equiv0$ and $\overline{\varphi }_{(0)-z}^{A}\Gamma _{5}\Gamma _{ij}\varphi
_{(0)-z}^{A}\equiv 0$ so that each term in the sum vanishes independently.
The only solution of the above equation  is $A_{(-1)z}=const$.

Moreover, as previously shown in the main text, we can choose a particular solution with
\begin{equation}\label{varphi1pmzero}
\varphi_{A(1)+\mu} = 0 \,.
\end{equation}
Consequently, taking $A_{(-1)z}=0$ and plugging \eqref{varphi1pmzero} into the last equation in (\ref{dA}), we are left with $A_{(0)z}=0$. On the other hand, if we take $A_{(-1)z} \neq 0$ and use \eqref{varphi0pmexpr} and \eqref{varphi1pmzero} into the last equation of (\ref{dA}), we obtain
\begin{equation}
  A_{(0)z} E^\lambda_{\ k} \, \overline{\varphi}^A_{(0)-z} \Gamma^k \varphi^A_{(0)-z} = 0 \,,
\end{equation}
which is identically satisfied since $\overline{\varphi}^A_{(0)-z} \Gamma^k \varphi^A_{(0)-z}=0$. In particular, this means that, in this case, the last equation in (\ref{dA}) is solved by \eqref{varphi0pmexpr} and \eqref{varphi1pmzero}, without forcing $A_{(0)z}$ to vanish.\footnote{Note that one could consistently assume that the relation of proportionality between $\varphi^A_{-z}$ and $A_z$ given by eq.~(\ref{solution}) holds at all orders, in the neighborhood of the boundary, imposing the stronger condition
\begin{equation}\label{strongercond}
\varphi_{(n)-z}^{A}=\frac{\ii }{6}\,A_{(n-1)z}\epsilon ^{AB}\,\Gamma^{i} \, E_{\ i}^{\mu } \,\varphi _{B(0)+\mu }\,,\quad \forall \,n \,,
\end{equation}
that is equivalent to
\begin{equation}\label{strongercondnew}
\varphi_{-z}^{A}=\frac{\ii }{6}\,A_{z}\epsilon ^{AB}\,\Gamma^{i} \, E_{\ i}^{\mu } \,\varphi _{B(0)+\mu } \,.
\end{equation}
One can then prove that, considering the divergent terms in the $z/\ell$ expansion of the outer components ($\hat{\tau}=z$) of the gravitini equations \eqref{gravitinieom4d}, that is $E^i_{\ [\mu} \Gamma_i \hat{\boldsymbol{\rho}}_{(-1/2)+A\nu] z}=0$ and in particular  using \eqref{strongercond} in the equation for $\hat{\boldsymbol{\rho}}_{(-1/2)+A\mu z}$ in \eqref{rho+scft}, one obtains
\begin{equation}
 \Gamma_i E^i_{\ [\mu} \left( A_{(-1)z} \epsilon_{AB} - 2\,\delta_{AB} \right) \varphi_{B(1)+\nu]} = 0 \,,
\end{equation}
which enforces the condition \eqref{varphi1pmzero}
to hold also in the case $\hat{A}_z \neq 0$, $\Psi_{z-} \neq 0$. If we now take $A_{(-1)z}=0$ and plug \eqref{varphi1pmzero} into the last equation of (\ref{dA}), we can see that, in this case, $A_{(0)z}=0$, $\varphi^A_{(1)+\mu}=0$ is actually the only solution to the aforesaid equation.}

Summing up the results, the following gauge fixings for $A_z$ and $\varphi_{-z}^A$  are allowed:
\begin{eqnarray}
A_{(-1)z} &=&0\,,\qquad A_{(0)z}=0\,,\qquad \varphi _{(1)+\mu }^{A}=0\,,\qquad \varphi _{(0)-z}^{A}=0\,,  \notag \\
A_{(-1)z} &=&0\,,\qquad A_{(0)z}\neq 0\,,\qquad \varphi _{(1)+\mu }^{A}=\frac 12\,A_{(0)z}\varphi _{(0)+\mu }^{B}\epsilon _{AB}\,,\quad \varphi
_{(0)-z}^{A}=0\,, \label{A(-1)z} \\
A_{(-1)z} &=&const\,,\quad A_{(0)z}\neq 0\,,\quad \, \varphi _{(1)+\mu}^{A}=0\,,\qquad \varphi _{(0)-z}^{A} =\frac{\ii}{6} \, A_{(-1)z}\Gamma^{\mu } \varphi^B_{(0)+\mu}\epsilon _{AB}\,,  \notag
\end{eqnarray}
where the first line can be seen as a special case of the general solution given in the second line. If one imposes the condition   $\Gamma^{\hat{\mu}} \Psi_{\hat{\mu}}=0$ as in \cite{Amsel:2009rr}, then eq.~(\ref{solution}) implies $\psi_{-z}=0$ and therefore $A_{(-1)z}=0$ as the only solution.

In this text, we mostly focus on the case $\varphi _{(0)-z}^{A}=0$. Then the gauge-fixing function $\Psi _{-z}^{A}$ becomes subleading and can be safely set to zero at all orders, as suggested by eq.~(\ref{strongercond}).

At the end, let us recall that, in our approach, the gauge-fixing functions are invariant under the gauge transformations ($\delta \hat{A}_{z}=0$).
Thus, the above solutions are consistent because, since $A_{(-1)z}$ is constant, it also implies $\delta A_{(-1)z}=0$ for the asymptotic transformations.

%%%%%%%%%%%%%%%%%%%%%%%%%%%%%%%%%%%%%%%%%%%%%%%%%%%%%%%%%%%%

\section{The rheonomic parametrizations}\label{appC}

In this section we present the asymptotic expansion of the rheonomic parametrizations $\tilde R^{ab}_{\ \ cd}$, $\tilde\rho^A_{ab}$ and $\tilde F_{ab}$. The procedure is the one described in the main text and the applied gauge fixing corresponds to $A_{(-1)z}=0$ and $\Psi^A_{z-}=0$.

We start from the {graviphoton} field strength
\begin{align}
{\bf\hat F}=\diff \hat A - \,\overline{\Psi}{}_A\Psi_B\epsilon^{AB}=\tilde F_{ab} V^a V^b \,.
\end{align}
By expanding both sides of this equation onto the basis $\diff x^{\hat\mu} \wedge \diff x^{\hat\nu}$, one can derive the explicit expression of the rheonomic parametrizations
\begin{align}\nonumber
\tilde F_{ij}&=\left(\frac{z}{\ell}\right)^3E^\mu_{[i}E^\nu_{j]}\left(\partial_\mu{A}_{(1)\nu}-2\epsilon_{AB}\overline\psi^A_{\mu+}\zeta^B_{\nu-}-2\epsilon_{AB}\overline\zeta^A_{\mu+}\psi^B_{\nu-}\right)+\mathcal O(z^4) \,,\\
2\tilde F_{i3}&=-\frac{1}{\ell} \left(\frac{z}{\ell}\right)^2{A}_{(1)\mu} E^\mu_i+\left(\frac{z}{\ell}\right)^3\left(\partial_\mu A_{(1)z}-\frac{2}{\ell}A_{(2)\mu}+2\epsilon_{AB}\overline\psi^A_{z+}\psi^B_{\mu-}\right)E^\mu_i+\mathcal O(z^4) \,,
\end{align}
where we have used that ${\bf\hat F}_{\mu\nu}=\mathcal O(z)$.\\ \par

We now focus on the {super}curvature of the gravitino {and conformino,}
\begin{align}\nonumber
\hat\rr^A&=\diff \Psi^A +\frac 14 \Gamma_{ab}\hat\omega^{ab}\Psi^A- \frac 1{ 2\ell}\hat A \epsilon^{AB} \Psi_B-    \frac {\ii}{2\ell}\Gamma_a \Psi^A V^a\\
&=\tilde\rho^A_{ab} V^a V^b - {\frac{\ii}{{2}}} \Gamma^a \Psi_B V^b \tilde F_{ab} \epsilon^{AB} - {\frac{1}{{4}}} \Gamma_5 \Gamma^a \Psi_B V^b \tilde F^{cd} \epsilon^{AB} \epsilon_{abcd}
\end{align}
and expand this relation onto the basis $\diff x^{\hat\mu}\wedge \diff x^{\hat\nu}$ to obtain
\begin{align}\nonumber
\tilde\rho^A_{ij+}&=\left(\frac{z}{\ell}\right)^{\frac{5}{2}}E^\mu_{[i}E^\nu_{j]}\bigg(\nabla_\mu\zeta^{A}_{\nu+}+\frac{\rm i}{\ell}E^k_\mu \gamma_k \zeta^{A}_{\nu-}+\frac{1}{4}{\omega}{}_{ (1) \mu}^{kl}\gamma_{kl}\psi^A_{\nu+}{-\frac{1}{{4}\ell}A_{(1)\mu} \psi_{\nu+B} \epsilon^{AB} }\\ \nonumber
&+{\frac{\ii}{{4}\ell}}\epsilon_{lmn}\gamma^l\psi_{B\mu+}E^m_\nu E^{\rho n}{A}_{(1)\rho}\epsilon^{AB} \bigg)+\mathcal O(z^{7/2})\,,\\ \nonumber
2\tilde\rho^A_{i3+}&=-\frac{1}{\ell} \left(\frac{z}{\ell}\right)^{\frac{3}{2}}E^\mu_i\zeta^A_{\mu+}+\left(\frac{z}{\ell}\right)^{\frac{5}{2}}E^\mu_i\bigg(\nabla_\mu\psi^A_{z+}-\frac{1}{4}w^{jk}_{(0)}\gamma_{jk}\psi^A_{\mu+}+\frac{1}{2\ell}\epsilon^{AB}A_{(1)z}\psi_{B\mu+}\\ \nonumber
&-\frac{2}{\ell}\Pi^A_{\mu+}\bigg)+\mathcal O(z^{7/2})\,,\\ \nonumber
\tilde\rho^A_{ij-}&=\left(\frac{z}{\ell}\right)^{\frac{5}{2}} E^\mu_{[i}E^\nu_{j]} \left(\nabla_\mu\psi^A_{\nu-}+\frac{\ii\ell}{2} \mathcal{S}^k_{\ \mu} \gamma_k\psi^A_{\nu+}\right)+\mathcal O(z^{7/2})\,,\\
2\tilde\rho^A_{i3-}&=-\frac 1\ell \left(\frac{z}{\ell}\right)^{\frac{5}{2}}E^\mu_i\left(\zeta^A_{\mu-}+\frac{1}{4}\epsilon^{AB}\gamma^j\psi_{B\mu+}{A}_{(1)\nu} E^\nu_j\right)+\mathcal O(z^{7/2})\,,
\end{align}
where we used $\hat\rr^A_{\mu\nu}=\mathcal O(z^{1/2})$.
This result allows to compute the spinor-tensor
\begin{equation}\label{thetadef}
    \Theta^{ab|c}_A=-2\ii\Gamma^{[a}\tilde\rho^{b]c}_A+\ii\Gamma^c\tilde\rho^{ab}_A
\end{equation}
{as an intermediate step necessary to find} the remaining parametrizations. In particular, we obtain
\begin{align}\nonumber
\Theta^{ij|k}_{A+}&=\ii\left(\frac{z}{\ell}\right)^{\frac{5}{2}}\left(-\gamma^iE^{[j\mu}E^{k]\nu}+\gamma^jE^{[i\mu}E^{k]\nu}+\gamma^k E^{[i\mu}E^{j]\nu}\right)\bigg(\nabla_\mu\psi_{A\nu-}+\frac{\ii\ell}{2} \mathcal{S}^l{}_\mu \gamma_l\psi_{A\nu+}\bigg)\\ \nonumber
&+\mathcal O(z^{7/2})\,,\\ \nonumber
\Theta^{ij|3}_{A+}&=-{\frac{\ii}{\ell}}\left(\frac{z}{\ell}\right)^{\frac{5}{2}}\gamma^{[i}E^{j]\mu}\left(\zeta_{A\mu-}+\frac{\ii}{4}\epsilon_{AB}\gamma^k\psi^B_{\mu+}{A}_{(1)\rho} E^\rho_k\right)-\left(\frac{z}{\ell}\right)^{\frac{5}{2}}E^{[i\mu}E^{j]\nu}\bigg(\nabla_\mu\zeta_{A\nu+}\\ \nonumber
&+\frac{\rm i}{\ell}E^k_\mu\gamma_k \zeta_{A\nu-}+ \frac{1}{4}{\omega}{}_{(1) \mu}^{kl}\gamma_{kl}\psi_{A\nu+}{-\frac{1}{{4}\ell}A_{(1)\mu} \psi_{\nu+B} \epsilon^{AB} }+\frac{\ii}{4\ell}\epsilon_{klm}\gamma^k\psi^B_{\mu+}E^l_\nu E^{\rho m}{A}_{(1)\rho}\epsilon_{AB} \bigg)\\ \nonumber
&+\mathcal O(z^{7/2})\,,\\ \nonumber
\Theta^{i3|j}_{A+}&={\frac{\ii}{\ell}}\left(\frac{z}{\ell}\right)^{\frac{5}{2}} \gamma^{(i}E^{j)\mu} \left(\zeta_{A\mu-}+ \frac{\ii}{4} \epsilon_{AB}\gamma^k \psi^B_{\mu+}{A}_{(1)\nu} E^\nu_k\right)-\left(\frac{z}{\ell}\right)^{\frac{5}{2}}E^{[i\mu}E^{j]\nu}\bigg(\nabla_\mu\zeta_{A\nu+}\\ \nonumber
&+\frac{\rm i}{\ell}E^k_\mu\gamma_k\zeta_{A\nu-}+ \frac{1}{4}{\omega}{}_{(1) \mu}^{kl}\gamma_{kl}\psi_{A\nu+}{-\frac{1}{{4}\ell}A_{(1)\mu} \psi_{\nu+B} \epsilon^{AB} }+\frac{\ii}{{4\ell}}\epsilon_{klm}\gamma^k\psi^B_{\mu+}E^l_\nu E^{\rho m}{A}_{(1)\rho}\epsilon_{AB} \bigg)\\ \nonumber
&+\mathcal O(z^{7/2})\,,\\ \nonumber
\Theta^{i3|3}_{A+}&=-\frac{1}{\ell} \left(\frac{z}{\ell}\right)^{\frac{3}{2}} \zeta_{A\mu+}E^{\mu i}+\left(\frac{z}{\ell}\right)^{\frac{5}{2}}E^{i\mu}\bigg(\nabla_\mu\psi_{Az+}-\frac{1}{4}w^{jk}_{(0)}\gamma_{jk}\psi_{A\mu+}+\frac{1}{2\ell}\epsilon_{AB}A_{(1)z}\psi^B_{\mu+}\\ \nonumber
&-\frac{2}{\ell}\Pi_{A\mu+}\bigg)+\mathcal O(z^{7/2})\,, \\ \nonumber
\Theta^{ij|k}_{A-}&=\ii\left(\frac{z}{\ell}\right)^{\frac{5}{2}}\left(-\gamma^iE^{[j\mu}E^{k]\nu}+\gamma^jE^{[i\mu}E^{k]\nu}+\gamma^k E^{[i\mu}E^{j]\nu}\right)\bigg(\nabla_\mu\zeta_{A\nu+}+\frac{\rm i}{\ell}E^l_\mu\gamma_l\zeta_{A\nu-}\\ \nonumber
&+\frac{1}{4}{\omega}{}_{(1) \mu}^{lm}\gamma_{lm}\psi_{A\nu+}{-\frac{1}{{4}\ell}A_{(1)\mu} \psi_{\nu+B} \epsilon^{AB} }+\frac{\ii}{{4}\ell}\epsilon_{lmn}\gamma^l\psi^B_{\mu+}E^m_\nu E^{\rho n}{A}_{(1)\rho}\epsilon_{AB} \bigg)+\mathcal O(z^{7/2})\,,\\ \nonumber
\Theta^{ij|3}_{A-}&=-\frac{\ii}{\ell} \left(\frac{z}{\ell}\right)^{\frac{3}{2}} \gamma^{[i}E^{j]\mu}\zeta_{A\mu+}+ \ii \left(\frac{z}{\ell}\right)^{\frac{5}{2}} \gamma^{[i}E^{j]\mu}\bigg(\nabla_\mu\psi_{Az+}-\frac{1}{4}w^{kl}_{(0)}\gamma_{kl}\psi_{A\mu+}\\ \nonumber
&+\frac{1}{2\ell}\epsilon_{AB}A_{(1)z}\psi^B_{\mu+}-\frac{2}{\ell}\Pi_{A\mu+}\bigg)+ \left(\frac{z}{\ell}\right)^{\frac{5}{2}}E^{[i\mu}E^{j]\nu} \bigg(\nabla_\mu\psi_{A\nu-}+\frac{\ii\ell}{2} \mathcal{S}^k{}_\mu \gamma_k\psi_{A\nu+}\bigg)+\mathcal O(z^{7/2})\,, \\ \nonumber
\Theta^{i3|j}_{A-}&=\frac{\ii}{\ell} \left(\frac{z}{\ell}\right)^{\frac{3}{2}} \gamma^{(i}E^{j)\mu} \zeta_{A\mu+}- \ii\left(\frac{z}{\ell}\right)^{\frac{5}{2}} \gamma^{(i}E^{j)\mu} \bigg(\nabla_\mu\psi_{Az+}- \frac{1}{4}w^{kl}_{(0)}\gamma_{kl}\psi_{A\mu+}+\frac{1}{2\ell}\epsilon_{AB}A_{(1)z}\psi^B_{\mu+}\\ \nonumber
&-\frac{2}{\ell}\Pi_{A\mu+}\bigg)+\left(\frac{z}{\ell}\right)^{\frac{5}{2}}E^{[i\mu}E^{j]\nu}\bigg(\nabla_\mu\psi_{A\nu-}+ \frac{\ii\ell}{2} \mathcal{S}^k{}_\mu \gamma_k\psi_{A\nu+}\bigg)+\mathcal O(z^{7/2})\,,\\ \nonumber
\Theta^{i3|3}_{A-}&={\frac 1\ell}\left(\frac{z}{\ell}\right)^{\frac{5}{2}}E^{i\mu}\left(\zeta_{A\mu-}+\frac{\ii}{4}\epsilon_{AB}\gamma^j\psi^B_{\mu+}{A}_{(1)\rho} E^\rho_j\right)+\mathcal O(z^{7/2}) \,.
\end{align}
 We are now ready to compute the rheonomic parametrization of the {super}curvature $\hat \RR^{ab}$. Since
\begin{align}\nonumber
\hat{\RR}^{ab}&=\diff \hat\omega^{ab} +\hat\omega^{ac}\hat\omega_c{}^b- \frac{1}{\ell^2} V^a V^b - \frac 1{2\ell} \overline{\Psi}{}^A \Gamma^{ab}\Psi_A\\
&={\tilde R^{ab}}_{\ \ cd} V^c V^d - \overline{\Theta}{}^{ab}_{A \vert c} \Psi_A V^c - {\frac{1}{{2}}}\overline{\Psi}{}_A \Psi_B \epsilon_{AB} \tilde F^{ab} - \frac{\ii}{{4}} \epsilon^{abcd} \overline{\Psi}{}_A \Gamma_5 \Psi_B \epsilon_{AB} \tilde F_{cd} \,,
\end{align}
{applying} the usual procedure yields
\begin{align}\nonumber
\tilde R^{i3}{}_{jk}&=\frac{\ii}{2\ell}\left(\frac{z}{\ell}\right)^2E^\mu_{[j}E^\nu_{k]}\overline\psi^A_{\mu+}\gamma^i\zeta_{A\nu+}+\frac{\ii}{2\ell}\left(\frac{z}{\ell}\right)^2E^\mu_{[j}E^\nu_{k]}\overline\psi^A_{\mu+}\gamma^l\zeta_{A\rho+}E_{l\nu}E^{i\rho}\\ \nonumber
&+\frac{1}{\ell}\left(\frac{z}{\ell}\right)^3E^\mu_{[j}E^\nu_{k]}\bigg\{-\mathcal D_\mu\tilde S^i_\nu+\omega^i_{(2)l\mu}E^l_\nu-\ii\overline\Pi^A_{\mu+}\gamma^i\psi_{A\nu+}-\frac{\ii}{2}\overline\zeta^A_{\mu+}\gamma^i\zeta_{A\nu+}\\ \nonumber
&+\frac{\ii}{2}\overline\psi^A_{\mu-}\gamma^i\psi_{A-\nu}+\overline\psi^A_{\mu+} E_{l \nu} \bigg[-\ii\gamma^{(i}E^{l)\rho}\bigg(\nabla_\rho\psi_{Az+}-\frac{1}{4}w^{mn}_{(0)}\gamma_{mn}\psi_{A\rho+}\\ \nonumber
&+\frac{1}{2\ell} \epsilon_{AB}A_{(1)z}\psi^B_{\rho+}-\frac{2}{\ell}\Pi_{A\rho+}\bigg)+E^{[i\rho}E^{l]\sigma}\bigg(\nabla_\rho\psi_{A\sigma-}+\frac{\ii\ell}{2} \mathcal{S}^m{}_\rho \gamma_m\psi_{A\sigma+}\bigg)\bigg] \bigg\} +\mathcal O(z^4)\,,\\ \nonumber
2\tilde R^{i3}{}_{j3}&=\left(\frac{z}{\ell}\right)^3E^\mu_j\bigg\{-\frac{1}{\ell}w^i_{(1)k}E^k_{\ \mu} +\frac{1}{\ell^2}\left(4\tilde{\tau}^i_\mu-\tau^i_\mu\right)-\frac{\ii}{\ell}\overline\zeta^A_{\mu+}\gamma^i\psi_{Az+}-\frac{\ii}{\ell}\overline\psi^A_{\mu+}\gamma^i\zeta_{Az+}\\
&+\frac{1}{\ell}\overline\psi^A_{\mu-}\zeta_{A\nu+}E^{\nu i}-\overline\psi^A_{\mu+}E^{i\nu}\left(\frac{1}{\ell}\zeta_{A\nu-}+{\frac{\ii}{{4}\ell}}\epsilon_{AB}\gamma^l\psi^B_{\nu+}{A}_{(1)\rho} E^\rho_l\right) \bigg\} +\mathcal O(z^4)\,,\\ \nonumber
\tilde R^{ij}{}_{kl}&=\left(\frac{z}{\ell}\right)^3E^\mu_{[k}E^\nu_{l]}\bigg\{\partial_\mu{\omega}{}^{ij}_{(1)\nu}+{\omega}{}^{i}_{(1)m\mu}\omega^{mj}{}_\nu+\omega^{i}{}_{m\mu}{\omega}{}^{mj}_{(1)\nu}-\frac{2}{\ell^2}(\tau^{[i}_\mu+2\tilde\tau^{[i}_\mu)E^{j]}_\nu\\ \nonumber
&-\frac{1}{\ell}\left(\overline\psi^{A}_{\mu+}\gamma^{ij}\zeta_{A\nu-}+\overline\zeta^{A}_{\mu+}\gamma^{ij}\psi_{A\nu-}\right)+\ii E_{m\nu} \overline\psi^A_{\mu+} \bigg(-\gamma^i E^{[j\rho}E^{m]\sigma}+\gamma^j E^{[i\rho}E^{m]\sigma}\\ \nonumber
&+\gamma^m E^{[i\rho}E^{j]\sigma}\bigg)\bigg(\nabla_\rho\zeta_{A\sigma+}+\frac{\rm i}{\ell}E^n_\rho\gamma_n\zeta_{A\sigma-}+\frac{1}{4}{\omega}{}_{(1) \rho}^{np}\gamma_{np} \psi_{A\sigma+}\\ \nonumber
&{-\frac{1}{{4}\ell}A_{(1)\rho} \psi^B_{\sigma+} \epsilon_{AB} }+\frac{\ii}{{4}\ell}\epsilon_{npq}\gamma^n\psi^B_{\rho+}E^p_\sigma E^{\lambda q}{A}_{(1)\lambda}\epsilon_{AB} \bigg) \bigg\}+\mathcal O(z^4)\,,\\ \nonumber
2\tilde R^{ij}{}_{k3}&=-\left(\frac{z}{\ell}\right)^2E^\mu_k\left(\frac{1}{\ell}{\omega}{}_{(1) \mu}^{ij}-\frac{\ii}{\ell} \overline\psi^A_{\mu+}\gamma^{[i}E^{j]\nu}\zeta_{A\nu+} \right)\\ \nonumber
&+\left(\frac{z}{\ell}\right)^3 E^\mu_k \bigg\{\partial_\mu w^{ij}- \frac{2}{\ell} \omega^{ij}_{(2) \mu}+ \omega^{i}{}_{l\mu} w^{lj}_{(0)}- w^i{}_{l} \omega^{lj}_{\ \ \mu}+ \frac{1}{\ell} \left( E^i_\mu w^j_{(0)}- w^i_{(0)}E^j_\mu\right)\\ \nonumber
&+\frac{1}{\ell} \overline\psi^A_{z+} \gamma^{ij} \psi_{A\mu-}-\overline\psi^A_{\mu+}\bigg[\ii\gamma^{[i}E^{j]\nu} \bigg(\nabla_\nu\psi_{Az+}-\frac{1}{4}w^{lm}_{(0)}\gamma_{lm}\psi_{A\nu+}+\frac{1}{2\ell}\epsilon_{AB}A_{(1)z}\psi^B_{\nu+}\\ \nonumber
&-\frac{2}{\ell}\Pi_{A\nu+}\bigg)+ E^{[i\nu}E^{j]\rho}\bigg(\nabla_\nu\psi_{A\rho-}+\frac{\ii\ell}{2} \mathcal{S}^l{}_\nu \gamma_l\psi_{A\rho+}\bigg)\bigg] \bigg\}+\mathcal O(z^4)\,.
\end{align}
{To obtain the above formulas,}
we used $\hat\RR^{ab}_{{\mu\nu}}=\mathcal O(z)$ and that the supertorsion is zero (see, in particular, \eqref{SuperT=0}).

%%%%%%%%%%%%%%%%%%%%%%%%%%%%

\end{document}